\documentclass[a4paper,11pt]{article}
\pdfoutput=1 % if your are submitting a pdflatex (i.e. if you have
\usepackage{jcappub} % for details on the use of the package, please
\usepackage{mathabx}
\usepackage{multirow}
\usepackage{slashed}
\usepackage{array}

\usepackage[T1]{fontenc} % if needed

\newcommand{\besc}{\beta_\mathrm{esc}}
\newcommand{\Eend}{E_{\mathrm{C}\nu\mathrm{B}}}
\newcommand{\mh}{m_{^3\mathrm{H}}}
\newcommand{\mhe}{m_{^3\mathrm{He}}}
\newcolumntype{M}[1]{>{\centering\arraybackslash}m{#1}}

\title{\boldmath Limits on the cosmic neutrino background}

\author[]{Martin Bauer and}
\author[]{Jack D. Shergold}

\affiliation[]{Institute for Particle Physics Phenomenology, Department of Physics,\\Durham University,\\Durham, UK}

\emailAdd{martin.m.bauer@durham.ac.uk}
\emailAdd{jack.d.shergold@durham.ac.uk}

\abstract{We present the first comprehensive discussion of constraints on the cosmic neutrino background (C$\nu$B) overdensity, including theoretical, experimental and cosmological limits for a wide range of neutrino masses and temperatures. Additionally, we calculate the sensitivities of future direct and indirect relic neutrino detection experiments and compare the results with the existing constraints, extending several previous analyses by taking into account that the C$\nu$B reference frame may not be aligned with that of the Earth. The Pauli exclusion principle strongly disfavours overdensities $\eta_\nu \gg 1$ at small neutrino masses, but allows for overdensities $\eta_{\nu}\lesssim 125$ at the KATRIN mass bound $m_{\nu} \simeq 0.8\,\mathrm{eV}$. On the other hand, cosmology strongly favours $0.2 \lesssim \eta_{\nu} \lesssim 3.5$ in all scenarios. We find that direct detection proposals are capable of observing the C$\nu$B without a significant overdensity for neutrino masses $m_{\nu} \gtrsim 50\,\mathrm{meV}$, but require an overdensity $\eta_{\nu} \gtrsim 3\times 10^5$ outside of this range. We also demonstrate that relic neutrino detection proposals are sensitive to the helicity composition of the C$\nu$B, whilst some may be able to distinguish between Dirac and Majorana neutrinos.}

\begin{document}
\maketitle
\flushbottom
\clearpage
\section{Introduction} 
Precision measurements of the Cosmic Microwave Background (CMB) underpin much of our current understanding of the evolution of the universe~\cite{Planck:2018vyg,WMAP:2012fli}. These measurements are soon to be complimented by those of the LISA space-based gravitational wave observatory~\cite{LISA:2017pwj}, which aims to detect the echoes of the Big Bang. Despite the remarkable achievements of modern cosmology, relic neutrinos from the Cosmic Neutrino Background (C$\nu$B) remain ever elusive, owing to their weakly interacting nature and low energy. As the C$\nu$B predates the CMB, its detection could give important insight into Big Bang nucleosynthesis (BBN), whilst simultaneously augmenting measurements made from the CMB. The successful detection of photons, gravitational waves and neutrinos from the early universe would truly signal the dawn of multi-messenger cosmology. 

\begin{figure}[tbp]
\centering
\includegraphics[width=.99\textwidth]{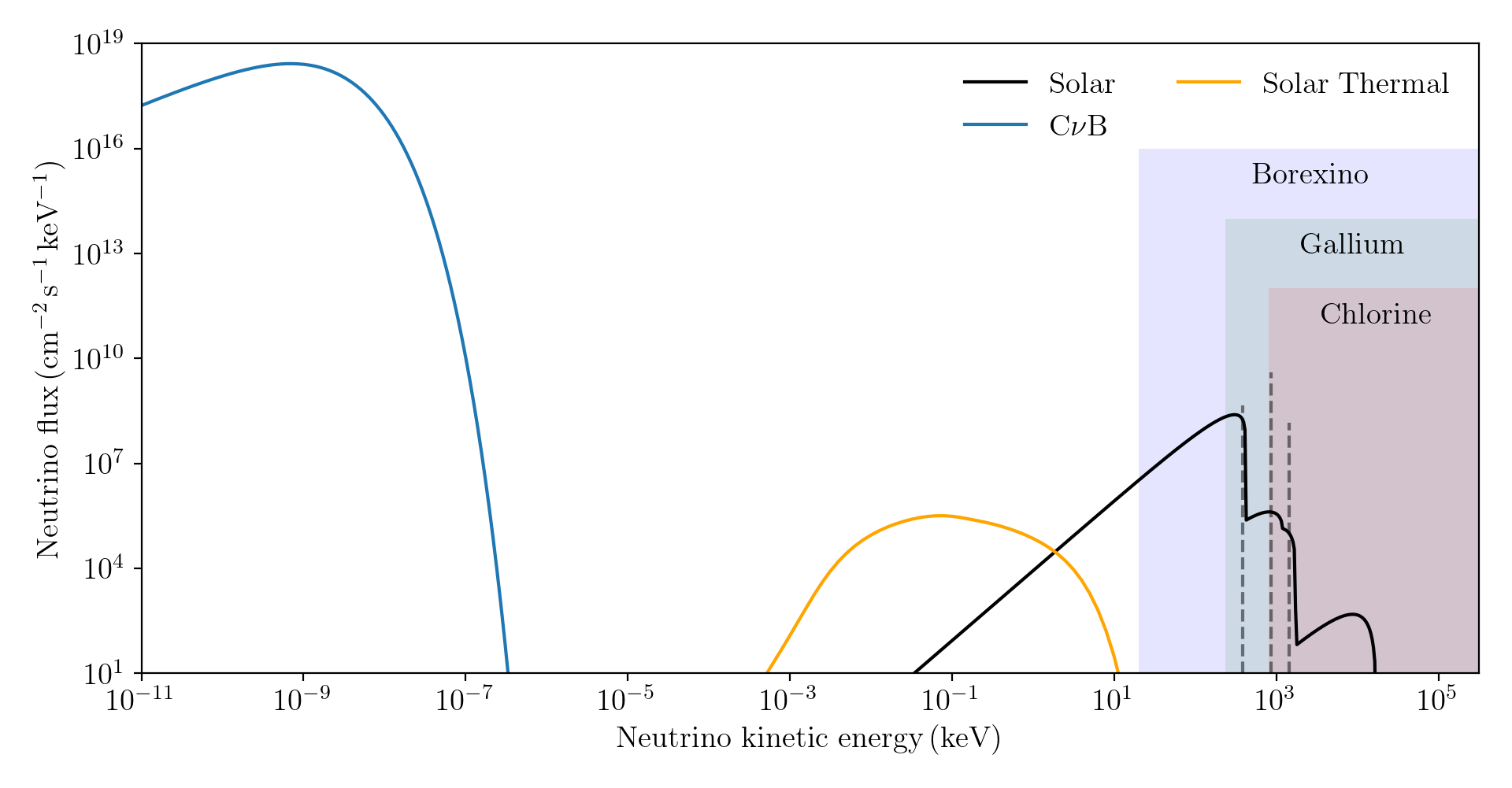}
\caption{\label{fig:nuFlux}Predicted electron neutrino flux from the C$\nu$B (blue), alongside those from nuclear (black) and thermal (orange) processes in the sun, assuming degenerate neutrino mass $m_\nu = 0.1\,\mathrm{eV}$. We also show the energy thresholds for neutrino detection at gallium and chlorine radiochemical experiments, alongside the threshold at Borexino~\cite{Borexino:2008gab}. For a comprehensive review of neutrino fluxes at all energies, see~\cite{Vitagliano:2019yzm}.}
\end{figure}

As shown in Figure~\ref{fig:nuFlux}, existing neutrino experiments have detection thresholds that are many orders of magnitude above the predicted C$\nu$B energy. Any experiment wishing to observe relic neutrinos therefore requires a complete re-imagination of neutrino detection. At present, there exist several proposals to detect the C$\nu$B using a wide range of techniques: capturing relic neutrinos on radioactive nuclei~\cite{Weinberg:1962zza, PTOLEMY:2018jst}; observing the annihilation of ultra-high energy cosmic ray neutrinos on the C$\nu$B at the $Z$-resonance~\cite{Eberle:2004ua}; exploiting neutrino capture resonances using an accelerator experiment~\cite{Bauer:2021uyj}; measuring tiny accelerations induced by elastic scattering of the relic neutrino wind on a test mass~\cite{Shergold:2021evs, Domcke:2017aqj, Duda:2001hd, Stodolsky:1974aq, Opher:1974drq,Lewis:1979mu,Shvartsman:1982sn,Smith:1983jj,Cabibbo:1982bb,Gelmini:2004hg,Ringwald:2004np,Vogel:2015vfa} and searching for modifications in atomic de-excitation spectra due to Pauli blocking~\cite{Yoshimura:2014hfa}. Many of these proposals require a local overdensity of neutrinos to make a discovery, the magnitude of which depends strongly on the properties of the C$\nu$B. In this paper we will attempt to place constraints on the present day local relic neutrino overdensity, where possible in a model independent way, exploring both the theoretical and experimental constraints for a range of neutrino masses and temperatures, as well as the constraints that could be set by each proposal to detect the C$\nu$B.

Several studies have already attempted to estimate the magnitude of the local relic neutrino overdensity using a variety of techniques. Estimates of the overdensity using simulations of relic neutrinos in the galactic gravitational field typically lie in the range $\eta_{\nu} \simeq 1-10$ in most conservative scenarios~\cite{Ringwald:2004np,deSalas:2017wtt,Mertsch:2019qjv,Yoshikawa:2020ehd}, assuming the standard cosmological evolution that we will present in Section~\ref{sec:thermalHist}. On the other hand, arguments based on the local baryon density predict overdensities as large as $\eta_{\nu} \simeq 10^3 - 10^6$~\cite{Lazauskas:2007da,Faessler:2014bqa}. These results depend on the choice of simulation method~\cite{deSalas:2019kpa}, as well the assumptions made regarding the density profile of the galaxy and evolution of the C$\nu$B, which could significantly differ from the standard scenario \textit{e.g.} in the presence of extra degrees of freedom coupling to neutrinos. As such, there is clearly a need for model independent constraints that can be placed on the C$\nu$B from theory and experiment. 

In Section~\ref{sec:existing}, we will explore the Pauli exclusion principle and the closely related Tremaine-Gunn bound, which provide theoretical upper limits on the C$\nu$B overdensity as a function of the neutrino mass and temperature. We will also calculate the bounds set by the existing experiment with the lowest neutrino energy threshold, Borexino. Cosmological constraints from Big Bang nucleosynthesis, the polarisation of the CMB, and the effective number of neutrino species, $N_\mathrm{eff}$, will be discussed in Section~\ref{sec:cosmology}. Throughout this work, we will label proposals to observe the C$\nu$B as either direct or indirect detection. Direct detection proposals directly observe the final state of an interaction with the C$\nu$B, which will be discussed in Section~\ref{sec:direct}. By contrast, indirect detection proposals are sensitive the effects of the C$\nu$B on other observable parameters, for example, signals that are reduced in a window of parameter space due to absorption by relic neutrinos. We will discuss indirect C$\nu$B detection proposals in Section~\ref{sec:indirect}. Finally, we will present our main results in Figures~\ref{fig:NH_future},~\ref{fig:IH_future} and~\ref{fig:temp_NH} and discuss them in Section~\ref{sec:discussion}, before concluding in Section~\ref{sec:conclusions}.
%The capability of each of these proposals depends on the neutrino mass, the strongest bound on which is currently set by the KATRIN experiment at $m_\nu \leq 0.8\,\mathrm{meV}$. In addition, many of these proposals require a significant local overdensity of relic neutrinos to make a discovery.  
%With the exception of the PTOLEMY experiment~\cite{PTOLEMY:2018jst}, all of these proposals lack the required sensitivity to probe the C$\nu$B, requiring either significant improvements to currently available technology or a local overdensity of relic neutrinos to make a discovery. 
%The latter of these remains an important question to those hunting for the C$\nu$B: \textit{what is the local density of relic neutrinos?} 

\section{Neutrino thermal history}\label{sec:thermalHist}

Neutrinos in the early universe remain in equilibrium with the Standard Model (SM) thermal bath through weak interactions, which proceed at a rate $\Gamma_{\nu} \sim G_F^2 T_{\nu}^5$, where $G_F$ is Fermi's constant and $T_{\nu}$ is the neutrino temperature. As the universe expands and cools, the rate of weak interactions slows to the point where the time between interactions is of order the age of the universe, $\Gamma_{\nu} \simeq H \sim  \sqrt{G_N} T_{\nu}^2$, where $H$ is the Hubble parameter and $G_N$ is the gravitational constant. Neutrinos therefore freeze-out at a temperature $T_\nu\simeq 1\,\mathrm{MeV}$, shortly before the temperature of the SM thermal bath drops sufficiently for electron-positron pair production to become kinematically unfavourable. The subsequent annihilation of electron-positron pairs must conserve entropy, reheating photons to a temperature $T_\gamma = (11/4)^{1/3}\,T_\nu$. Using the present day CMB temperature, $T_{\gamma,0} = 0.235\,\mathrm{meV}$~\cite{Fixsen:2009ug}, this relation gives a present day neutrino temperature $T_{\nu,0} \simeq 0.168\,\mathrm{meV}$. We additionally expect that in the absence of significant interactions or clustering since decoupling, relic neutrinos should follow a redshifted equilibrium distribution~\cite{Gnedin:1997vn},
\begin{equation}\label{eq:fermiDirac}
    f_{\nu_i}(|\vec{\widetilde p}_{\nu_i}|) = \frac{g_{\nu_i}}{\exp(|\vec{\widetilde p}_{\nu_i}|/T_{\nu_i})+1},
\end{equation}
with $\vec{\widetilde p}_{\nu_i}$ the momentum of neutrinos in the C$\nu$B reference frame\footnote{Going forward, we will reserve tildes for quantities specific to the C$\nu$B frame.}, whilst $g_{\nu_i}$ is the degeneracy which may itself be a function of the neutrino momentum. By integrating~\eqref{eq:fermiDirac} over all momenta at a temperature $T_{\nu_i} = T_{\nu,0}$, we find a present day neutrino number density per degree of freedom 
\begin{equation}\label{eq:eqScaling}
    n_{\nu,0} = \frac{1}{g_{\nu_i}}\int \frac{d^3\widetilde{p}_{\nu_i}}{(2\pi)^3} f_{\nu_i}(|\vec{\widetilde p}_{\nu_i}|) = \frac{3T_{\nu,0}^3 \zeta(3)}{4\pi^2} \simeq 56\,\mathrm{cm}^{-3},
\end{equation}
as well as a mean momentum $|\vec{\widetilde p}_{\nu,0}| \simeq 3.15\,T_{\nu,0}$. Combined with the results of neutrino oscillation experiments, which set lower bounds on the neutrino masses of $m_{\nu_2} \gtrsim 8.6\,\mathrm{meV}$ and $m_{\nu_3} \gtrsim 50.1\,\mathrm{meV}$ in the normal mass hierarchy (NH) and $m_{\nu_1} \gtrsim 49.9\,\mathrm{meV}$ and $m_{\nu_2} \gtrsim 50.6\,\mathrm{meV}$ in the inverted mass hierarchy (IH)~\cite{Esteban:2020cvm}, we conclude that at least two out of three mass eigenstates must be non-relativistic today provided that $T_\nu = T_{\nu,0}$.

Before continuing, we make two important observations following the arguments of~\cite{Long:2014zva}. First, whilst neutrinos are produced as flavour eigenstates, coherent superpositions of the mass eigenstates, they have long since decohered to mass eigenstates. Secondly, as helicity and chirality coincide for ultra-relativistic particles and only left-chiral neutrino fields exist in the SM, we expect that all neutrinos will be left-helicity at freeze-out. Further, since helicity is a good quantum number, all neutrinos should remain left-helicity until the present day provided that they do not interact or cluster significantly since decoupling. By a similar argument, there should be an equal abundance of right-helicity antineutrinos today and an absence of left-helicity antineutrinos. As a result, we expect that $g_{\nu_i} = 1$ in~\eqref{eq:fermiDirac}, and the predicted number densities for Dirac neutrinos today are\footnote{Here and in what follows, we will use the subscripts $i\in\{1,2,3\}$ and $\alpha \in \{e,\mu,\tau\}$ to denote quantities that differ between neutrino mass or flavour eigenstates respectively. Where appropriate, we will also use the subscript $s\in\{L,R\}$ to denote quantities differing between neutrinos with left or right helicity. Finally, we will use the superscripts $D$ or $M$ when referencing quantities specific to Dirac or Majorana neutrinos respectively, whilst $\sum_{\nu}$ is the instruction to sum over neutrinos and antineutrinos.} 

\begin{equation}\label{eq:diracAbundance}
    \begin{alignedat}{3}
    \widetilde n_{\nu}(\nu_{i,L}^D) &= n_{\nu,0}, \quad &\widetilde n_{\nu}(\nu_{i,R}^D) &\simeq 0, \\
    \widetilde n_{\nu}(\bar\nu_{i,L}^D) &\simeq 0, \quad &\widetilde n_{\nu}(\bar\nu_{i,R}^D) &= n_{\nu,0}.
    \end{alignedat}
\end{equation}

If instead neutrinos are Majorana fermions, then we are unable to distinguish between neutrino and antineutrino. In this case, the expected abundances are 
\begin{equation}\label{eq:majoranaAbundance}
    \widetilde n_{\nu}(\nu_{i,L}^M) = n_{\nu,0}, \quad \widetilde n_{\nu}(\nu_{i,R}^M) = n_{\nu,0}.
\end{equation}

After summing over helicity and mass eigenstates, the total predicted neutrino number density for both Dirac and Majorana neutrinos is $6n_{\nu,0}$. For the remainder of the paper, we will refer to the scenario with temperature $T_{\nu_i} = T_{\nu,0}$ and the ratios of abundances given in~\eqref{eq:diracAbundance} and~\eqref{eq:majoranaAbundance} as the standard scenario. 

Of course, it is entirely possible for the true neutrino number densities to differ from those presented in~\eqref{eq:diracAbundance} and~\eqref{eq:majoranaAbundance}. For example, the addition of extra degrees of freedom with late decays to neutrinos~\cite{Chacko:2018uke} alters the relation between the CMB and C$\nu$B temperature $T_{\nu_i}$, leading to a modified number density $(T_{\nu_i}/T_{\nu,0})^3\,n_{\nu,0}$. There is also no reason that the new temperature should be shared by all three neutrino mass eigenstates. If the extra degrees of freedom decay exclusively to a single mass eigenstate, then only those neutrinos will be reheated. Other scenarios including mass dependent clustering and neutrino decay could lead to similar situations in which the relic density differs on a per-eigenstate basis. For this reason, we will consider the overdensity $\eta_{\nu}=  n_{\nu}/n_{\nu,0}$ separately for each neutrino and antineutrino eigenstate and helicity, as well as for Dirac and Majorana neutrinos. This will be particularly important when considering the constraints from experiments looking to detect the C$\nu$B, whose sensitivities often differ depending on the properties of the neutrino being considered.
\subsection{Kinematics}\label{sec:kinematics}
In general, it cannot be assumed that the C$\nu$B reference frame coincides with that of the Earth. As such, the momentum distribution~\eqref{eq:fermiDirac} only applies in the C$\nu$B frame and we must necessarily transform the neutrino momentum into the Earth's reference frame to make accurate lab frame calculations. If the C$\nu$B is isotropic in its own reference frame, the momentum vector of any given relic neutrino is
\begin{equation}\label{eq:pCnuB}
    \vec{\widetilde p}_{\nu_i} = |\vec{\widetilde p}_{\nu_i}|
    \left(\begin{array}{c}
        \cos\widetilde\phi \sin\widetilde\theta\\
        \sin\widetilde\phi \sin\widetilde\theta\\
        \cos\widetilde\theta
    \end{array}\right),
\end{equation}
where $\widetilde\phi \in [0,2\pi]$ and $\widetilde\theta\in [0,\pi]$. Supposing that the Earth travels along the $z$-axis at speed $\beta_\Earth$ with respect to the C$\nu$B frame, the true lab frame momentum of any neutrino can be found through a simple Lorentz transformation
\begin{equation}\label{eq:labFrameSingle}
    \vec{p}_{\nu_i,\mathrm{true}} = 
    \left(\begin{array}{c}
        |\vec{\widetilde p}_{\nu_i}|\cos\widetilde\phi \sin\widetilde\theta\\
        |\vec{\widetilde p}_{\nu_i}|\sin\widetilde\phi \sin\widetilde\theta\\
        \gamma_\Earth(|\vec{\widetilde p}_{\nu_i}|\cos\widetilde\theta + \beta_\Earth \widetilde{E}_{\nu_i})
    \end{array}\right),
\end{equation}
where $\widetilde E_{\nu_i}$ is the energy of relic neutrinos in the C$\nu$B frame and $\gamma_\Earth$ is the Lorentz factor of the frame transformation. Unfortunately, as we cannot know the orientation of every neutrino in the C$\nu$B, it is difficult to perform calculations using~\eqref{eq:labFrameSingle}. Instead, we should use averaged quantities, however we must be careful when doing so.

\begin{figure}[tbp]
\centering 
\includegraphics[width=.45\textwidth]{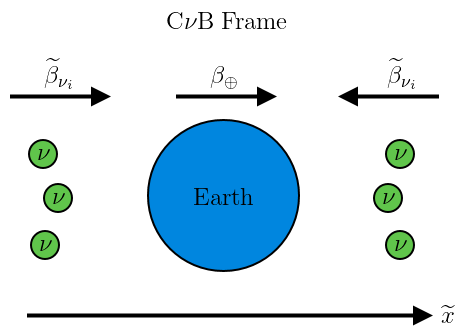}
\hfill
\includegraphics[width=.45\textwidth]{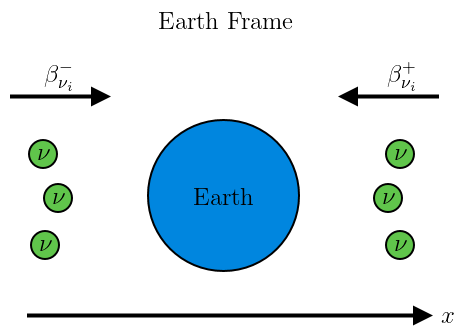}
\caption{\label{fig:frames}1-dimensional setup used to estimate the average momentum transfer by each scattering event, $\Delta p_{\nu_i}$, in~\cite{Shergold:2021evs}. Left: The Earth moves at velocity $\beta_\Earth$ relative to the C$\nu$B frame, where neutrinos have mean velocity $\widetilde \beta_{\nu_i}$. Right: In the Earth's reference frame, neutrinos move with velocity $\beta^+_{\nu_i}$ or $\beta^-_{\nu_i}$, generating a lab frame asymmetry in the C$\nu$B.}
\end{figure}

Consider, for example, the average momentum transfer $\Delta p_{\nu_i}$ to a test mass over several neutrino scattering events. In the C$\nu$B frame where relic neutrinos are isotropic, we expect that $\Delta p_{\nu_i} = 0$. However, in the laboratory frame, the relative motion of the Earth induces an asymmetry in the C$\nu$B, leading to a small momentum transfer proportional to $\beta_\Earth$. For clarity, we sketch the simple 1-dimensional setup in Figure~\ref{fig:frames}. This quantity, averaged over many scattering events, is therefore sensitive to the orientation of relic neutrinos. On the other hand, cross sections depend only on the momentum of a single neutrino. This leads us to define two averaged quantities 
\begin{align}
    |\vec{p}_{\nu_i}| &\equiv \langle |\vec{p}_{\nu_i,\mathrm{true}}| \rangle = \frac{\int (\vec{\beta}_{\nu_i,\mathrm{true}}\cdot \vec{n}_\Earth)\,|\vec{p}_{\nu_i,\mathrm{true}}|\, d\widetilde\Omega}{\int (\vec{\beta}_{\nu_i,\mathrm{true}}\cdot \vec{n}_\Earth) d\widetilde\Omega} = |\vec{\widetilde{p}}_{\nu_i}| + \mathcal{O}(\beta_{\Earth}^2),\label{eq:av1}\\
    \Delta p_{\nu_i} &\equiv |\langle\vec{p}_{\nu_i,\mathrm{true}}\rangle| = \left|\frac{\int (\vec{\beta}_{\nu_i,\mathrm{true}}\cdot \vec{n}_\Earth)\,\vec{p}_{\nu_i,\mathrm{true}} \,d\widetilde\Omega}{\int (\vec{\beta}_{\nu_i,\mathrm{true}}\cdot \vec{n}_\Earth) d\widetilde\Omega}\right| = \frac{\beta_\Earth}{3\widetilde E_{\nu_i}}\left(4\widetilde E_{\nu_i}^2 - |\vec{\widetilde p}_{\nu_i}|^2\right) + \mathcal{O}(\beta_\Earth^2)\label{eq:av2},
\end{align}
which importantly are not equal. Here, the factor $(\vec{\beta}_{\nu_i,\mathrm{true}}\cdot \vec{n}_\Earth)$, where $\vec{n}_\Earth$ is the normal vector to the Earth in the lab frame, accounts for the increased flux of neutrinos in the path of the Earth, compared to those in its wake. We additionally define the average lab frame neutrino energy and velocity, $E_{\nu_i} = \sqrt{|\vec{p}_{\nu_i}|^2 + m_{\nu_i}^2}$ and $\beta_{\nu_i} = |\vec{p}_{\nu_i}|/E_{\nu_i}$, respectively. Going forward, we will use $|\vec{p}_{\nu_i}|$ and $E_{\nu_i}$ when a quantity depends only on the dynamics of a single neutrino (\textit{e.g.} for calculating cross sections), and $\vec{p}_{\nu_i,\mathrm{true}}$ for those which depend on the dynamics of many neutrinos. The latter should then be flux-averaged using the same procedure as in~\eqref{eq:av1} and~\eqref{eq:av2}. Additionally, as $\beta_{\nu_i}$, $|\vec{p}_{\nu_i}|$ and $E_{\nu_i}$ are all equal to their C$\nu$B frame counterparts to leading order in $\beta_\Earth$, we will not distinguish between the two in what follows. 

There are two natural choices for the C$\nu$B frame. If relic neutrinos are unclustered, the C$\nu$B reference frame should coincide with that of the CMB, in which case we know from measurements of the CMB dipole that $\beta_\Earth = \beta_\Earth^{\mathrm{CMB}} \simeq 10^{-3}$~\cite{Amendola:2010ty,Ferreira:2020aqa}. On the other hand, if relic neutrinos are clustered then they should share a reference frame with the Milky Way (MW), allowing us to set $\beta_\Earth = \beta_\Earth^{\mathrm{MW}} = 7.6\times10^{-4}$~\cite{Vera:2020}.  We additionally assume that the velocity dispersion of clustered neutrinos is similar to that of objects in the MW, and so we set $\beta_{\nu_i} = \beta_\Earth^\mathrm{MW}$.

To cluster, the velocity of neutrinos must not exceed the escape velocity of the galaxy, $\besc = 1.8\times10^{-3}$~\cite{Kafle:2014xfa}. This in turn allows us to find the minimum mass above which neutrinos will cluster for a given C$\nu$B temperature, $m_{\nu_i} \gtrsim 1.75\times 10^{3}\,T_{\nu_i}$. For $T_{\nu_i} = T_{\nu,0}$, we find that neutrinos only cluster with masses $m_{\nu_i} \gtrsim 0.29\,\mathrm{eV}$, which lies below the upper bound on the effective neutrino mass $m_{\nu}\simeq \sqrt{\sum_i |U_{ei}|^2 m_{\nu_i}^2}~\lesssim 0.8\,\mathrm{eV}$ set by KATRIN~\cite{Aker:2021gma}, where $U_{ei}$ is an element of the Pontecorvo–Maki–Nakagawa–Sakata (PMNS) neutrino mixing matrix. As such, we will consider both clustered and unclustered neutrino scenarios in what follows.
\section{Present day constraints}\label{sec:existing}
Placing model-independent constraints on the neutrino overdensity in the present epoch is challenging. For example, a CMB constraint on the overdensity at recombination may not still be valid today due to late decays of dark matter into neutrinos, or the decay of neutrinos themselves. To that end, constraints on the C$\nu$B overdensity must be derived either from present day observables or theory.
\subsection{Pauli exclusion principle}\label{sec:pauli}
As neutrinos are fermions, their local number density is bounded above by the Pauli exclusion principle. This effect is particularly pronounced for relic neutrinos, which are expected to have macroscopic wavelengths $\lambda_\nu = 2\pi/|\vec{p}_{\nu,0}| \simeq 2.3\,\mathrm{mm}$. To find this bound, we note that each neutrino moving in an infinite square potential well of volume $V$ occupies a volume in momentum space $V_p = \pi^3/V$. The total volume in momentum space available to be filled by neutrinos is set by the Fermi momentum,
\begin{equation}
    p_{f,i} = \frac{\besc}{\sqrt{1-\besc^2}} m_{\nu_i} \simeq \besc m_{\nu_i},
\end{equation}
which differs for each mass eigenstate. For a system of $N_\nu^c$ clustered neutrinos, we therefore have the condition satisfied by each neutrino degree of freedom
\begin{equation}
    N^{c}_\nu\left(\nu_{i,s}\right)\left(\frac{\pi^3}{V}\right) \leq \frac{1}{8}\left(\frac{4}{3}\pi p_{f,i}^3\right),
\end{equation}
where the factor $1/8$ arises as we restrict ourselves to positive absolute momenta, whilst the superscript $c$ refers to the fact that we are only considering clustered neutrinos inside of the potential well. We can translate this to a limit on the neutrino overdensity
\begin{equation}\label{eq:pauliLimit}
    \begin{split}
        \eta_\nu\left(\nu_{i,s}\right) &\leq \frac{2}{9\zeta(3)}\left(\frac{p_{f,i}}{T_{\nu,0}}\right)^3 + \frac{2}{3\zeta(3)}\frac{1}{T_{\nu,0}^3}\int\displaylimits_{p_{f,i}}^\infty \frac{x^2\,dx}{\exp(x/T_{\nu_i})+1}\\
        &\simeq \begin{cases}
        1.82 \left(\frac{m_{\nu_i}}{0.2\,\mathrm{eV}}\right)^3 + \mathcal{O}\!\left(\frac{T_{\nu_i}}{p_{f,i}}\right),\quad & T_{\nu_i} \ll p_{f,i},\\
        \left(\frac{T_{\nu_i}}{T_{\nu,0}}\right)^3 + \mathcal{O}\!\left(\frac{p_{f,i}}{T_{\nu_i}}\right),\quad & T_{\nu_i} \gg p_{f,i},\\
        \end{cases}
    \end{split}
    %&\lesssim 1 + 5.39\times10^{-10}\left(\frac{m_{\nu_i}}{T_{\nu_i}}\right)^3, {\color{red}\text{Non-perturbative above $m_\nu = 0.1\,$eV.}}
\end{equation}
where the second term in the first line is the contribution from unclustered neutrinos above the Fermi momentum. We note that the expression~\eqref{eq:pauliLimit} naturally approaches equilibrium scaling, $\eta_{\nu} = (T_{\nu_i}/T_{\nu,0})^3$, as either $\besc\to 0$ or $m_{\nu_i} \to 0$ and neutrinos are unable to cluster.
\subsection{Tremaine-Gunn bound}
For completeness, we note that there exists a similar bound on the neutrino overdensity. Suppose that on macroscopic scales, clustered relic neutrinos are described by some coarse-grained distribution $\bar f_{\nu_{i,s}}$ satisfying
\begin{equation}
    \int d^3p_{\nu_i} \,\bar f_{\nu_{i,s}}(|\vec{p}_{\nu_i}|) = n^{c}_\nu(\nu_{i,s}),
\end{equation}
where once again the superscript $c$ refers to the clustered component of the C$\nu$B. From the requirement that the maximum of $\bar f_{\nu_{i,s}}$ does not exceed the maximum Fermi-Dirac phase space density, we find that
\begin{equation}\label{eq:maxPS}
    \max\left\{\bar f_{\nu_{i,s}}\right\} \leq \max\left\{\frac{1}{(2\pi)^3}\frac{1}{\exp(|\vec{p}_{\nu_i}|/T_{\nu_i})+1}\right\} = \frac{1}{2}\frac{1}{(2\pi)^3}.
\end{equation}
Defining the normalised coarse-grained phase space distribution satisfying $\bar f_{\nu_{i,s}} = n^{c}_\nu(\nu_{i,s}) \bar f_{\nu_{i,s}}^N$, the condition~\eqref{eq:maxPS} gives the constraint on the relic neutrino overdensity

%from requiring that the maximum phase space density of the coarse-grained neutrino momentum distribution, $\bar f_i$, be less than the maximum of the Fermi-Dirac distribution~\eqref{eq:fermiDirac}. This is commonly known as the Tremaine-Gunn bound~\cite{Tremaine:1979we}. Without making any assumptions about the form of $\bar f_i$ other than that it is normalised to unity, we find an upper limit on the overdensity from the Tremaine-Gunn bound
\begin{equation}\label{eq:tremaineGunn}
    \eta_\nu\left(\nu_{i,s}\right) \leq \frac{1}{12\pi\zeta(3)}\frac{1}{T_{\nu,0}^3}\frac{1}{\mathrm{max}\{\bar f_{\nu_{i,s}}^N\}}+ \frac{2}{3\zeta(3)}\frac{1}{T_{\nu,0}^3}\int\displaylimits_{p_{f,i}}^\infty \frac{x^2\,dx}{\exp(x/T_{\nu_i})+1}, 
\end{equation}
where once more the second term is the contribution from unclustered neutrinos. This is commonly known as the Tremaine-Gunn bound~\cite{Tremaine:1979we}. Supposing that the coarse-grained distribution is Maxwell-Boltzmann with velocity dispersion $\besc$, we find that the Tremaine-Gunn bound is weaker than~\eqref{eq:pauliLimit} by a factor $\sim 1.88$ for strongly clustered neutrinos. In general, the Tremaine-Gunn bound is stronger than~\eqref{eq:pauliLimit} for clustered neutrinos if
\begin{equation}
    \mathrm{max}\left\{\bar f_{\nu_{i,s}}^N\right\} \geq \frac{3}{8\pi p_{f,i}^3}.
\end{equation}
Similar to the Pauli limit, the first term in~\eqref{eq:tremaineGunn} should vanish as either $\besc \to 0$ or $m_{\nu_i} \to 0$. This places an additional constraint on the coarse-grained distribution $\bar f_{\nu_{i,s}}$.
\subsection{Borexino}
If the C$\nu$B is sufficiently energetic and dense, relic neutrinos could be visible at existing neutrino experiments. Of these, Borexino is the experiment capable of probing the lowest neutrino energies, with sensitivity down to $E_\nu \simeq 20\,\mathrm{keV}$~\cite{Borexino:2008gab}.

Recently, Borexino has made the first measurement of solar neutrinos from the CNO cycle~\cite{BOREXINO:2021mga}, which dominate  the solar neutrino flux in the energy range (range) $420\,\mathrm{keV}\leq E_\nu \leq 1.73\,\mathrm{MeV}$. The measured CNO flux at Borexino, $\phi_\mathrm{CNO} = (7.5^{+3.0}_{-2.0})\times 10^8\,\mathrm{cm}^{-2}\,\mathrm{s}^{-1}$, lies slightly above the predictions from theory in both the low (LZ) and high metallicity (HZ) Standard Solar Models, for which the predicted fluxes are $\phi_{\mathrm{CNO}}^\mathrm{LZ} \simeq (3.51\pm0.35)\times 10^8\,\mathrm{cm}^{-2}\,\mathrm{s}^{-1}$ and $\phi_{\mathrm{CNO}}^\mathrm{HZ} = (4.88\pm0.54)\times 10^8\,\mathrm{cm}^{-2}\,\mathrm{s}^{-1}$, respectively~\cite{Vinyoles:2016djt}. We can use these results to constrain the C$\nu$B overdensity and temperature, assuming $|\vec{p}_{\nu_i}| \simeq 3.15\,T_{\nu_i}$. Supposing that the entire difference between the observed and predicted fluxes is due to the capture of energetic relic neutrinos, we require that
\begin{equation}
    \sum_{\nu,i,s}|U_{ei}|^2 \beta_{\nu_i} n_{\nu}(\nu_{i,s}) \leq \phi_\mathrm{CNO} - \phi_{\mathrm{CNO}}^\mathrm{LZ}, \quad  133\,\mathrm{keV} \leq T_{\nu_i}\leq 550\,\mathrm{keV},
\end{equation}
where we have chosen to use the LZ flux as it gives the most conservative limits for CNO cycle neutrinos. At the high energies required for relic neutrinos to mimic solar neutrinos, $\beta_{\nu_i} \simeq 1$, giving a temperature dependent constraint on the overdensity
%\begin{equation}\label{eq:solarConstraints}
%    \begin{split}
%    \sum_{\nu,i,s}|U_{ei}|^2 \eta_{\nu}(\nu_{i,s}) &\lesssim \frac{4\pi^2}{3\zeta(3) T_{\nu,0}^3}\left(\phi_\mathrm{CNO} - \phi_{\mathrm{CNO}}^\mathrm{LZ}\right) \\
%    &\simeq (2.4^{+1.7}_{-1.2})\times 10^{-4}, \quad   133\,\mathrm{keV} \leq T_{\nu_i}\leq 550\,\mathrm{keV}.
%    \end{split}
%\end{equation}
\begin{equation}\label{eq:solarConstraints}
    \begin{aligned}
    \sum_{\nu,i,s}|U_{ei}|^2 \eta_{\nu}(\nu_{i,s}) &\lesssim \frac{4\pi^2}{3\zeta(3) T_{\nu,0}^3}\left(\phi_\mathrm{CNO} - \phi_{\mathrm{CNO}}^\mathrm{LZ}\right) \\
     &\simeq (2.4^{+1.7}_{-1.2})\times 10^{-4}, 
    \end{aligned}
     \quad   133\,\mathrm{keV} \leq T_{\nu_i}\leq 550\,\mathrm{keV}.
\end{equation}
If the neutrino temperature lies outside the above range then relic neutrinos will appear in other parts of the spectrum, for which different constraints will apply. By following this procedure for other parts of the solar neutrino spectrum we can obtain similar constraints on the overdensity, which we tabulate in Table~\ref{tab:solarConstraints}.

\begin{table}[]
\centering
\begin{tabular}{c|c|c|c}
Flux              & $T_{\nu_i,\mathrm{min}}\,(\mathrm{keV})$ & $T_{\nu_i,\mathrm{max}}\,\,(\mathrm{keV})$ & $\eta_{\nu,\mathrm{max}}$ \\
\hline\hline
$pp$              & 6.35                                     & 133                                        & $(0^{+3}_{-4})\times 10^{-3}$\\%$(0.4^{+3.4}_{-4.2})\times 10^{-3}$                     \\
${^7\mathrm{Be}}$ & 274                                      & 274                                        & $(3\pm2)\times 10^{-4}$\\%$(2.88_{-1.77}^{+1.76})\times 10^{-4}$\\
$pep$             & 457                                      & 457                                        & $(0\pm1)\times 10^{-5}$ \\%$(-0.42_{-1.32}^{+1.22})\times 10^{-5}$                         \\
CNO             & 133                                      & 550                                        & $(2.4^{+1.7}_{-1.2})\times 10^{-4}$\\%$(2^{+2}_{-1})\times 10^{-4}$                     \\
${^8\mathrm{B}}$ & 550                                      & 5400                                       & $(7\pm4)\times 10^{-7}$    %$(6.94^{+3.94}_{-3.97})\times 10^{-7}$   
%${\mathrm{He}p}$  & 5400                                     & 5970                                       & $1.3\times 10^{-5}$      
\end{tabular}
\caption{Constraints on the C$\nu$B overdensity from the solar neutrino spectrum as measured by Borexino, where the parameter $\eta_{\nu,\mathrm{max}}$ is understood to replace the RHS of~\eqref{eq:solarConstraints}. In all cases, we use the theoretical flux values from the LZ model and measured fluxes from~\cite{BOREXINO:2021mga,Vinyoles:2016djt,BOREXINO:2018ohr}.}
\label{tab:solarConstraints}
\end{table}

We can make strong arguments about the present day neutrino temperature using the results in Table~\ref{tab:solarConstraints}. Increasing the temperature of relic neutrinos typically increases the number density, either through equilibrium number density scaling, $n_\nu(\nu_{i,s}) \sim T_{\nu_i}^3$, or via late time decays of additional degrees of freedom into neutrinos, where the increase in neutrino temperature is due to entropy conservation. One would therefore naively expect that in the case where $T_{\nu_i} \gg T_{\nu,0}$, we would also have $\eta_\nu(\nu_{i,s}) \gg 1$, in contrast to the constraints presented in Table~\ref{tab:solarConstraints}. As a result, it is reasonable to suggest that $T_{\nu_i} \lesssim 5\,\mathrm{keV}$ in order to safely avoid these constraints. 

\subsection{Other constraints}\label{sec:otherExisting}
The bounds~\eqref{eq:pauliLimit} and~\eqref{eq:tremaineGunn} are the only model-independent constraints that can placed on the relic neutrino overdensity from theory. However, these can be supplemented with bounds on the neutrino mass to further restrict the allowed parameter space. For stable neutrinos, the strongest bounds on the neutrino mass come from cosmology~\cite{Planck:2018vyg}, which require that $\sum_i m_{\nu_i} \leq 0.12\,\mathrm{eV}$. This constraint is relaxed to $\sum_i m_{\nu_i} \lesssim 1\,\mathrm{eV}$ if neutrinos are allowed to decay~\cite{Escudero:2020ped}. An additional constraint on the mass of Majorana neutrinos comes from experiments searching for neutrinoless double beta decay (0$\nu\beta\beta$), for which the rate is proportional to the magnitude of effective Majorana mass, $m_{\beta\beta} = \sum_i U_{ei}^2 m_{\nu_i}$~\cite{Jones:2021cga}. The strongest bound on the neutrino mass using $0\nu\beta\beta$ is set by KamLAND-Zen, $|m_{\beta\beta}| \leq 0.17\,\mathrm{eV}$~\cite{KamLAND-Zen:2016pfg}. By 2024, the KATRIN collaboration expects to be sensitive to effective neutrino masses $m_\nu \leq 0.2\,\mathrm{eV}$~\cite{KATRIN:2022ayy}, which will become the strongest constraint on the mass of unstable Dirac neutrinos. Future experiments such as Project 8~\cite{Project8:2022wqh}, HOLMES~\cite{Faverzani:2017ycg} and ECHo~\cite{Gastaldo:2013wha} aim to improve this bound further, with the potential to probe effective neutrino mass scales as low as $m_\nu \leq 40~\mathrm{meV}$. The Simons Observatory will improve upon the cosmological neutrino mass constraints with a goal sensitivity of $\sum_i m_{\nu_i} \leq 90\,\mathrm{meV}$~\cite{SimonsObservatory:2018koc}, capable of ruling out the inverted mass hierarchy. 

Neutrino mass experiments utilising tritium beta decay such as KATRIN and those at Troitsk~\cite{Belesev:1995sb} and Los Alamos~\cite{Robertson:1991vn} are also able to place model-independent constraints on the relic neutrino overdensity by searching for the capture process ${^3}\mathrm{H} + \nu_e \to {^3}\mathrm{He}^{+} + e^-$. We discuss this at length in Section~\ref{sec:ptolemy}. To a good approximation\footnote{The parameter combination constrained by neutrino mass experiments is given on the LHS of~\eqref{eq:ptolemySignificance}.
This can be re-expressed in terms of the C$\nu$B frame overdensities using~\eqref{eq:flipL} and~\eqref{eq:flipR}, and then solved for $\widetilde\eta_\nu(\nu_{i,L}^D)$ in the standard scenario with $\widetilde\eta_{\nu}(\nu_{i,R}^D) = 0$, assuming the same overdensity for all three mass eigenstates. For Majorana neutrinos, the tritium experiments approximately constrain $\sum_s \widetilde\eta_{\nu}(\nu_{i,s}^M)$ instead.}, assuming the standard cosmological history presented in Section~\ref{sec:thermalHist}, the overdensity is constrained to $\widetilde\eta_{\nu}(\nu_{i,L}) \lesssim 1.3\times 10^{11}$, $8.9\times 10^{13}$ and $1.8\times 10^{14}$ by KATRIN~\cite{KATRIN:2022kkv}, Troitsk~\cite{Lobashev:1999dv} and Los Alamos~\cite{Robertson:1991vn} respectively. These bounds do not apply for antineutrinos and become a constraint on the combined left and right helicity densities under the Majorana neutrino hypothesis, still assuming the standard scenario. In this case, with $\widetilde\eta_{\nu}(\nu_{i,L}^M) = \widetilde\eta_{\nu}(\nu_{i,R}^M)$, the constraints are stronger by a factor of two.

\begin{figure}[tbp]
    \centering
    \includegraphics[width =.813\textwidth]{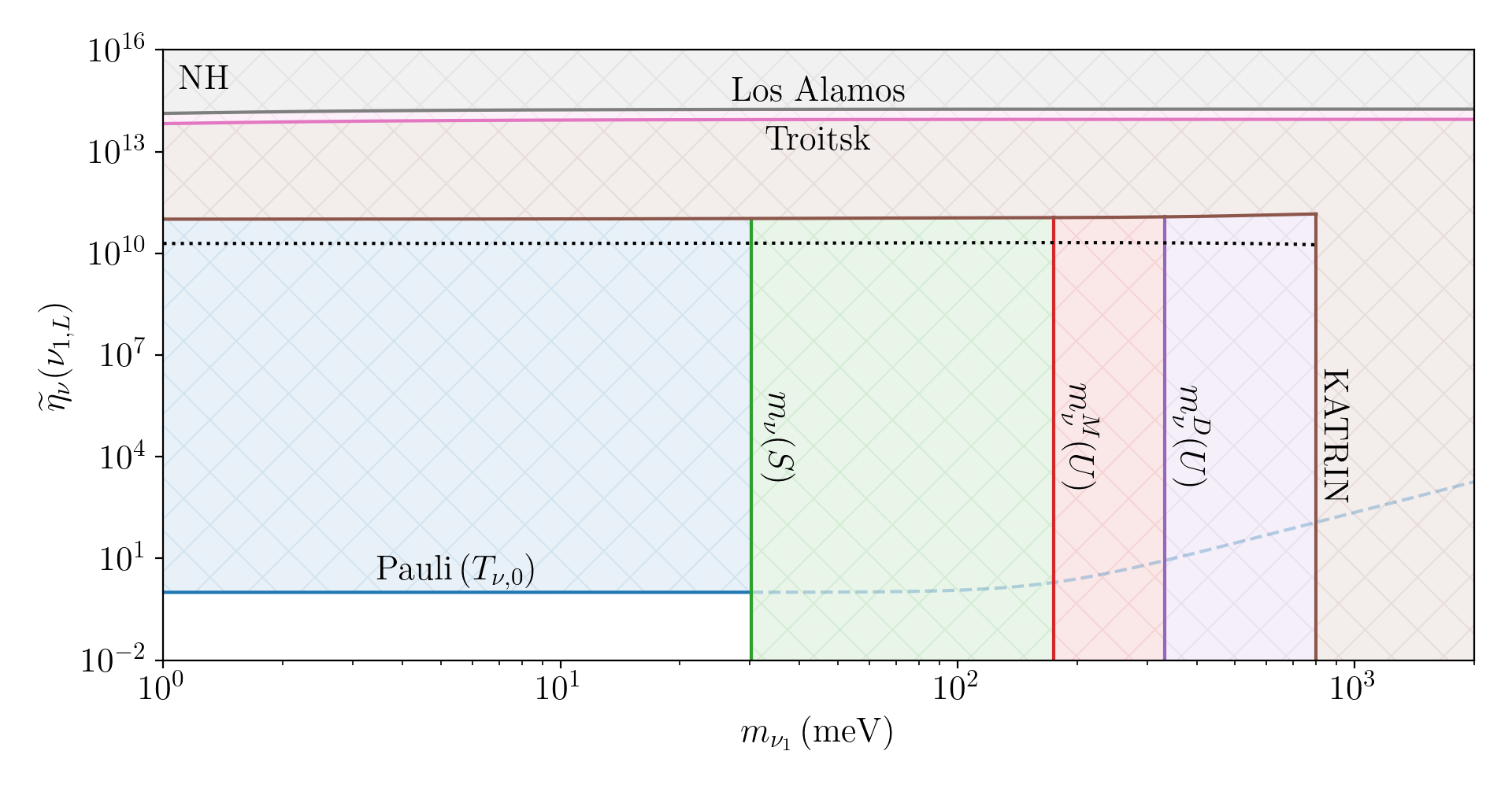}
    \hfill
    \includegraphics[width=.813\textwidth]{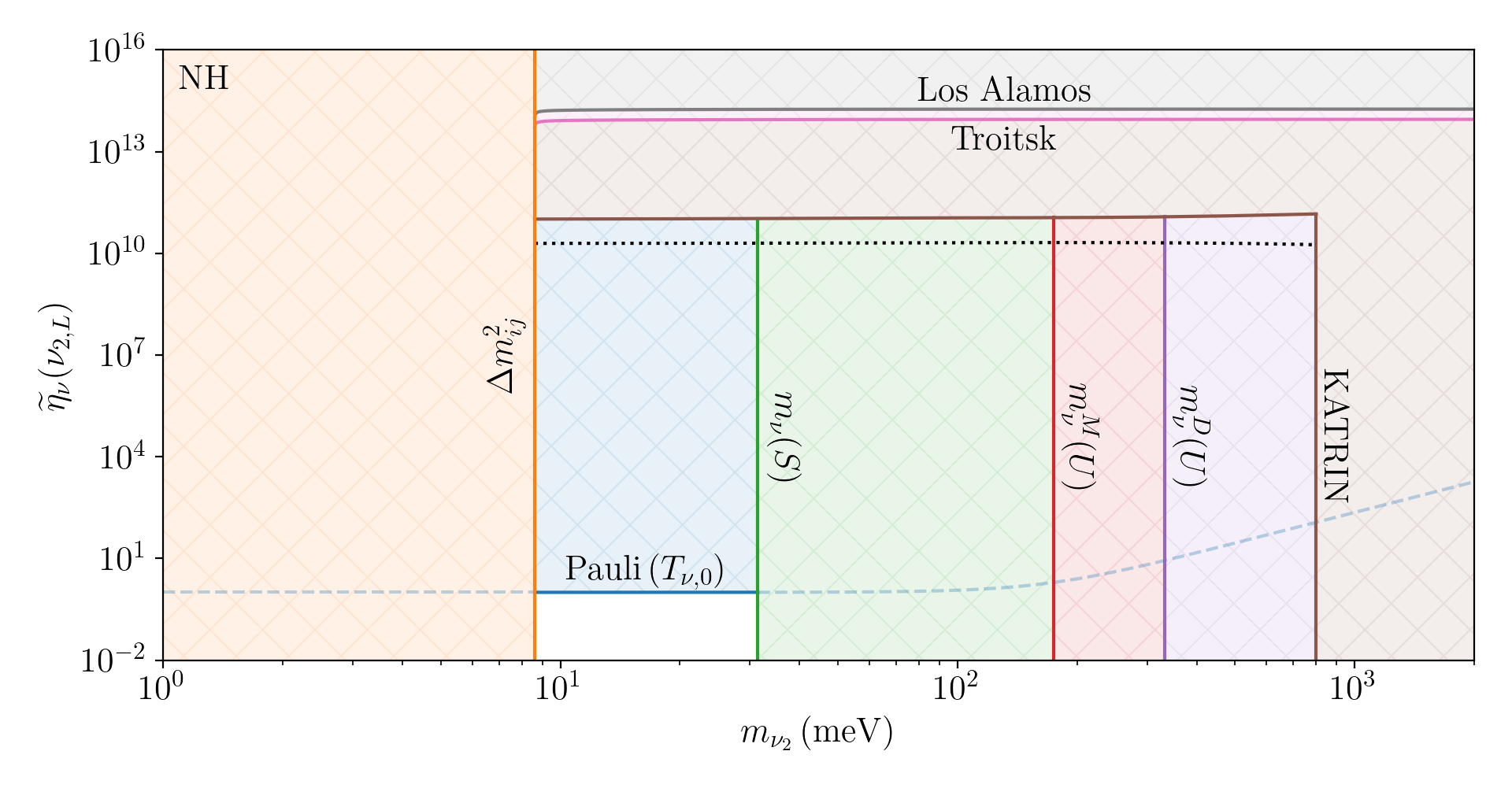}
    \hfill
    \includegraphics[width=.813\textwidth]{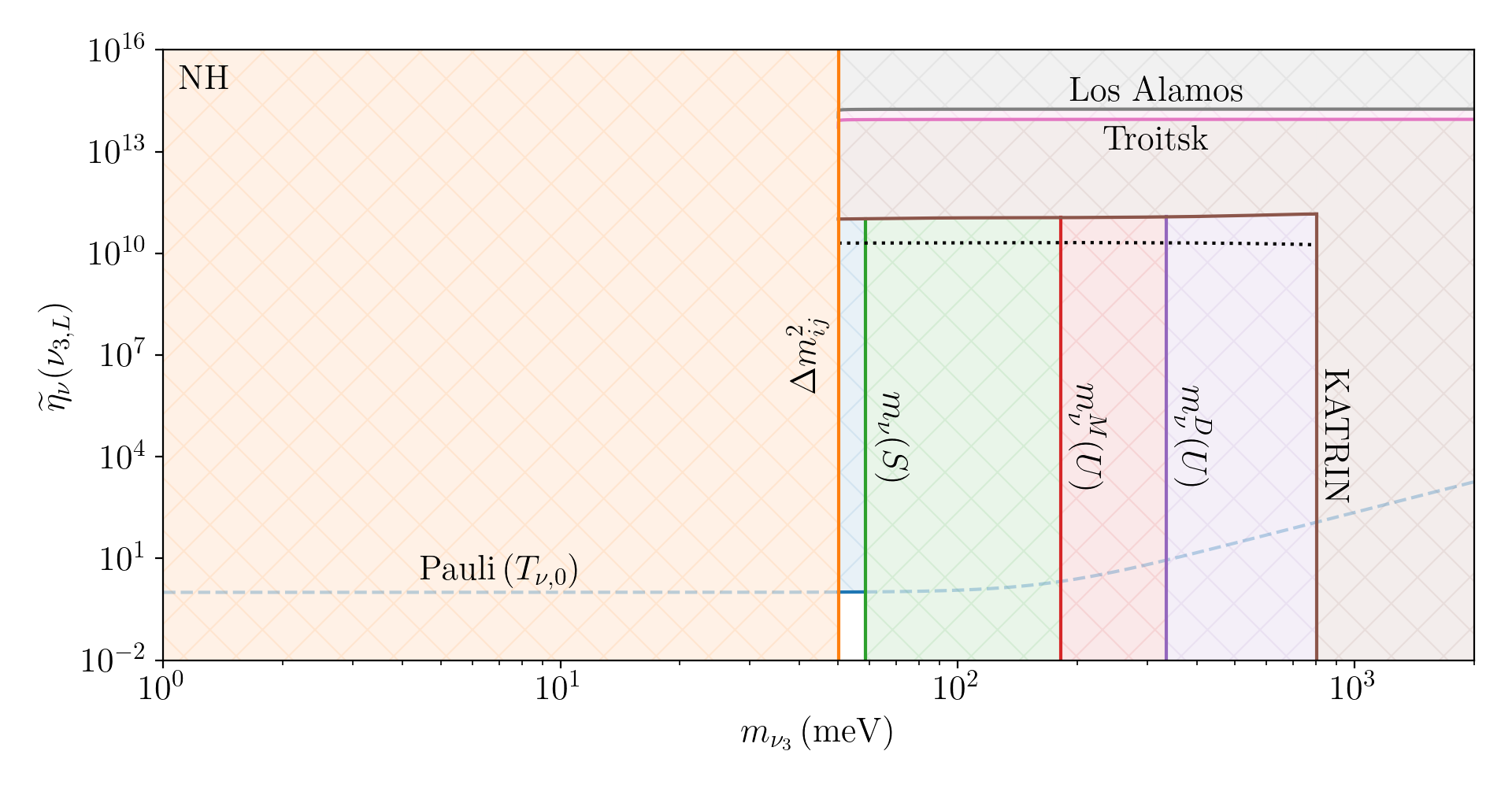}
    \caption{Present day constraints on the C$\nu$B frame relic neutrino overdensity in the normal mass hierarchy, assuming the standard scenario and $\beta_\Earth = \beta_\Earth^\mathrm{CMB}$. The KATRIN, Troitsk and Los Alamos overdensity bounds shown assume Dirac neutrinos, and are stronger by a factor of two under the Majorana neutrino hypothesis. See the text for a full description of the figure.} 
    \label{fig:NH_existing}
\end{figure}

\begin{figure}[t]
    \centering
    \includegraphics[width=.813\textwidth]{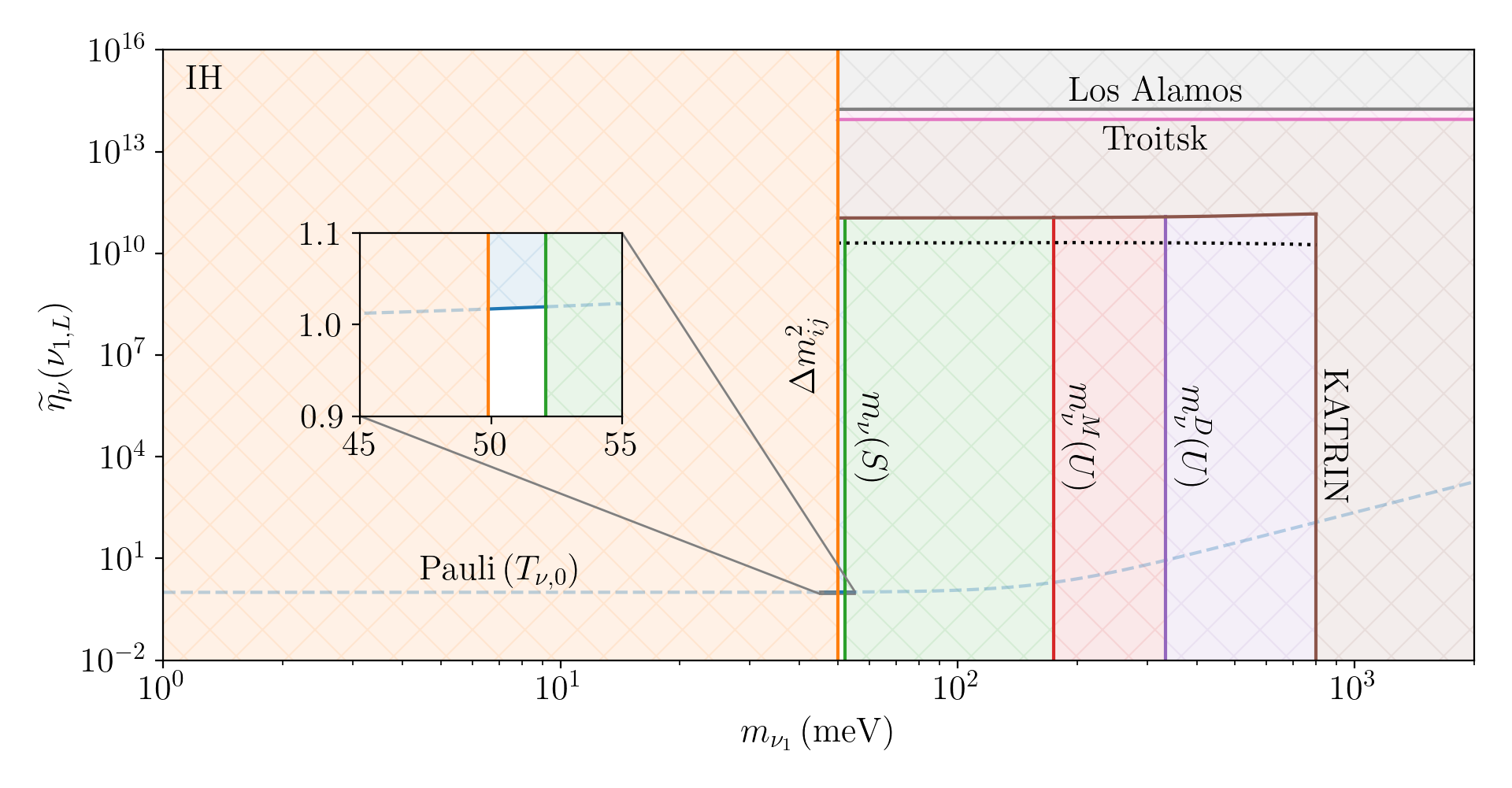}
    \hfill
    \includegraphics[width=.813\textwidth]{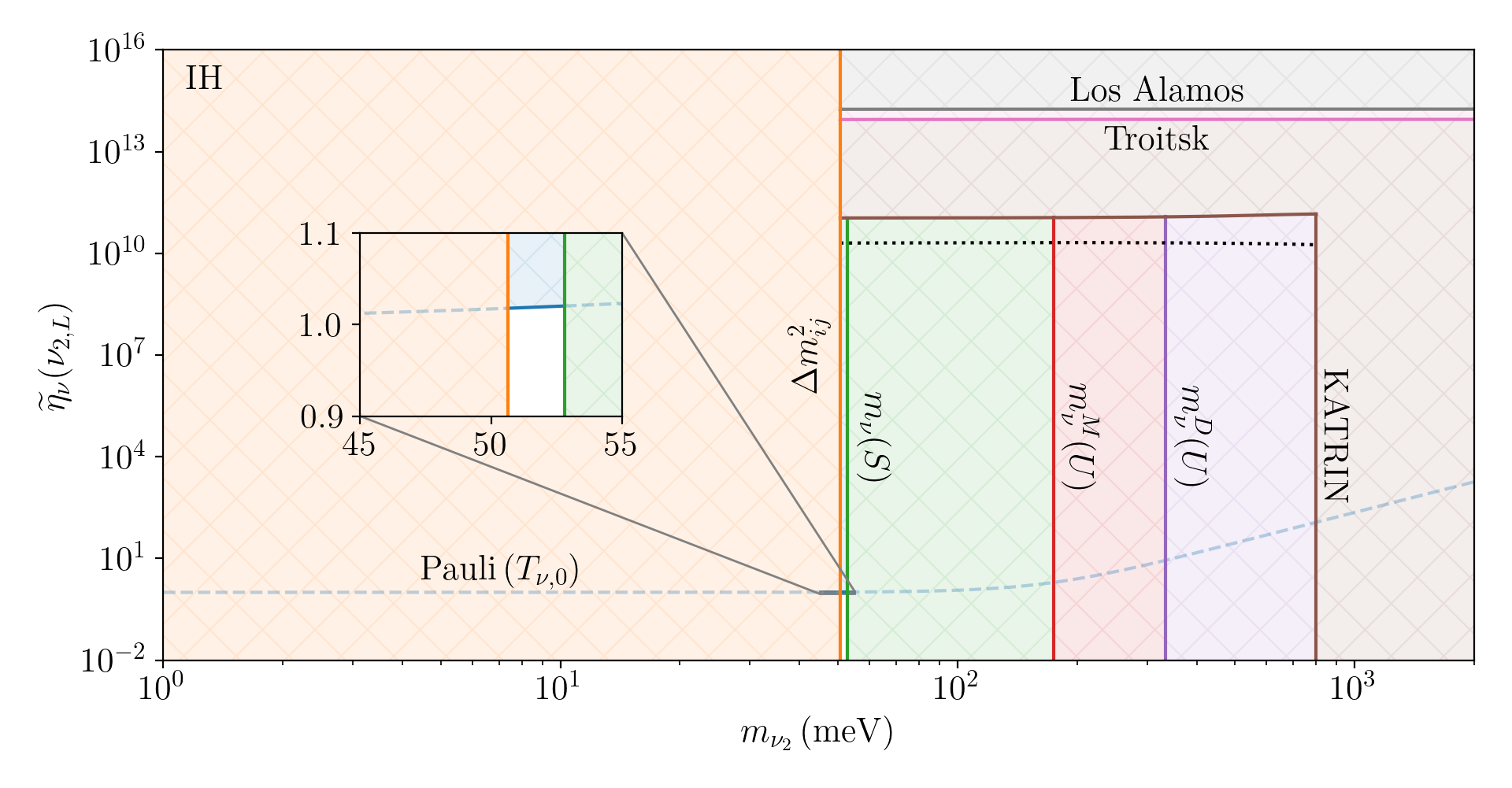}
    \hfill
    \includegraphics[width=.813\textwidth]{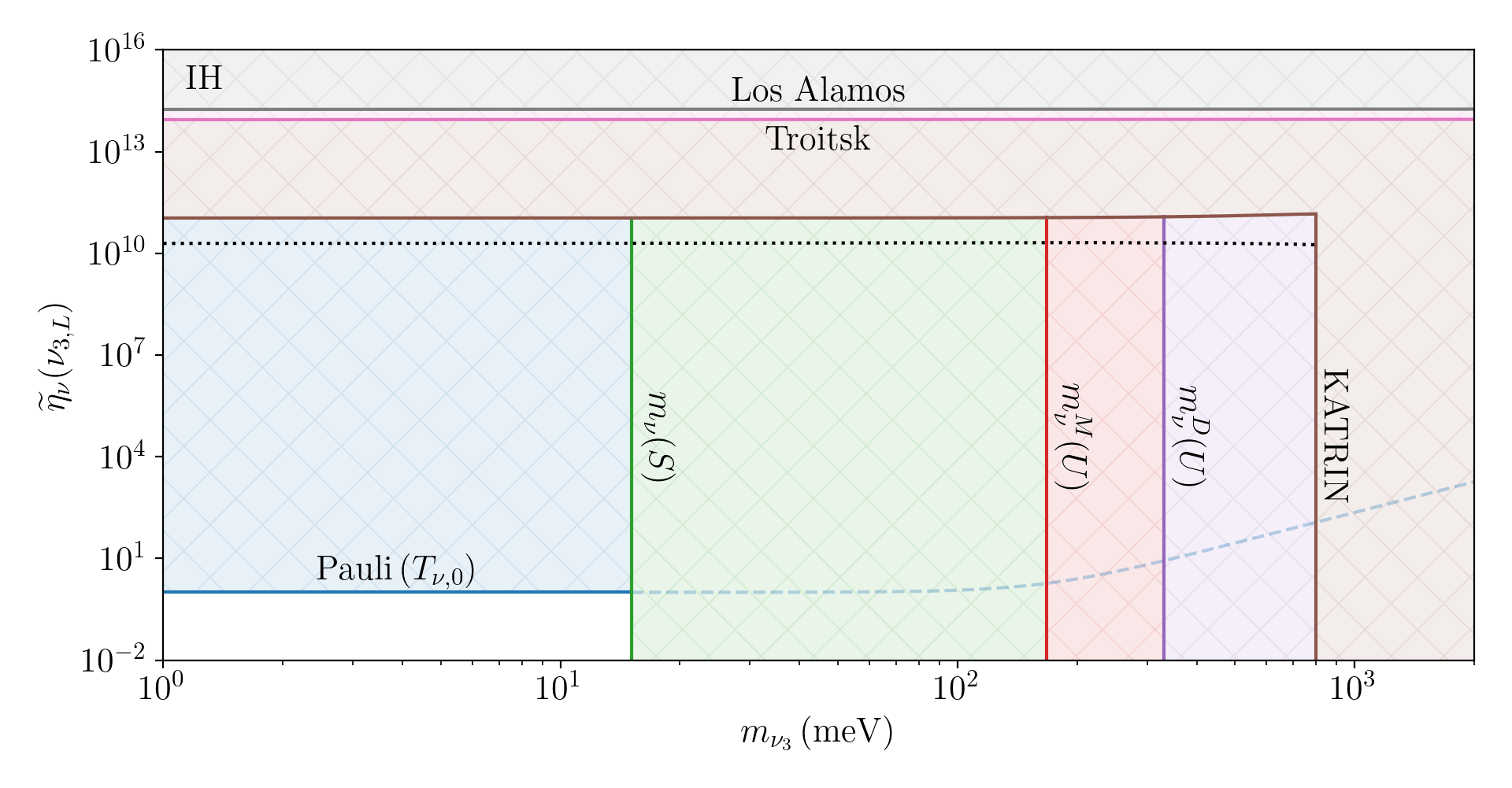}
    \caption{Present day constraints on the C$\nu$B frame relic neutrino overdensity in the inverted mass hierarchy, assuming the standard scenario and $\beta_\Earth = \beta_\Earth^\mathrm{CMB}$. The KATRIN, Troitsk and Los Alamos overdensity bounds shown assume Dirac neutrinos, and are stronger by a factor of two under the Majorana neutrino hypothesis. See the text for a full description of the figure.} 
    \label{fig:IH_existing}
\end{figure}

We plot all present day constraints on the C$\nu$B overdensity in Figures~\ref{fig:NH_existing} and~\ref{fig:IH_existing} assuming the standard scenario and $\beta_\Earth = \beta_\Earth^{\mathrm{CMB}}$, where we note that the choice $\beta_\Earth = \beta_\Earth^{\mathrm{MW}}$ makes very little difference. The limits labelled $\Delta m_{ij}^2$ (orange) refer to the minimum mass constraints from oscillation experiments, whilst those labelled $m_{\nu}(S/U)$ refer to the maximum mass allowed by cosmology (green, purple) and KamLAND-Zen (red), for stable/unstable neutrinos. The dotted line shows the KATRIN projection for Dirac neutrinos at three years, which is a factor of two stronger under the Majorana neutrino hypothesis. It is immediately obvious that large overdensities $\widetilde\eta_\nu(\nu_{i,s}) \gg 1$ are completely ruled out for stable neutrinos by the exclusion principle, whilst for unstable neutrinos the combination of constraints requires that $\widetilde\eta_\nu(\nu_{i,s}) \lesssim 10$. For warmer neutrinos, these constraints become weaker by a factor $\sim (T_{\nu_i}/T_{\nu,0})^3$, which could still allow for significant overdensities. To our best knowledge, there are no present day constraints that can be placed on the relic neutrino temperature. However, it is reasonable to suggest that the C$\nu$B temperature should be $T_{\nu_i} \lesssim 5\,\mathrm{keV}$, given the strength of the constraints on the overdensity set by Borexino above this temperature.
\clearpage
\clearpage
\section{Cosmological constraints}\label{sec:cosmology}
The presence of a relic neutrino overdensity at large redshifts could significantly modify the cosmological evolution of the universe. As such, if relic neutrinos do not interact strongly since decoupling and as a result maintain a similar distribution today, cosmology could provide strong constraints on the present day C$\nu$B overdensity. In this section we review the constraints on the C$\nu$B overdensity from cosmology, which may still hold today.

These constraints can be modelled by assuming a neutrino degeneracy parameter,\footnote{This is not to be confused with $g_{\nu_i}$, which is the number of neutrinos per momentum state.} $\xi_{\nu_i}$, proportional to the chemical potential. This leads to a modified relic neutrino number density, as well as a neutrino-antineutrino asymmetry through the modified distribution functions
\begin{align}
    f_{\nu_i}(|\vec{p}_{\nu_i}|) \to f_{\nu_i}^\xi(|\vec{p}_{\nu_i}|) = \frac{g_{\nu_i}}{\exp(|\vec{p}_{\nu_i}|/T_{\nu_i} - \xi_{\nu_i})+1},\\
    f_{\bar\nu_i}(|\vec{p}_{\nu_i}|) \to f_{\bar\nu_i}^\xi(|\vec{p}_{\nu_i}|) = \frac{g_{\nu_i}}{\exp(|\vec{p}_{\nu_i}|/T_{\nu_i} + \xi_{\nu_i})+1},
\end{align}
for neutrinos and antineutrinos, respectively. As we cannot distinguish between neutrino and antineutrino for Majorana fermions, only Dirac neutrinos can have non-zero chemical potential. The resulting overdensities are given by
\begin{equation}\label{eq:odDegeneracy}
    \begin{split}
    \sum_{s} \widetilde\eta_{\nu}(\nu_{i,s}^D) &= \frac{4\pi^2}{3\zeta(3)T_{\nu,0}^3 }\int \frac{d^3\vec{p}_{\nu_i}}{(2\pi)^3} f_{\nu_i}^\xi(|\vec{p}_{\nu_i}|) \\
    &= -\frac{4g_{\nu_i}}{3\zeta(3)}\left(\frac{T_{\nu_i}}{T_{\nu,0}}\right)^3 \mathrm{Li}_3\left(-e^{\xi_{\nu_i}}\right) \\
    &= g_{\nu_i}\left(\frac{T_{\nu_i}}{T_{\nu,0}}\right)^3\left(1 + \frac{\pi^2}{9\zeta(3)}\xi_{\nu_i} + \mathcal{O}\left(\xi_{\nu_i}^2\right)\right),
    \end{split} 
\end{equation}
where the antineutrino overdensity is found by making the replacement $\xi_{\nu_i} \to -\xi_{\nu_i}$ and $\mathrm{Li}_k(z)$ denotes the polylogarithm, defined by
\begin{equation}
    \mathrm{Li}_k(z) = \sum_{n=1}^\infty\frac{z^n}{n^k}.
\end{equation}
Introducing a chemical potential also modifies the fit to the neutrino masses, so the mass bounds from cosmology given in Section~\ref{sec:existing} do not necessarily apply here. Where appropriate, we will give the neutrino mass bounds for each fit. 

We also note that a large degeneracy parameter can modify the neutrino decoupling temperature due to Pauli blocking suppressing certain interactions. For a significantly large chemical potential, $\xi_{\nu_i}\gtrsim 14$, neutrinos decouple before muon-antimuon pair production becomes kinematically unfavourable~\cite{Kang:1991xa}, leading to an extra reheating of the photon thermal bath relative to the neutrinos. As a result, the ratio $T_{\gamma} = (11/4)^{1/3}\, T_{\nu}$ no longer holds, and we expect $T_{\nu_i} < T_{\nu,0}$. This scenario becomes more extreme as the chemical potential increases further and the decoupling temperature crosses more annihilation thresholds.
\subsection{Big Bang nucleosynthesis}
During the radiation-dominated era, protons and neutrons are kept in equilibrium through weak interactions until they freeze-out at a temperature $T_{\mathrm{dec}} \simeq 0.7\,\mathrm{MeV}$. Due to the presence of energetic photons, these are unable to form stable nuclei until the temperature drops below $T_{\mathrm{BBN}}\simeq 0.07\,\mathrm{MeV}$, at which point almost all neutrons become locked up in ${{^4}\mathrm{He}}$. In the intermediate phase, neutrons decay to protons with lifetime $\tau_{n}\simeq 879\,\mathrm{s}$, decreasing the neutron-proton ratio from its value at freeze-out. As a result, modifying the time between decoupling and Big Bang nucleosyntheis (BBN) will affect the neutron-proton ratio and in turn the primordial element abundances.

It should be clear from~\eqref{eq:odDegeneracy} that the introduction of a chemical potential increases the energy density of relic neutrinos, appearing as a contribution to the effective number of neutrino species, $N_\mathrm{eff}$, at order $\xi_{\nu_i}^2$. At early times this drives the expansion and cooling of the universe, reducing the time between freeze-out and BBN and subsequently increasing the ${{^4}\mathrm{He}}$ mass fraction, $Y_p$. An enhanced expansion rate could also modify structure formation, as density perturbations will not grow enough to form galaxies in a universe that expands too quickly~\cite{Kang:1991xa}. However, assuming the same temperature for neutrinos and antineutrinos, these contributions enter at $\mathcal{O}(\xi_{\nu_i}^2)$, which are largely irrelevant for $\xi_{\nu_i}\ll 1$. 

A much more significant effect occurs due to a neutrino-antineutrino asymmetry during equilibrium. Protons and neutrons are held in equilibrium through the processes $p+\bar{\nu}_e \leftrightarrow n + e^+$ and $n + \nu_e \leftrightarrow p + e^-$, which proceed at significantly different rates for non-zero electron neutrino chemical potential, $\xi_{\nu_e}$. The result is a neutron-proton fraction that depends on both magnitude and sign of $\xi_{\nu_e}$ through~\cite{Kang:1991xa}
\begin{equation}
    \frac{n_n}{n_p}\Bigg|_{\mathrm{eq}} \simeq \exp\left(-\frac{m_n - m_p}{T_\mathrm{SM}} - \xi_{\nu_e}\right),
\end{equation}
where $T_{\mathrm{SM}}$ is the temperature of the SM thermal bath. Between decoupling and the onset of BBN, neutrons are allowed to decay. By using the temperature-time relation $t_1/t_2 = (T_2/T_1)^2$ that holds during radiation-domination, we find that the neutron fraction at the start of BBN satisfies
\begin{equation}
    R_{\mathrm{BBN}}(\xi_{\nu_e}) \equiv \exp\left(-\frac{m_n - m_p}{T_\mathrm{dec}} - \xi_{\nu_e} -\frac{t_\mathrm{dec}}{\tau_n}\left[1-\left(\frac{T_\mathrm{dec}}{T_\mathrm{BBN}}\right)^2\right]\right) \simeq 0.141 e^{-\xi_{\nu_e}},
\end{equation}
where $t_\mathrm{dec}\simeq 1\,\mathrm{s}$ is the time at weak interaction freeze-out. The degeneracy parameter can therefore have a profound effect on the neutron fraction, which assuming that all neutrons are locked up in ${{^4}\mathrm{He}}$ during BBN translates into the helium mass fraction
\begin{equation}
    Y_p(\xi_{\nu_e}) = \frac{2R_{\mathrm{BBN}}(\xi_{\nu_e})}{1+R_{\mathrm{BBN}}(\xi_{\nu_e})} = 0.247 - 0.216\,\xi_{\nu_e} + \mathcal{O}(\xi_{\nu_e}^2).
\end{equation}
Given that present day measurements find $Y_p = 0.2449\pm 0.0002$~\cite{Aver:2015iza}, a large chemical potential is strongly disfavoured, preferring $\xi_{\nu_e} \sim \mathcal{O}(10^{-2})$.

The authors of~\cite{Nunes:2017xon} use a combination of data from Planck 2015~\cite{Planck:2015fie}, baryon acoustic oscillation measurements (BAO)~\cite{Beutler:2011hx,Ross:2014qpa,BOSS:2013rlg,BOSS:2013igd}, the local value of the Hubble parameter~\cite{Riess:2016jrr} and the abundance of galaxy clusters (GC)~\cite{Tinker:2011pv,DSDD:2009php,Henry:2008cg,Benson:2011uta,Hajian:2013rhm,Mantz:2009fw,Planck:2013lkt,Vikhlinin:2008ym,Heymans:2013fya} to determine $\xi_{\nu_i}$ under the assumptions that $T_{\nu_i} = T_{\nu,0}$, $g_{\nu_i} = 1$ and $\xi_{\nu_i} = \xi_\nu$ for all three mass eigenstates. We present their findings in Table~\ref{tab:degen}, with and without GC data which are known to be in tension with CMB data~\cite{Nunes:2017xon}, for both their Model I and II along with the constraint on the C$\nu$B overdensity derived using~\eqref{eq:odDegeneracy}. As expected, the fits strongly favour $\xi_{\nu_i} \sim \mathcal{O}(10^{-2})$ and subsequently $\eta_{\nu}^D \sim \mathcal{O}(1)$. Similar bounds $-0.018 \leq \xi_{\nu_e} \leq 0.008$ are found in~\cite{Mangano:2011ip}.

\begin{table}[tbp]
\centering
\begin{tabular}{c|c|c|c|c}
                                    & \multicolumn{2}{|c}{Model I}     & \multicolumn{2}{|c}{Model II}    \\
                                    \hline\hline
Parameter                           & w/o GC         & w/ GC          & w/o GC         & w/ GC          \\
\hline
$\sum_i m_{\nu_i}\,(\mathrm{eV})$   & $<0.24$        & $<0.64$        & $<0.18$        & $<0.52$        \\
$\xi_{\nu}$                       & $0.10\pm 0.54$ & $0.02\pm 0.50$ & $0.05\pm0.56$  & $-0.02\pm0.51$ \\
$\sum_{s}\widetilde\eta_{\nu}(\nu_{i,s}^D)$     & $1.10\pm 0.67$ & $1.02\pm 0.57$ & $1.05\pm 0.68$ & $0.98\pm 0.57$ \\
$\sum_{s}\widetilde\eta_{\nu}(\bar\nu_{i,s}^D)$ & $0.91\pm 0.36$ & $0.98\pm 0.36$ & $0.96\pm 0.39$ & $1.02\pm 0.38$
\end{tabular}
\caption{Constraints on the relic neutrino overdensity resulting from the introduction of a degeneracy parameter $\xi_{\nu_i}$ using the fits performed in~\cite{Nunes:2017xon}.}
\label{tab:degen}
\end{table}

\begin{figure}[tbp]
\centering 
\includegraphics[width=.495\textwidth]{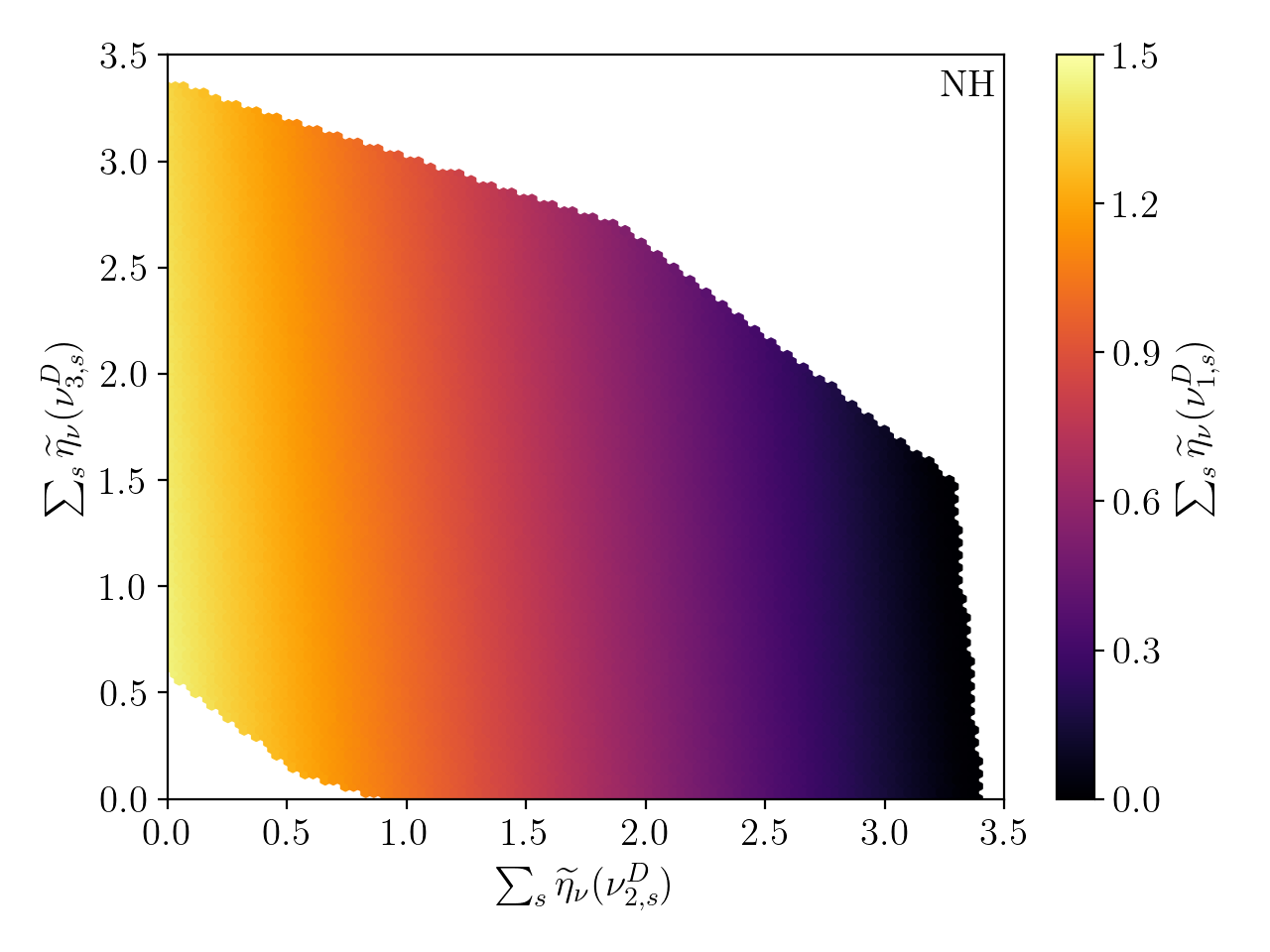}
\hfill
\includegraphics[width=.495\textwidth]{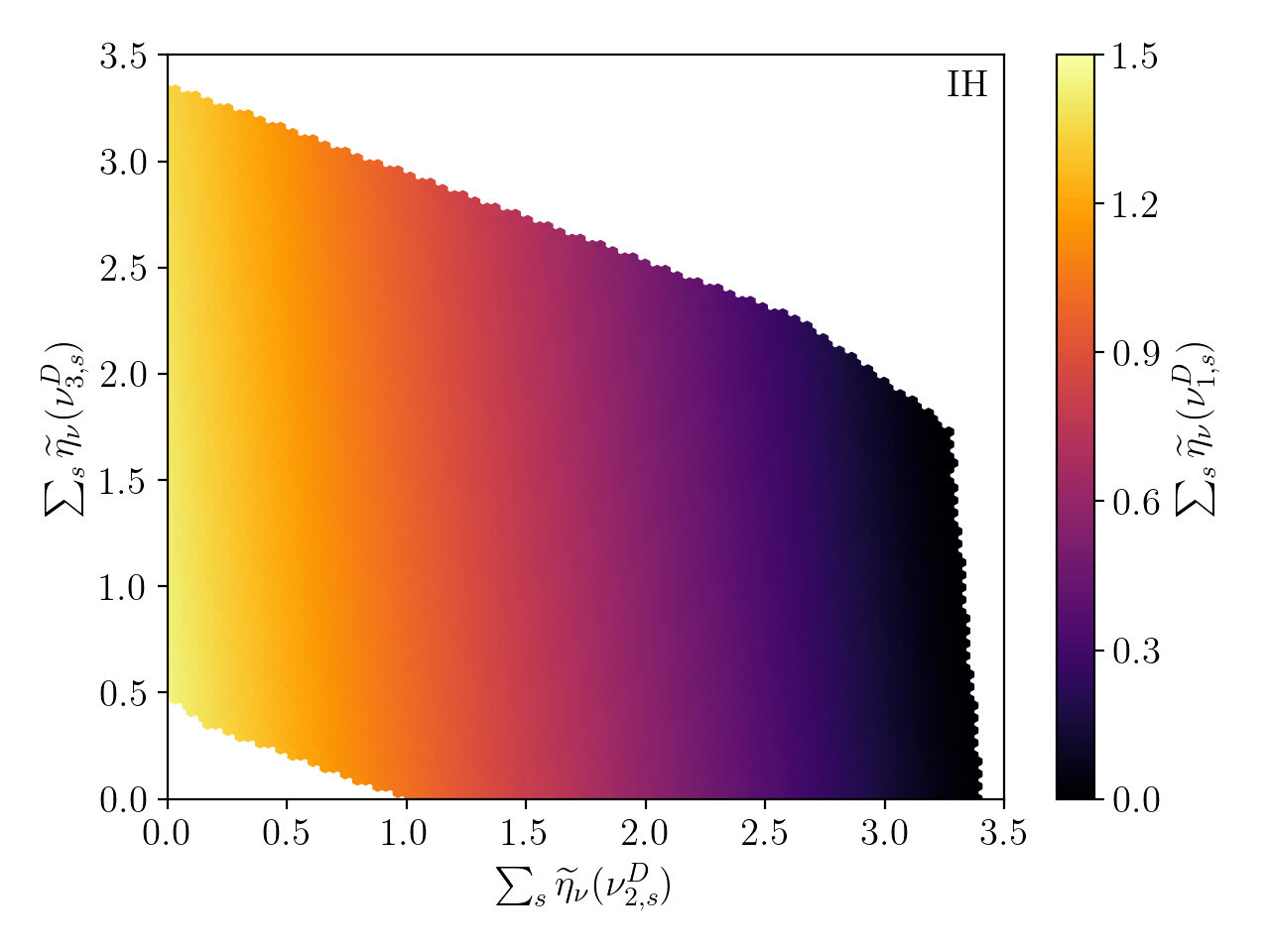}
\caption{\label{fig:cosConstraints}Cosmological constraints on the relic neutrino overdensity due to a degeneracy parameter $\xi_{\nu_i}$ in the $\sum_s \widetilde\eta_{\nu}(\nu_{2,s})$ - $\sum_s \widetilde\eta_{\nu}(\nu_{3,s})$ plane, where the colour shows the mean allowed magnitude of $\sum_s \widetilde\eta_{\nu}(\nu_{1,s})$. Any point in the white regions is excluded. Left: In the normal mass hierarchy. Right: In the inverted mass hierarchy.}
\end{figure}

It has been demonstrated in~\cite{Savage:1990by} that neutrino oscillations reduce asymmetry in the degeneracy parameter between the three neutrino states, such that $\xi_\nu$ should take a similar value for $\nu_\mu$ and $\nu_\tau$. However, later studies~\cite{Barenboim:2016shh,Barenboim:2016lxv} have argued that a full treatment of neutrino oscillations still allows for large $\xi_{\nu_\mu}$ and $\xi_{\nu_\tau}$ in spite of a small $\xi_{\nu_e}$. The strongest constraint on $\xi_{\nu_\mu}$ and $\xi_{\nu_\tau}$ therefore comes from $\Delta N_{\mathrm{eff}}$. In the standard scenario, the contribution of $\xi_{\nu_\alpha}$ to $\Delta N_\mathrm{eff}$ is given by
\begin{equation}
    \Delta N_\mathrm{eff}^\xi = \sum_\alpha\left[ \frac{30}{7}\left(\frac{\xi_{\nu_\alpha}}{\pi}\right)^2 + \frac{15}{7}\left(\frac{\xi_{\nu_\alpha}}{\pi}\right)^4 + \mathcal{O}(\xi_{\nu_\alpha}^6)\right],
\end{equation}
which using the 95\% CL result $\Delta N_\mathrm{eff} < 0.30$ from Planck 2018~\cite{Planck:2018vyg} translates to $|\xi_{\nu_\alpha}| \lesssim 0.82$. Including the BBN constraint on $\xi_{\nu_e}$ and using $n_{\nu}(\nu_{\alpha,s}) = \sum_i |U_{i\alpha}|^2 n_{\nu}(\nu_{i,s})$, we find the constraints on the relic neutrino overdensity
\begin{align}
    0.984 \leq &\sum_{i,s} |U_{ei}|^2 \widetilde\eta_{\nu}\left(\nu_{i,s}^D\right) \leq 1.007,\label{eq:c1Cos}\\
    0.46 \leq &\sum_{i,s} |U_{\alpha i}|^2 \widetilde\eta_{\nu}\left(\nu_{i,s}^D\right) \leq 2.06, \quad \alpha = \mu,\tau,\label{eq:c2Cos}
\end{align}
By substituting in the values of the PMNS matrix, the constraints~\eqref{eq:c1Cos} and~\eqref{eq:c2Cos} can be used to constrain the individual overdensities. We show the allowed region in Figure~\ref{fig:cosConstraints}. Clearly, the constraints from BBN and $\Delta N_\mathrm{eff}$ are strongest for $\nu_1$ due to its large overlap with $\nu_e$. For the remaining two states, however, overdensities as large as $\sum_s \widetilde\eta_{\nu}\sim\mathcal{O}(3)$ are permitted in both mass hierarchies. Several other works, \textit{e.g.}~\cite{Burns:2022hkq,Escudero:2022okz}, find bounds on the neutrino degeneracy parameters, and subsequently the overdensities, that are of the same order of magnitude.

For completeness, we also note that $\Delta N_\mathrm{eff}$ gives constraints on the relic neutrino temperature during decoupling. Assuming no chemical potential, $g_{\nu_i} = 1$ and constant temperature $T_{\nu}$ for all three mass eigenstates, $N_\mathrm{eff}$ scales with the neutrino temperature as $N_\mathrm{eff}\sim T_{\nu}^4$. We therefore find the relation
\begin{equation}
    \frac{N_\mathrm{eff}}{N_{\mathrm{eff},0}} = \left(\frac{T_{\nu}}{T_{\nu,0}}\right)^4,
\end{equation}
where $N_{\mathrm{eff},0} = 3.044$ is the predicted value of $N_\mathrm{eff}$ in standard cosmology, taking both neutrino oscillations and finite-temperature quantum electrodynamic effects into account~\cite{Akita:2020szl,Froustey:2020mcq,Bennett:2019ewm,Bennett:2020zkv}, and whose value is largely insensitive to the CP-violating phase~\cite{Froustey:2021azz}. Defining $\Delta N_\mathrm{eff} = N_\mathrm{eff} - N_{\mathrm{eff},0}$, we find
\begin{equation}
    \frac{T_{\nu}}{T_{\nu,0}} \leq \left(1+\frac{\Delta N_\mathrm{eff}}{N_{\mathrm{eff},0}}\right)^{\frac{1}{4}} \simeq 1.024,
\end{equation}
or equivalently $T_{\nu} \leq 0.172\,\mathrm{meV}$. Assuming equilibrium number density scaling~\eqref{eq:eqScaling}, we can also translate this to the overdensity constraint $\sum_s \widetilde\eta_{\nu}(\nu_{i,s}) \leq 1.073$. If the degenerate temperature constraint is relaxed, however, a combination of states with $T_{\nu_i} < T_{\nu,0}$ and $T_{\nu_i} > T_{\nu,0}$ could still reproduce the measured value of $N_\mathrm{eff}$. In the most extreme case with two neutrinos at temperature $T_{\nu_i} = 0$ and a third, hot, neutrino state, still with $g_{\nu_i} = 1$, the temperature bound becomes $T_{\nu_i} \leq 0.227\,\mathrm{meV}$. This corresponds to $\sum_s \widetilde\eta_{\nu}(\nu_{i,s}) \leq 2.47$.

Despite their strength, we once again stress that these constraints only hold if the C$\nu$B is largely unmodified between the early universe and today. Extended scenarios, \textit{e.g.} late time decays to or of neutrinos, could significantly alter the C$\nu$B from its profile in the early universe.

\subsection{Baryon acoustic oscillations}
The presence of relativistic, weakly interacting degrees of freedom, such as neutrinos, in the early universe has profound effects on the primordial photon-baryon plasma. Due to a lack of interactions, hot neutrinos free-stream with speed $\beta_{\nu_i} \simeq 1$, whilst sound waves in the plasma propagate at a speed $\beta_s \simeq 1/\sqrt{3}$. Neutrinos therefore travel ahead of the sound horizon, leaving metric perturbations in their wake that are felt by the succeeding sound waves~\cite{Baumann:2019keh,Baumann:2017gkg,Bashinsky:2003tk}. The result is a phase shift in the BAO spectrum that depends on the wavenumber, $k_s$, which can be parameterised as~\cite{Baumann:2019keh,Baumann:2017gkg}
\begin{equation}\label{eq:BAOshift}
    \phi_\mathrm{BAO}(k_s) \equiv b_\mathrm{BAO} \mathcal{F}(k_s),
\end{equation}
where $b_\mathrm{BAO}$ is the amplitude of the phase shift and $\mathcal{F}(k_s)$ denotes its wavenumber dependence. The amplitude of the phase shift depends on $N_\mathrm{eff}$, and is normalised such that $b_\mathrm{BAO} = 1$ corresponds to the SM prediction, $N_{\mathrm{eff}} = 3.046$, whilst $b_\mathrm{BAO} = 0$ and the limit $b_\mathrm{BAO} \to 2.45$ correspond to $N_\mathrm{eff} = 0$ and $N_\mathrm{eff} \to \infty$, respectively. Attributing the phase shift to neutrinos, the amplitude of the phase shift can be written as
\begin{equation}\label{eq:baoForm}
    b_\mathrm{BAO} = \frac{1}{\varepsilon_\mathrm{fid}} \frac{\rho_\nu}{\rho_\nu + \rho_\gamma},
\end{equation}
where $\rho_{\nu} = \sum_{\nu,i,s} |\vec{p}_{\nu_i}| \widetilde n_{\nu}(\nu_{i,s})$ and $\rho_\gamma$ are the total energy density in neutrinos and photons, respectively, whilst $\varepsilon_\mathrm{fid} \simeq 0.407$ is the SM prediction for fractional energy density in neutrinos during the radiation-dominated era.

\begin{equation}\label{eq:odBAO}
    \sum_{\nu,i,s} \frac{|\vec{p}_{\nu_i}|}{|\vec{p}_{\nu,0}|} \widetilde \eta_{\nu}(\nu_{i,s}) = \frac{16}{7} \left(\frac{T_{\gamma,0}}{T_{\nu,0}}\right)^4 \frac{b_\mathrm{BAO} \varepsilon_\mathrm{fid}}{1 - b_\mathrm{BAO} \varepsilon_\mathrm{fid}},
\end{equation}
where we have used the present day photon energy density, $\rho_\gamma = \pi^2 T_{\gamma,0}^4/15$ and we remind the reader that $|\vec{p}_{\nu,0}| \simeq 3.15\,T_{\nu,0}$. As should be expected, the right hand side of~\eqref{eq:odBAO} gives a value of six for $b_\mathrm{BAO} = 1$.

Using the BOSS DR12 dataset~\cite{BOSS:2016wmc} and without making any assumptions about the underlying cosmology, the authors of~\cite{Baumann:2019keh} find the value $b_\mathrm{BAO} = 1.2\pm 1.8$. The central value of this measurement predicts a set of overdensities satisfying $\eta_{\mathrm{tot}} \simeq 8.3$, however, the $1\sigma$ error bounds allow for the full range of values $\eta_\mathrm{tot}\in [0,\infty]$, where $\eta_{\mathrm{tot}}$ is understood to be the left hand side of~\eqref{eq:odBAO}. By instead assuming a $\Lambda\mathrm{CDM}$ cosmology, for which the standard scenario applies with $|\vec{p}_{\nu,i}| = |\vec{p}_{\nu,0}|$ and the number density ratios given in~\eqref{eq:diracAbundance} and~\eqref{eq:majoranaAbundance}, the same study~\cite{Baumann:2019keh} finds a more restricted value $b_\mathrm{BAO} = 2.22\pm 0.75$. At $99\%$ significance, this gives the bound on a common overdensity for the six populated neutrino states of
\begin{equation}\label{eq:baoBound99}
    \widetilde\eta_\nu(\nu_{i,s}) \geq 0.19,
\end{equation}
which allows for scenarios with significantly diminished cosmic neutrino backgrounds. However, the same result excludes $\eta_\nu(\nu_{i,s}) = 0$ at $99.69\%$ CL or $2.96\sigma$. Importantly,~\eqref{eq:baoBound99} represents the strongest lower bound on the Majorana neutrino overdensity, as the bounds given in~\eqref{eq:c1Cos} and~\eqref{eq:c2Cos} only apply to Dirac neutrinos. As with the other cosmology bounds, however,~\eqref{eq:baoBound99} only applies to scenarios in which the C$\nu$B is largely unmodified between radiation-domination and the present era.
\subsection{CMB polarisation}
Photons scatter on neutrinos at a rate proportional to $(\alpha G_F)^2$, which is enhanced in the early radiation-dominated universe where both the relic neutrino and photon number densities are large. As relativistic neutrinos are almost exclusively left helicity, the CMB photons that scatter on relic neutrinos are polarised~\cite{Bakhti:2014zya}. Several studies~\cite{Mohammadi:2013dea,Khodagholizadeh:2014nfa,Mohammadi:2016bxl} suggest that this could contribute to the B-mode power spectrum of the CMB at large multipole moments $50\lesssim l \lesssim 200$, modifying the ratio of tensor-scalar ratio $r_\mathrm{TS}$. Assuming the standard C$\nu$B scenario, the authors of~\cite{Khodagholizadeh:2014nfa} estimate the contribution to $r_\mathrm{TS}$ from this effect to be $\sim 0.025$, which is currently constrained using combined measurements from Planck~\cite{Planck:2020olo} and BICEP~\cite{BICEP:2021xfz} to $r_\mathrm{TS}<0.032$~\cite{Tristram:2021tvh}. Both the Simons Observatory and CMB-S4 forecast sensitivity to $r_\mathrm{TS} \sim \mathcal{O}(10^{-3})$~\cite{SimonsObservatory:2018koc,Thorne:2019mrd,CMB-S4:2020lpa}, allowing them to place constraints on this effect.

The magnitude of the contribution to the B-mode spectrum depends on the averaged relic neutrino number density,
\begin{equation}\label{eq:avND}
    \bar n_{\nu}(\nu_{i,s}) = \int\displaylimits_0^{z_\mathrm{LSS}}  \frac{\widetilde n_{\nu}(\nu_{i,s})(z)}{(1+z)H(z)} \,dz,
\end{equation}
where $z_\mathrm{LSS}\simeq 1100$ is the redshift at the last scattering surface and $H(z)$ is the Hubble parameter. Under the assumption that relic neutrinos do not interact strongly since decoupling, the neutrino number density scales as $\widetilde n_{\nu}(\nu_{i,s})(z) = (1+z)^3 \widetilde n_{\nu}(\nu_{i,s})$. In this case, $\bar n_{\nu}$ will be proportional to the present day number densities, allowing us to place constraints. However, as the integrand of~\eqref{eq:avND} is likely to peak strongly at large $z$, if we relax the assumption of minimally interacting neutrinos since decoupling then the CMB polarisation provides very little insight into the present day number density. Perhaps more interestingly, measurements indicating no contribution from this effect would indicate a lack of polarisation in the C$\nu$B, particularly at early times when they are relativistic. As relativistic neutrinos are expected to be exclusively left helicity, this would require significant new physics. Finally, we note that this effect is expected to be twice as large for Majorana neutrinos than for Dirac neutrinos~\cite{Bakhti:2014zya}. 
\section{Direct detection proposals}\label{sec:direct}
There exist several unique proposals to hunt for the C$\nu$B, despite the multitude of difficulties in observing relic neutrinos. Each of these is sensitive to different regions of the temperature, mass and overdensity parameter space, with some capable of offering additional information about the Dirac or Majorana nature of neutrinos. Here we discuss direct detection proposals, where the product of a relic neutrino interaction is directly observed.

\subsection{PTOLEMY}\label{sec:ptolemy}
The PTOLEMY experiment aims to detect the C$\nu$B by capturing electron neutrinos on a $100\,\mathrm{g}$ tritium target in the process ${^3}\mathrm{H} + \nu_e \to {^3}\mathrm{He}^{+} + e^-$~\cite{PTOLEMY:2018jst}, as first proposed by Weinberg in 1963~\cite{Weinberg:1962zza} and later explored alongside several other candidate targets in~\cite{Cocco:2007za}. Importantly, this process has no energy threshold, making the capture of relic neutrinos possible independently of their mass and temperature. The signature at PTOLEMY is an electron emitted with energy $E_{\mathrm{C}\nu\mathrm{B},i} = K_\mathrm{end} + m_e + m_{\nu_l} + E_{\nu_i}$~\cite{Long:2014zva}, where $m_e$ and $m_{\nu_l}$ are the electron and lightest neutrino mass, respectively. Including the effects of nuclear recoil, the endpoint kinetic energy\footnote{Due to nuclear recoil, $K_\mathrm{end}$ is smaller than the $Q$-value of tritium $Q_\mathrm{H}\simeq 18.59\,\mathrm{keV}$ by $\sim 3.4\,\mathrm{eV}$~\cite{Long:2014zva, Purcell:2015gtm}. As this difference is larger than the neutrino mass, we use $K_\mathrm{end}$ in our analysis instead of $Q_\mathrm{H}$.} of electrons emitted in tritium $\beta$-decay is given by
\begin{equation}
    K_\mathrm{end} = Q_\mathrm{H} - \frac{m_e Q_\mathrm{H}}{m_{^3\mathrm{H}}} - \frac{Q_\mathrm{H}^2}{2 m_{^3\mathrm{He}}}.
\end{equation}
This takes the approximate value $K_\mathrm{end} \simeq 18.59\,\mathrm{keV}$, and is given in terms of the energy release $Q_\mathrm{H} = m_{^3\mathrm{H}} - m_{^3\mathrm{He}} - m_e - m_{\nu_l}$, where $m_{^3\mathrm{H}}$ and $m_{^3\mathrm{He}}$ denote the nuclear masses of tritium and helium-3 in turn. An excess of electrons with energies $m_{\nu_l} + E_{\nu_i}$ beyond the tritium $\beta$-decay endpoint energy, $K_\mathrm{end} + m_e$, would therefore signal the capture of low energy neutrinos, such as those from the C$\nu$B.

Following the formalism of~\cite{Long:2014zva}, the neutrino capture rate on tritium per mass eigenstate is
\begin{equation}\label{eq:ptolemyRate}
    \Gamma_{\mathrm{C}\nu\mathrm{B}}(\nu_{i,s}) =N_T|U_{ei}|^2\bar\sigma(E_{\mathrm{C}\nu\mathrm{B},i})\mathcal{A}_s(\beta_{\nu_i})n_\nu(\nu_{i,s}),
\end{equation}
where $N_T \simeq 2\times 10^{25}$ is the number of active tritium atoms in the target and $\bar\sigma$ is the neutrino capture cross section. The function $\mathcal{A}_s$ encodes the helicity dependence of the cross section
\begin{equation}\label{eq:aSpin}
    \mathcal{A}_s(\beta_{\nu_i}) = \begin{cases}
       1-s\beta_{\nu_i}, \quad & \nu = \nu^D,\\
       1+s\beta_{\nu_i}, \quad & \nu = \bar\nu^D,
    \end{cases}
\end{equation} where $s = \pm 1$ for right (+) and left (-) helicity neutrinos, respectively. For Majorana neutrinos, $\mathcal{A}_s$ should be chosen according to the equivalent Dirac neutrino process. We immediately see from~\eqref{eq:ptolemyRate} that PTOLEMY is sensitive to the helicity composition of the C$\nu$B. On the contrary, as tritium can only be used to capture neutrinos, PTOLEMY is unable to place any constraints on antineutrinos.

The capture cross section is given in terms of the final state electron energy and 3-momentum, $E_e$ and $\vec{p}_e$, by
\begin{equation}\label{eq:nucapXsec}
    \bar\sigma(E_{e}) = \frac{G_F^2}{2\pi}|V_{ud}|^2 F(Z,E_e) \frac{m_{^3\mathrm{He}}}{m_{^3\mathrm{H}}} C(|\vec{q}|^2) E_e |\vec{p}_e|,
\end{equation}
where $|V_{ud}|\simeq 0.974$ is an element of the Cabibbo-Kobayashi-Maskawa (CKM) quark mixing matrix~\cite{ParticleDataGroup:2020ssz}, $|\vec{q}|^2$ is the squared momentum transfer and $C(|\vec{q}|^2 \simeq 0) \simeq 5.49$ contains details of nuclear structure~\cite{Long:2014zva}. The Fermi function accounts for electromagnetic interactions between the final state electron and a nucleus with atomic number $Z$, and is given by
\begin{equation}
    F(Z,E_e) = \frac{2(1+S_\beta)}{\Gamma(1+2S_\beta)^2}\left(2 |\vec{p}_e|\rho_N\right)^{2S_\beta-2}e^{-\pi\eta_\beta}|\Gamma(S_\beta-i\eta_\beta)|^2,
\end{equation}
where $\eta_\beta = -Z\alpha E_e/|\vec{p}_e|$, $S_\beta = \sqrt{1 - (\alpha Z)^2}$ depend on the fine-structure constant $\alpha$, and $\rho_N \simeq 1.2 A^{1/3}\,\mathrm{fm}$ is the nuclear radius, which depends on the final state mass number $A$. At the endpoint, the cross section~\eqref{eq:nucapXsec} takes the constant value $\bar\sigma(E_{\mathrm{C}\nu\mathrm{B}})\simeq 3.84\times 10^{-45}\,\mathrm{cm}^2$ provided that $E_{\nu_i} \ll K_\mathrm{end}$. %$F(Z,\Eend)\simeq 1.19$

By summing over the mass eigenstates\footnote{If PTOLEMY is able to resolve the individual mass splittings, $\Delta m_{ij}^2$, then we do not perform this sum. Resolving the mass splittings requires an energy resolution $\Delta \ll \sqrt{\Delta m_{ij}^2}$, whilst PTOLEMY is expected to achieve an energy resolution $\Delta \simeq 0.05\,\mathrm{eV}$~\cite{Apponi:2021hdu,Betti:2018bjv}. As such, we will retain the sum for the remainder of this work.} and neglecting the neutrino energy dependence of the capture cross section, we find that PTOLEMY will be able to set the C$\nu$B overdensity constraint
\begin{equation}\label{eq:ptolemySignificance}
        \sum_{i,s} |U_{ei}|^2\mathcal{A}_s(\beta_{\nu_i})\eta_\nu(\nu_{i,s}) \leq \frac{4\pi^2}{3\zeta(3)T_{\nu,0}^3}\frac{N}{N_T}\frac{1}{t}\frac{1}{\bar\sigma(\Eend)} \simeq 0.244 N \left(\frac{1\,\mathrm{y}}{t}\right),
\end{equation}
after a runtime $t$, if $N$ events are required for statistical significance. As the counting error increases as $\sqrt{N}$, the significance scales like $N/\sqrt{N} = \sqrt{N}$. We therefore require $N\simeq 25$ events for a 5$\sigma$ discovery of the C$\nu$B. Interestingly, whilst the capture rate~\eqref{eq:ptolemyRate} does not explicitly depend on the Dirac or Majorana nature, the standard scenario predicts that the capture rate at PTOLEMY will differ between Dirac and Majorana neutrinos. As we expect only left helicity Dirac neutrinos in the standard scenario, but an additional right helicity abundance if neutrinos are Majorana in nature, the event rate at PTOLEMY should be twice as large for Majorana neutrinos as it is for Dirac neutrinos. This distinction vanishes for large neutrino velocities as the right helicity Majorana neutrino flux becomes non-interacting.

In order to place any constraints at all, however, PTOLEMY needs sufficient energy resolution to distinguish between $\beta$-decay and relic neutrino capture electrons. This roughly corresponds to an energy resolution requirement $\Delta \lesssim E_{\nu_i} + m_{\nu_l}$ to resolve the signal associated with neutrino mass eigenstate $\nu_i$. At PTOLEMY, the energy resolution goal is $\Delta = 0.05\,\mathrm{eV}$~\cite{Apponi:2021hdu,Betti:2018bjv}, such that for $E_{\nu_i} \simeq m_{\nu_i}$ the neutrino capture signal due to the heaviest neutrino state will only be resolvable if the lightest neutrino mass satisfies
\begin{equation}\label{eq:roughSens}
    m_{\nu_l} \gtrsim \frac{\Delta^2 - \Delta m_{hl}^2}{2\Delta},
\end{equation}
where $\Delta m_{hl}^2 = \Delta m_{31}^2$ (NH) or $ \Delta m_{21}^2 - \Delta m_{31}^2$ (IH) is the squared mass splitting between the heaviest and lightest neutrino mass eigenstates. Below this minimum mass threshold, no signal will be seen at all. With $\Delta = 0.05\,\mathrm{eV}$, the right hand side of~\eqref{eq:roughSens} is negative in both the NH and IH scenarios, such that with this naive estimate we expect that PTOLEMY should always be able to resolve at least some signal neutrinos. 

More rigorously, events at PTOLEMY will be collected in histogram bins of finite width $\Delta$, each centred on energy $E_c$. In order to see the signal from relic neutrino capture for a given bin, PTOLEMY requires a signal-noise ratio
\begin{equation}\label{eq:rsn}
    r_\mathrm{SN} = \frac{\Gamma_{\mathrm{C}\nu\mathrm{B}}^s(E_c,\Delta)}{\Gamma_\beta^s(E_c,\Delta)} \gtrsim r_{\mathrm{SN},0},
\end{equation}
where $\Gamma_\nu^s(E_c,\Delta)$ and $\Gamma_\beta^s(E_c,\Delta)$ are the finite-energy-resolution-smeared neutrino capture and tritium $\beta$-decay rates in the bin centred on $E_c$, respectively, and $r_{\mathrm{SN},0}$ is the minimum signal noise ratio required for a discovery, which we leave as a free parameter. The smeared capture rates are in turn defined by~\cite{Long:2014zva,PTOLEMY:2019hkd}
\begin{align}
    \Gamma_{\mathrm{C}\nu\mathrm{B}}^s(E_c,\Delta) &= \frac{1}{\sqrt{2\pi}\sigma}\int\displaylimits_{E_c-\frac{\Delta}{2}}^{E_c+\frac{\Delta}{2}}dE'_e \left(\sum_{i,s} \Gamma_{\mathrm{C}\nu\mathrm{B}}(\nu_{i,s})\exp\left[-\frac{(E'_e - E_{\mathrm{C}\nu\mathrm{B},i})^2}{2\sigma^2}\right]\right),\label{eq:smearedCapture}\\
    \Gamma_\beta^s(E_c,\Delta) &= \frac{1}{\sqrt{2\pi}\sigma}\int\displaylimits_{E_c-\frac{\Delta}{2}}^{E_c+\frac{\Delta}{2}}dE'_e\int\displaylimits_{m_e}^{E_\mathrm{end}}dE_e \left(\frac{d\Gamma_\beta}{dE_e} \exp\left[-\frac{(E'_e - E_e)^2}{2\sigma^2}\right]\right),\label{eq:smearedBeta}
\end{align}
where $\sigma = \Delta/\sqrt{8\ln2}$ is the standard deviation of the Gaussian smearing envelope. The $\beta$-decay spectrum is~\cite{Masood:2007rc}
\begin{equation}
    \frac{d\Gamma_\beta}{dE_e} \simeq \frac{1}{\pi^2} N_T \sum_i |U_{ei}|^2 \bar\sigma(E_{\mathrm{C}\nu\mathrm{B}}) H_\beta(E_e, m_{\nu_i}), 
\end{equation}
with the shape function defined by
\begin{equation}
    \begin{split}
    H_{\beta}(E_e, m_{\nu_i})=\frac{1-m_e^2/(E_e \mh)}{(1-2E_e/\mh+m_e^2/\mh^2)^2}&\sqrt{y\left(y+\frac{2m_{\nu_i}\mhe}{\mh}\right)}\\
    &\times \left(y+\frac{m_{\nu_i}}{\mh}(\mhe + m_{\nu_i})\right),
    \end{split}
\end{equation}
where $y = E_\mathrm{end} - E_e$ and  $E_\mathrm{end} = K_\mathrm{end} + m_e$. The largest signal-noise ratio will be found in a bin centred on the most energetic mass eigenstate, $E_c = E_m \equiv \mathrm{max}\left\{E_{\mathrm{C}\nu\mathrm{B},i}\right\}$, for which the smeared neutrino capture rate~\eqref{eq:smearedCapture} reduces to
\begin{equation}
    \Gamma_{\mathrm{C}\nu\mathrm{B}}^s(E_m,\Delta) = \sum_{i,s} \Gamma_{\mathrm{C}\nu\mathrm{B}}(\nu_{i,s})G(\Delta E_{\nu_i},\Delta),
\end{equation}
where $\Delta E_{\nu_i} = E_{\nu_i} - E_{\nu,\mathrm{max}}$ is the difference in energy between mass eigenstate $i$ and the most energetic C$\nu$B neutrino state, which for non-relativistic neutrinos will be of order the mass splittings. The integral function $G$ is defined by  
\begin{equation}\label{eq:Gnu}
    G(\Delta E_{\nu_i},\Delta) = \frac{1}{2}\left\{\mathrm{erf}\left[\left(1 + \frac{2\Delta E_{\nu_i}}{\Delta}\right)\sqrt{\ln 2}\right]+\mathrm{erf}\left[\left(1 - \frac{2\Delta E_{\nu_i}}{\Delta}\right)\sqrt{\ln 2}\right]\right\},
\end{equation}
which takes the approximately constant value $G(\Delta E_{\nu_i},\Delta) \simeq 0.761$ for $\Delta E_i \ll \Delta$. With the same choice, $E_c = E_m$, we can perform the integral over $E'_e$ in~\eqref{eq:smearedBeta}, yielding
\begin{equation}
    \Gamma^s_{\beta}(E_m,\Delta) = \int\displaylimits_{m_e}^{E_\mathrm{end}} dE_e \frac{d\Gamma_\beta}{dE_e} G(\Delta E_m, \Delta),
\end{equation}
where $\Delta E_m = E_m - E_e$. The energy resolution requirement therefore corresponds to the complimentary constraint on the C$\nu$B overdensity
\begin{equation}\label{eq:ptolemyResolution}
    \begin{split}
        \sum_{i,s} |U_{ei}|^2\mathcal{A}_s(\beta_{\nu_i})\eta_\nu(\nu_{i,s}) G(\Delta E_{\nu_i},\Delta)\leq \frac{4 r_{\mathrm{SN},0}}{3\zeta(3)}\frac{1}{T_{\nu,0}^3} \int\displaylimits_{m_e}^{E_\mathrm{end}}dE_e \sum_i |U_{ei}|^2 H_\beta(E_e, m_{\nu_i})\\
        \times G(\Delta E_m,\Delta).
     \end{split}
\end{equation}
As both of the conditions $\sum_{i,s} \Gamma_{\mathrm{C}\nu\mathrm{B}}(\nu_{i,s})t \geq N$ and~\eqref{eq:rsn} need to be satisfied to make a statistically significant discovery of the C$\nu$B, the constraint on the relic neutrino overdensity for a given set of input parameters should be chosen as the weakest bound of~\eqref{eq:ptolemySignificance} and~\eqref{eq:ptolemyResolution}. The efficacy of PTOLEMY has also been explored in~\cite{Alvey:2021xmq} for several specific C$\nu$B scenarios.

The energy resolution strongly limits the range of neutrino masses that could be observed at PTOLEMY, with the signal-noise ratio rapidly diminishing for $m_{\nu_i} \ll \Delta$. To that end, a more recent work~\cite{Akhmedov:2019oxm} has suggested hunting for relic neutrinos using angular correlations in neutrino capture on $\beta$-decaying nuclei. By considering the polarisation of the target nucleus, along with the polarisation of the outgoing electron, the authors of~\cite{Akhmedov:2019oxm} find additional terms proportional to products of $\vec{\beta}_{\nu_i}$, $\vec{\beta}_{e}$, $\vec{n}_N$, $\vec{n}_{\nu_i}$, $\vec{n}_e$ that contribute to~\eqref{eq:aSpin}, where $\vec{n}$ denotes the direction of the particle spin in its own reference frame. 

As a result of the periodic motion of the Earth with respect to the C$\nu$B rest frame, arising from the rotation of the Earth about the Sun and its own axis, these quantities all have a time dependence. This leads to a time dependent capture rate, which could help to distinguish electrons originating from neutrino capture from those emitted in $\beta$-decays. For a peculiar velocity $\beta_\Earth \simeq 10^{-3}$ and neutrino masses $m_{\nu_i} \lesssim 0.05\,\mathrm{eV}$, below the energy resolution of PTOLEMY, the authors predict that the capture rate will vary by $\sim 0.1\%$. Given that for the standard scenario without overdensities, approximately four events are expected per year for Dirac neutrinos, and eight for Majorana neutrinos~\cite{Long:2014zva}, this small variation will have little to no effect on the capture rate at PTOLEMY. 

To observe a consistent variation of one event per year due to this effect would require overdensities $\eta_\nu(\nu_{i,s}^D) \gtrsim 250$, or $\eta_\nu(\nu_{i,s}^M) \gtrsim 125$, corresponding to a few thousand neutrino captures per year. As we will show in Section~\ref{sec:discussion}, these overdensities lie below those required for the standard PTOLEMY setup to be sensitive to the C$\nu$B in the region where $m_{\nu_i} \ll \Delta$, such that this method could improve the efficacy of PTOLEMY. More concerning, however, is that variations in the stochastic background of $\beta$-decay electrons will far exceed variations due to the time dependent signal. This is also taken into account in~\cite{Akhmedov:2019oxm}, where the authors estimate that with appropriate signal processing techniques, a signal-noise ratio of
\begin{equation}\label{eq:timeDepRSN}
    r_\mathrm{SN} \simeq \frac{9}{40}\frac{1}{\Gamma_\beta^s(E_c,\Delta)} \left(A_{\mathrm{C}\nu\mathrm{B}} \Gamma_{\mathrm{C}\nu\mathrm{B}}^s(E_c,\Delta)\right)^2 t\gtrsim r_{\mathrm{SN},0},
\end{equation}
can be achieved, where $A_{\mathrm{C}\nu\mathrm{B}}$ is the amplitude of the time variation and therightmost inequality denotes the requirement to make a discovery using this technique. Note that unlike the standard approach to PTOLEMY, this signal-noise ratio increases with experimental runtime as well as the number of targets, $N_T$, through the additional factor of the neutrino capture rate. By substituting the smeared capture rates~\eqref{eq:smearedCapture} and~\eqref{eq:smearedBeta} into~\eqref{eq:timeDepRSN}, we find the limit that can be set on the overdensity using this method
\begin{equation}\label{eq:ptolemyResolutionTime}
    \begin{split}
        \sum_{i,s} |U_{ei}|^2\mathcal{A}_s(\beta_{\nu_i})\eta_\nu(\nu_{i,s}) G(\Delta E_{\nu_i},\Delta)&\leq \frac{8\sqrt{10}\pi}{9A_{\mathrm{C}\nu\mathrm{B}}\zeta(3)}\frac{1}{T_{\nu,0}^3}\sqrt{\frac{r_{\mathrm{SN},0}}{N_T \bar\sigma(E_{\mathrm{C}\nu\mathrm{B}}) t}} \\ & \times \left[\int\displaylimits_{m_e}^{E_\mathrm{end}}dE_e \sum_i |U_{ei}|^2 H_\beta(E_e, m_{\nu_i})
        G(\Delta E_m,\Delta)\right]^{\frac{1}{2}}.
     \end{split}
\end{equation}
As stated previously, we also need sufficient events to observe a time variation at all, which in line with~\eqref{eq:ptolemySignificance} will correspond to the complimentary constraint
\begin{equation}\label{eq:ptolemySignificanceTime}
    \sum_{i,s} |U_{ei}|^2\mathcal{A}_s(\beta_{\nu_i})\eta_\nu(\nu_{i,s}) \leq 0.244 \left(\frac{N}{A_{\mathrm{C}\nu\mathrm{B}}}\right) \left(\frac{1\,\mathrm{y}}{t}\right).
\end{equation}
In line with this reasoning, the limit on overdensity that can be set using this method will be the weakest of the bounds~\eqref{eq:ptolemyResolutionTime} and~\eqref{eq:ptolemySignificanceTime}. In practice, the value of $A_{\mathrm{C}\nu\mathrm{B}}$ will depend on several properties including the neutrino mass, temperature and the peculiar velocity of the Earth. For simplicity, however, we will use the constant value $A_{\mathrm{C}\nu\mathrm{B}} = 0.001$ for the remainder of this paper, which holds in the low mass regime where this method is expected to be most effective. Clearly, this method offers an additional window through which the C$\nu$B may be detected, which with an appropriate choice of target may be able to set strong bounds on relic neutrinos in the regions of parameter space where the finite energy resolution of PTOLEMY becomes problematic. Finally, we note that this result may be further improved with more advanced signal processing techniques~\cite{Samsing:2019,Akhmedov:2019oxm}.  

A similar technique to PTOLEMY using neutrino capture on both $\beta^{+}$ and electron-capture-decaying (EC) nuclei has been explored in~\cite{Cocco:2007za} and~\cite{Cocco:2009rh}, which could instead be used to detect antineutrinos in the C$\nu$B. Here, the signal is an excess of final state positrons with energies $m_{\nu_l} + E_{\nu_i}$ beyond the endpoint energy of the decay process, analogous to that of PTOLEMY. In addition to the decay positrons, however, there will also be a background of photons originating from the de-exciting EC daughter nuclei, which may complicate detection \textit{e.g.} through scattering on an outgoing positron. Nevertheless, this remains an alternative method through which the C$\nu$B could be detected using a thresholdless process. We also note that~\cite{Cocco:2009rh} makes a very important point regarding neutrino capture on nuclei at rest. If the target is stable but has a decay threshold smaller than twice the neutrino mass, neutrinos of all energies can be captured without background. This would constitute an unparalleled technique to detect the C$\nu$B if a suitable target could be found.
\subsection{Stodolsky effect}\label{sec:stodolsky}
Another widely discussed proposal to detect the C$\nu$B uses the elastic scattering of relic neutrinos on macroscopic targets. This can be roughly decomposed into two effects. The Stodolsky effect~\cite{Domcke:2017aqj, Duda:2001hd, Stodolsky:1974aq}, in which the presence of a neutrino background acts as a potential that changes the energy of atomic electron spin states, analogous to the Zeeman effect in the presence of a magnetic field. The second uses neutral current scattering of relic neutrinos on a test mass~\cite{Shergold:2021evs, Domcke:2017aqj, Duda:2001hd,Opher:1974drq,Lewis:1979mu,Shvartsman:1982sn,Smith:1983jj,Cabibbo:1982bb,Gelmini:2004hg,Ringwald:2004np,Vogel:2015vfa}, which is considerably enhanced by a coherence factor due to the macroscopic de Broglie wavelength of relic neutrinos~\cite{Shergold:2021evs, Domcke:2017aqj, Duda:2001hd, Opher:1974drq, Shvartsman:1982sn, Smith:1983jj, Lewis:1979mu}, $\lambda_\nu \sim \mathcal{O}(\mathrm{mm})$. Both of these effects may be observed from the small momenta that they impart to the target. 

We begin by focusing on the Stodolsky effect. At low energies, the Hamiltonian density for neutrino-electron interactions in the flavour basis is
\begin{equation}\label{eq:neutrinoHamiltonianf}
    \mathcal{H}(x) = \frac{G_F}{\sqrt{2}}\left[\sum_\alpha \,\bar{\nu}_\alpha\gamma_\mu(1-\gamma^5)\nu_\alpha\,\bar{e}\gamma^\mu (g_V^e-g_A^e \gamma^5)e
    +\bar{\nu}_e\gamma_\mu(1-\gamma^5)e\,\bar{e}\gamma^\mu (1-\gamma^5)\nu_e\right],
\end{equation}
where $g_V^e = -1/2 + 2\sin^2\theta_W$ and $g_A^e = -1/2$ are the electron vector and axial-vector couplings to the $Z$-boson, respectively, given in terms of the Weinberg angle $\theta_W$. The first term in~\eqref{eq:neutrinoHamiltonianf} contains flavour diagonal neutral current interactions, whilst the second term accounts for charged current interactions, in which only electron neutrinos can partake.  It is instructive to switch to the mass basis as we are interested in relic neutrinos, which have long since decohered to mass eigenstates. To do so, we note that $\nu_{\alpha} = \sum_\alpha U_{\alpha i} \nu_i$ and use the unitarity of the PMNS matrix, $\sum_\alpha U_{\alpha i}^* U_{\alpha j} = \delta_{ij}$, to find
\begin{equation}\label{eq:neutrinoHamiltonianm}
    \mathcal{H}(x) = \frac{G_F}{\sqrt{2}}\sum_{i,j} \,\bar{\nu}_i\gamma_\mu(1-\gamma^5)\nu_j\,\bar{e}\gamma^\mu (V_{ij}-A_{ij} \gamma^5)e,
\end{equation}
where we have introduced $V_{ij} = \delta_{ij} g_V^e + U_{ei}^* U_{ej}$ and $A_{ij} = \delta_{ij} g_A^e + U_{ei}^* U_{ej}$ for brevity. In going from~\eqref{eq:neutrinoHamiltonianf} to~\eqref{eq:neutrinoHamiltonianm}, we have also applied a Fierz transformation to the charged current to separate the neutrino and electron currents, allowing for both the charged and neutral currents to be combined into a single term.

To leading order in $\mathcal{H}(x)$, the energy shift of an electron with spin $s_e$ and momentum $\vec{p}_e$ in the presence of a neutrino background with approximately uniform momentum $\vec{p}_{\nu_k}$ is given by
\begin{equation}\label{eq:energyShiftFull}
    \Delta E_e(\vec{p}_e, s_e) = \sum_{\nu,k,s} N_{\nu}(\nu_{k,s}) \int {d^3x}\frac{\langle e_{s_e},\nu_{k,s}| \mathcal{H}(x)|e_{s_e},\nu_{k,s}\rangle}{\langle e_{s_e},\nu_{k,s}|e_{s_e},\nu_{k,s}\rangle},
\end{equation}
where normal ordering is implied and we have summed over all neutrinos and antineutrinos, mass eigenstates and helicities.

We use relativistic normalisation for the external states
\begin{align}
        \left|e_{s_e}\right\rangle &= \sqrt{2E_e}a^\dagger_e(\vec{p}_e,s_e)\left|0\right\rangle,\\ \left|\nu_{k,s}\right\rangle &= \sqrt{2E_{\nu_k}}a^\dagger_\nu(\vec{p}_{\nu_k},s)\left|0\right\rangle,\\
        \left|\bar\nu_{k,s}\right\rangle &= \sqrt{2E_{\nu_k}}b^\dagger_\nu(\vec{p}_{\nu_k},s)\left|0\right\rangle,
\end{align}
where $a^\dagger(\vec{p},s)$ and $b^{\dagger}(\vec{p},s)$ are the creation operators for particles and antiparticles with momentum $\vec{p}$ and helicity $s$ respectively. Along with their respective annihilation operators, $a(\vec{p},s)$ and $b(\vec{p},s)$, these satisfy the standard anticommutation relations
\begin{equation}
    \left\{a_i(\vec{p},r),a^{\dagger}_j(\vec{q},s)\right\} = \left\{b_i(\vec{p},r),b^{\dagger}_j(\vec{q},s)\right\} = (2\pi)^3 \delta^{(3)}(\vec{p} - \vec{q}) \delta_{rs}\delta_{ij},
\end{equation}
with all other anticommutators vanishing identically. With these definitions, the denominator of~\eqref{eq:energyShiftFull} trivially evaluates to
\begin{equation}\label{eq:normalisationStodolsky}
    \langle e_{s_e},\nu_{k,s}|e_{s_e},\nu_{k,s}\rangle = 4 E_{e}E_{\nu_k} V^2,
\end{equation}
where $V = (2\pi)^3 \delta^{(3)}(\vec{0})$ is the volume. This is formally infinite, but we will see that the volume factors cancel later on. To evaluate the numerator of~\eqref{eq:energyShiftFull}, we first define the field operators
\begin{align}\label{eq:fieldDef1}
        \psi^D(x) = \int \frac{d^3 p}{(2\pi)^3}\frac{1}{\sqrt{2E_p}}\sum_s\left(a(\vec{p},s)u(p,s)e^{-ip\cdot x} + b^\dagger(\vec{p},s)v(p,s)e^{ip\cdot x}\right),\\
        \bar{\psi}^D(x) = \int \frac{d^3p}{(2\pi)^3}\frac{1}{\sqrt{2E_p}}\sum_s\left(a^\dagger(\vec{p},s)\bar{u}(p,s)e^{ip\cdot x} + b(\vec{p},s)\bar{v}(p,s)e^{-ip\cdot x}\right),\label{eq:fieldDef2}
\end{align}
in terms of the positive and negative frequency spinors $u$ and $v$. For Majorana fields, the $b$ and $b^\dagger$ operators appearing in~\eqref{eq:fieldDef1} and~\eqref{eq:fieldDef2} should be replaced by $a$ and $a^\dagger$ respectively. After a little work, the numerator of~\eqref{eq:energyShiftFull} for Dirac neutrino fields evaluates to
\begin{align}
    \langle e_{s_e},\nu_{k,s}| \mathcal{H}^D|e_{s_e},\nu_{k,s}\rangle &= \frac{G_F}{\sqrt{2}}\bar{u}(p_{\nu_k},s)\gamma_\mu (1-\gamma^5) u(p_{\nu_k},s) j^\mu_k,\label{eq:nuME}\\
    \langle e_{s_e},\bar\nu_{k,s}| \mathcal{H}^D|e_{s_e},\bar\nu_{k,s}\rangle &= -\frac{G_F}{\sqrt{2}}\bar{v}(p_{\nu_k},s)\gamma_\mu (1-\gamma^5) v(p_{\nu_k},s) j^\mu_k,\label{eq:antinuME}
\end{align}
for external neutrino and antineutrino states, respectively, where
\begin{equation}
    j^\mu_k = \bar{u}(p_e,s_e)\gamma^\mu(V_{kk}-A_{kk}\gamma^5)u(p_e,s_e),
\end{equation}
is the electron current. For Majorana fields, we instead have that
\begin{equation}\label{eq:majnuME}
    \begin{split}
        \langle e_{s_e},\nu_{k,s}| \mathcal{H}^M|e_{s_e},\nu_{k,s}\rangle &= \langle e_{s_e},\nu_{k,s}| \mathcal{H}^D|e_{s_e},\nu_{k,s}\rangle + \langle e_{s_e},\bar\nu_{k,s}| \mathcal{H}^D|e_{s_e},\bar\nu_{k,s}\rangle\\
        &= -\sqrt{2}G_F\bar{u}(p_{\nu_k},s)\gamma_\mu \gamma^5 u(p_{\nu_k},s)j^\mu_k,
    \end{split}
\end{equation}
where in going from the first line to the second we have used the Majorana condition to make the replacement $v(p,s) = \mathcal{C}\bar{u}(p,s)^T$, with $\mathcal{C}$ the charge conjugation matrix. This change transforms the $V-A$ vertex to a purely axial one\footnote{A neutral current vertex of the form $\bar u \Gamma_\mu u$ for Dirac fermions transforms to $\bar u (\Gamma_\mu + \mathcal{C}\Gamma_{\mu}^T \mathcal{C}^{-1}) u$ for Majorana fermions as a result of the Majorana condition~\cite{Gluza:1991wj}.}, as Majorana fermions cannot carry charge. 

If the experiment is set up such that in the laboratory frame the electrons are at rest, $|\vec{p}_e| = 0$ and $E_e = m_e$. On the other hand, due to the relative motion of the Earth to C$\nu$B, relic neutrinos have a momentum given by~\eqref{eq:labFrameSingle}. The resulting energy splitting of the electron spin states is found by taking the difference between the energy shift~\eqref{eq:energyShiftFull} for each spin state, which should then be flux-averaged to yield (see Appendix~\ref{sec:energyShift} for details of the calculation)
%\begin{equation}\label{eq:energySplitDirac}
%    \begin{split}
%        \Delta E^D_e = \frac{\sqrt{2}G_F}{(2\pi)^6}\sum_i A_{ii}\Big [\beta_{\nu_i}&\left(n_{\nu}(\nu_{i,L}) + n_{\nu}(\nu_{i,R}) - n_{\nu}(\bar\nu_{i,L}) - n_{\nu}(\bar\nu_{i,R})\right)\\
%        &+\left(n_{\nu}(\nu_{i,L}) - n_{\nu}(\nu_{i,R}) + n_{\nu}(\bar\nu_{i,L}) - n_{\nu}(\bar\nu_{i,R})\right)\Big],
%    \end{split}
%\end{equation}
%\begin{equation}\label{eq:energySplitDirac}
%    \Delta E^D_e = \sqrt{2}G_F\sum_i A_{ii}\Big [\beta_{\nu_i}\sum_s\left(n_{\nu}(\nu_{i,s}^D) - n_{\nu}(\bar\nu_{i,s}^D)\right)+\sum_\nu\left(n_{\nu}(\nu_{i,L}^D) - n_{\nu}(\nu_{i,R}^D)\right)\Big],
%\end{equation}
\begin{equation}\label{eq:energySplitDirac}
    \begin{split}
    \Delta E_e^D &= \frac{\sqrt{2}G_F}{3}\beta_\Earth\sum_i A_{ii}\Big[2\sum_{s}(2-\beta_{\nu_i}^2)(n_\nu(\nu_{i,s}^D)-n_\nu(\bar\nu_{i,s}^D)) \\
    &+ \frac{1}{\beta_{\nu_i}}\left(3 -\beta_{\nu_i}^2\right)(n_\nu(\nu_{i,L}^D)-n_\nu(\nu_{i,R}^D) + n_\nu(\bar\nu_{i,R}^D)-n_\nu(\bar\nu_{i,L}^D))\Big],
    \end{split}
\end{equation}
for Dirac neutrinos. We see immediately that there are two terms that may contribute to the Stodolsky effect. The first term, which was identified by Stodolsky~\cite{Stodolsky:1974aq}, requires a difference in the number of relic neutrinos and antineutrinos to be non-zero. The second term is only non-vanishing if there is a net helicity asymmetry in the C$\nu$B; this effect was first identified in~\cite{Duda:2001hd} and appears to diverge as $\beta_{\nu_i}\to 0$. This is an artefact of the transformation between the C$\nu$B and laboratory frames, and we will soon show that there is no real divergence in this limit. It should be noted that this is the only mechanical effect that scales linearly in $G_F$~\cite{Cabibbo:1982bb}, avoiding the brutal $G_F^2$ suppression that typical neutrino cross sections face.
%This is the case where $\beta_{\Earth} = 0$ and the C$\nu$B frame, where neutrinos are expected to be isotropic, coincides with the Earth reference frame. In this scenario, the average momentum of relic neutrinos $\vec{p}_{\nu_i} = (0,0,0)$ and we subsequently set $\beta_{\nu_i} = 0$ in~\eqref{eq:energySplitDirac}. 

The result~\eqref{eq:energySplitDirac} also has several pleasing features. First, whilst the energy shifts~\eqref{eq:energyShiftFull} depend on the spin-insensitive vector couplings $V_{ii}$, their difference only depends on the axial couplings $A_{ii}$, as should be expected. Second, all terms proportional to $n_\nu(\nu_{i,R})$ and $n_{\nu}(\bar\nu_{i,L})$ vanish in the ultrarelativistic limit $\beta_{\nu_i} \to 1$ when helicity and chirality coincide. This is also to be expected, as right chiral neutrinos and left chiral antineutrinos are sterile. For Majorana neutrinos we find the similar result
%\begin{equation}\label{eq:energySplitMajorana}
%        \Delta E^M_e = 2\sqrt{2}G_F\sum_i A_{ii}\left(n_{\nu}(\nu_{i,L}^M) - n_{\nu}(\nu_{i,R}^M)\right),
%\end{equation}
\begin{equation}\label{eq:energySplitMajorana}
    \Delta E_e^M = \frac{2\sqrt{2}G_F}{3}\beta_\Earth \sum_{i}\frac{A_{ii}}{\beta_{\nu_i}}\left(3 -\beta_{\nu_i}^2\right)(n_\nu(\nu_{i,L}^M)-n_\nu(\nu_{i,R}^M)),
\end{equation}
which naturally only contains the term requiring a helicity asymmetry. Both~\eqref{eq:energySplitDirac} and~\eqref{eq:energySplitMajorana} are signed quantities, which could provide extra information about the C$\nu$B if measured. In the case of~\eqref{eq:energySplitDirac}, it is also possible that the energy splitting due to a neutrino-antineutrino asymmetry could cancel with that from a helicity asymmetry. Similarly, since $A_{11} < 0$, whilst $A_{22}, A_{33} > 0$, for the right combination of neutrino masses and temperatures the contributions from each mass eigenstate could sum to zero. Finally, we note that the standard scenario predicts no neutrino-antineutrino asymmetry for Dirac neutrinos, such that the effect will be dominated by the helicity asymmetry term. On the other hand, for Majorana neutrinos there should be no helicity asymmetry and consequently no Stodolsky effect from the C$\nu$B. Nevertheless, there are several mechanisms (\textit{e.g.} finite chemical potential, non-standard neutrino interactions, gravitational potentials) through which either asymmetry could develop. 

\begin{figure}
    \centering
    \includegraphics[width=\textwidth]{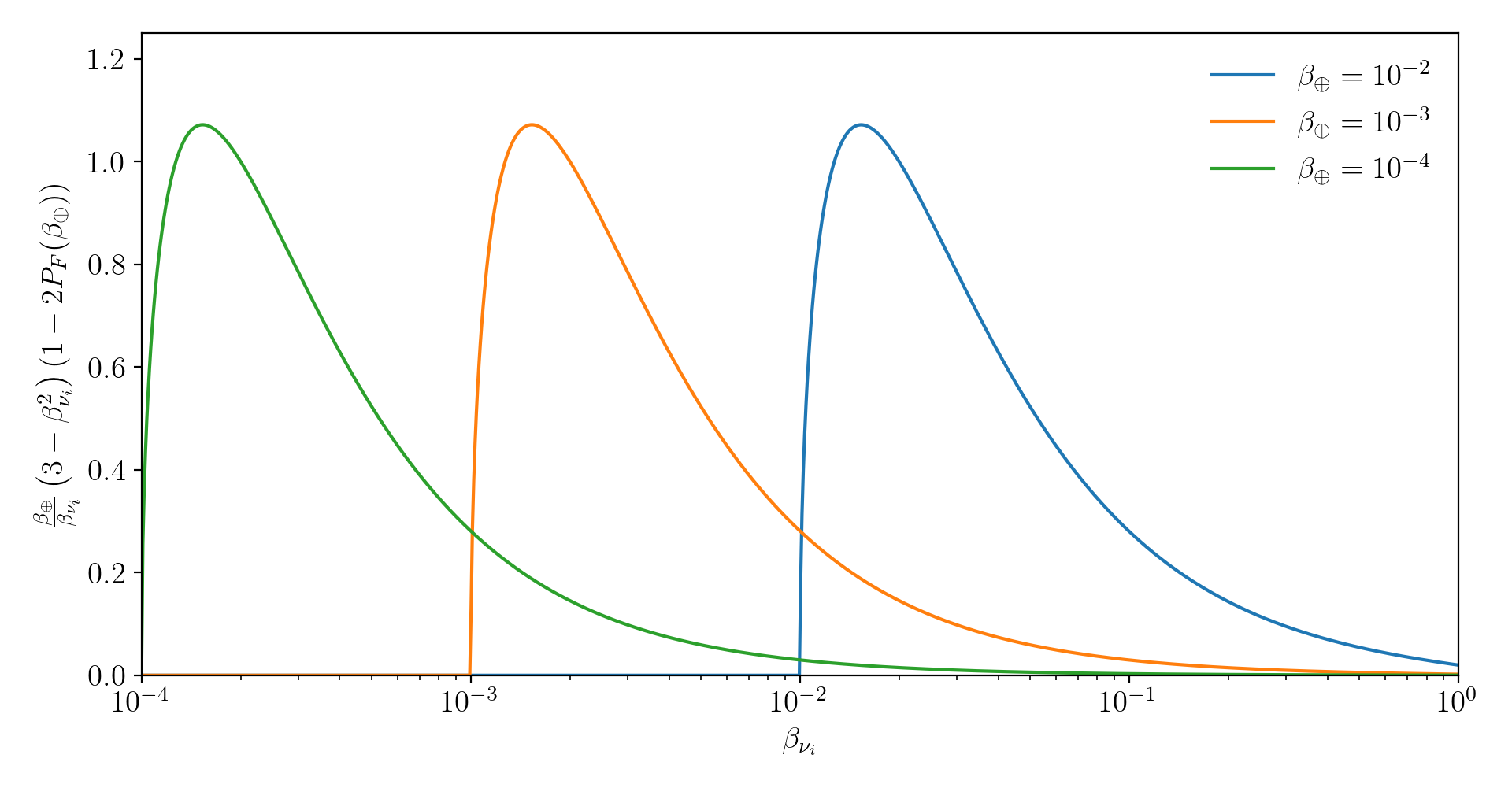}
    \caption{Helicity term appearing in the Stodolsky effect energy shifts for a range of reference frame velocities, $\beta_\Earth$. As the neutrino velocity, $\beta_{\nu_i}$, approaches the Earth's velocity, the term vanishes identically.}
    \label{fig:helicityAsymmetry}
\end{figure}

Before continuing, we make some important comments about the helicity asymmetry term appearing in both~\eqref{eq:energySplitDirac} and~\eqref{eq:energySplitMajorana}, and address the apparent singularity. As helicity is not a Lorentz invariant quantity, an asymmetry in the C$\nu$B rest frame is not necessarily indicative of one in the laboratory frame. It is entirely possible that if the relative motion of the Earth far exceeds the velocity of neutrinos in the C$\nu$B frame then the helicity asymmetry can be washed out entirely. Additionally, the relative motion of the Earth cannot generate helicity asymmetry if there is none in the C$\nu$B frame. To prove these statements, suppose that in going between frames the helicity of relic neutrinos is flipped with a velocity dependent probability $P_F(\beta_{\Earth})$. In this case, the number densities in the two frames are related by
\begin{align}
    n_{\nu}(\nu_{i,L}) &= \gamma_\Earth \left\{ P_F(\beta_\Earth)\,\widetilde n_{\nu}(\nu_{i,R}) + (1-P_F(\beta_\Earth))\,\widetilde n_{\nu}(\nu_{i,L})\right\},\label{eq:flipL}\\
    n_{\nu}(\nu_{i,R}) &= \gamma_\Earth \left\{ P_F(\beta_\Earth)\,\widetilde n_{\nu}(\nu_{i,L}) + (1-P_F(\beta_\Earth))\,\widetilde n_{\nu}(\nu_{i,R})\right\},\label{eq:flipR}
\end{align}
where the Lorentz factor $\gamma_\Earth$ appears due to length contraction along the direction of motion, which increases the number density of relic neutrinos. The helicity difference is therefore
\begin{equation}\label{eq:helicityAsymmetry}
    n_{\nu}(\nu_{i,L}) - n_{\nu}(\nu_{i,R}) = \gamma_\Earth \left\{\widetilde n_{\nu}(\nu_{i,L}) - \widetilde n_{\nu}(\nu_{i,R}) + 2 P_F(\beta_\Earth) (\widetilde n_{\nu}(\nu_{i,R}) - \widetilde n_{\nu}(\nu_{i,L}))\right\},
\end{equation}
which is identically zero if $\widetilde n_{\nu}(\nu_{i,L}) = \widetilde n_{\nu}(\nu_{i,R})$ independently of $P_F(\beta_\Earth)$, showing that the relative motion of the Earth cannot generate a helicity asymmetry. Next, we note that for initially isotropic relic neutrinos in the C$\nu$B frame (see Appendix~\ref{sec:helicity})
\begin{equation}\label{eq:helicityFlip}
    P_F(\beta_\Earth) =
    \begin{cases}
       \frac{1}{\pi}\arcsin\left(\frac{\beta_\Earth}{\beta_{\nu_i}}\right), \quad & \beta_\Earth < \beta_{\nu_i} , \\
        \frac{1}{2},\quad & \beta_\Earth \geq \beta_{\nu_i},
    \end{cases}
\end{equation}
such that the asymmetry~\eqref{eq:helicityAsymmetry} vanishes for $\beta_\Earth \geq \beta_{\nu_i}$, where we have used $\beta_{\nu_i} \simeq \widetilde{\beta}_{\nu_i}$. This demonstrates that a sufficiently large relative velocity between the two frames equalises the number of left and right helicity neutrinos in the laboratory frame, regardless of the initial distribution. The same arguments can be applied to the antineutrino helicity distributions. This also resolves the singularity as $\beta_{\nu_i} \to 0$; since $\beta_\Earth > 0$, the helicity asymmetry will tend to zero before the $1/\beta_{\nu_i}$ term diverges. This is demonstrated in Figure~\ref{fig:helicityAsymmetry}. With this in mind, the Stodolsky effect is expected to vanish completely for Majorana neutrinos if $\beta_\Earth \geq \beta_{\nu_i}$. 

The energy splitting induces a torque with magnitude $\tau_e \simeq |\Delta E_e|$ on each electron, such that a ferromagnet with $N_e$ polarised electrons in the presence of the C$\nu$B experiences a total torque
\begin{equation}
    N_e \tau_e \simeq \frac{N_A}{m_A}\frac{Z}{A}M|\Delta E_e|,
\end{equation}
where $N_A$ is Avogadro's number, $Z$ and $A$ are the atomic and mass number of the target material respectively, $M$ is the total target mass and we have introduced the ``Avogadro mass'' $m_A = 1\,\mathrm{g}\,\mathrm{mol}^{-1}$. A ferromagnet with spatial extent $R$ and moment of inertia $I = I_0 M R^2$ will therefore experience a linear acceleration
\begin{equation}
    a \simeq \frac{N_A}{m_A}\frac{Z}{A}\frac{1}{R} I_0 |\Delta E_e|. 
\end{equation}
As our reference scenario we consider a torsion balance consisting of $N_m$ spherical and uniformly dense ferromagnets of mass $M$, each a distance $R$ from a common central axis. The ferromagnets should be oriented such that the polarisation of those on opposing sides of the central axis are antiparallel in order to maximise the net torque on the system. In this scenario $I_0 = N_m$. 

Assuming that accelerations as small as $a_0$ are measurable, by plugging in our expressions for the energy splittings we find the overdensities that can be constrained by the Stodolsky effect
\begin{equation}\label{eq:stodLimitsD}
    \begin{split}
        \bigg|\sum_i A_{ii}\Big[2\sum_{s}(2-\beta_{\nu_i}^2)(&\eta_{\nu}(\nu_{i,s}^D)-\eta_{\nu}(\bar\nu_{i,s}^D))+\sum_{\nu}\frac{1}{\beta_{\nu_i}}\left(3-\beta_{\nu_i}^2\right)(\eta_{\nu}(\nu_{i,L}^D)-\eta_{\nu}(\nu_{i,R}^D))\Big]\bigg|  \\
        &\leq \frac{2\sqrt{2}\pi^2}{\zeta(3) T_{\nu,0}^3 G_F}\frac{m_A}{N_A}\frac{A}{Z}\frac{R}{N_m}\frac{a_0}{\beta_\Earth}\\
        &\simeq (2.18\times 10^{11})\frac{A}{Z}\left[\frac{R}{1\,\mathrm{cm}}\right]\left[\frac{2}{N_m}\right]\left[\frac{a_0}{10^{-15}\,\mathrm{cm}\,\mathrm{s}^{-2}}\right]\left[\frac{\beta_\Earth^{\mathrm{CMB}}}{\beta_\Earth}\right],
    \end{split}
\end{equation}
for Dirac neutrinos, whilst for Majorana neutrinos
\begin{equation}\label{eq:stodLimitsM}
    \begin{split}
    \bigg|\sum_{i}\frac{A_{ii}}{\beta_{\nu_i}}\left(3 -\beta_{\nu_i}^2\right)(\eta_{\nu}&(\nu_{i,L}^M)-\eta_{\nu}(\nu_{i,R}^M))\bigg| \\
    &\leq \frac{\sqrt{2}\pi^2}{\zeta(3) T_{\nu,0}^3 G_F}\frac{m_A}{N_A}\frac{A}{Z}\frac{R}{N_m}\frac{a_0}{\beta_\Earth}\\
    &\simeq (1.09\times 10^{11})\frac{A}{Z}\left[\frac{R}{1\,\mathrm{cm}}\right]\left[\frac{2}{N_m}\right]\left[\frac{a_0}{10^{-15}\,\mathrm{cm}\,\mathrm{s}^{-2}}\right]\left[\frac{\beta_\Earth^{\mathrm{CMB}}}{\beta_\Earth}\right],
    \end{split}
\end{equation}
where we have chosen $a_0 = 10^{-15}\,\mathrm{cm}\,\mathrm{s}^{-2}$ as our reference sensitivity, which has recently been achieved in tests of the weak equivalence principle using Cavendish-style torsion balances~\cite{Wagner:2012ui}. Torsion balances utilising test masses suspended by superconducting magnets have also been considered in~\cite{Hagmann:1999kf}, which have the potential to probe accelerations as small as $a_0 \simeq 10^{-23}\,\mathrm{cm}\,\mathrm{s}^{-2}$. Such an experiment would be able to set constraints on C$\nu$B overdensities that are competitive with the PTOLEMY proposal. Due to their helicity dependence, the constraints that can be set using the Stodolsky effect are naturally complimentary to those set by PTOLEMY, as together they can give an insight into the helicity composition of the C$\nu$B.  
\subsection{Coherent scattering}\label{sec:coherent}
We now turn our attention to the detection of relic neutrinos using coherent neutral current scattering. This section will largely follow the formalism of~\cite{Shergold:2021evs}, with some exceptions. To avoid the introduction of ill-defined quantities such as $m_{\nu_\alpha}$ and $T_{\nu_\alpha}$, particularly at small neutrino masses, we will work in the mass basis throughout. Additionally, we will work with polarised cross sections, and by introducing structure factors we will more rigorously introduce macroscopic coherence, allowing us to extend the proposal to a system of more than one coherent scattering volume. Finally, we will address the contribution from coherent neutrino-electron scattering in more detail than in previous works~\cite{Shergold:2021evs,Domcke:2017aqj}. 

For neutrino energies much less than the nuclear mass, the cross sections for coherent neutrino-nucleus scattering are (see Appendix~\ref{sec:polarisedXsecs})
\begin{align}
    \sigma_N(\nu_{i,s}^D) &= \frac{G_F^2}{8\pi}(Q_V^2 + 3Q_A^2)\mathcal{A}_s(\beta_{\nu_i})E_{\nu_i}^2,\label{eq:nuNuclexSecD}\\
    \sigma_N(\bar\nu_{i,s}^M) &= \frac{G_F^2}{4\pi}\left(\beta_{\nu_i}^2 Q_V^2 + 3(2-\beta_{\nu_i}^2)Q_A^2\right)E_{\nu_i}^2,\label{eq:nuNuclexSecM}
\end{align}
where $Q_V = A - 2Z(1-2\sin^2\theta_W) \simeq A-Z$ is the vector charge of the nucleus and $Q_A = A -2Z$ is its axial charge, given in terms of its mass and atomic numbers $A$ and $Z$, respectively. For a typical nucleus $Q_V \gg Q_A$, such that the term proportional to $Q_A$ is typically neglected~\cite{Formaggio:2012cpf}. As a result, in previous works~\cite{Domcke:2017aqj,Duda:2001hd}, including a paper by one of the present authors~\cite{Shergold:2021evs}, it was stated that the Majorana neutrino scattering cross section was $\beta_{\nu_i}^2$ suppressed compared to the equivalent Dirac neutrino cross section. From~\eqref{eq:nuNuclexSecM}, it is clear that this is only true for symmetric nuclei, for which $A = 2Z$.

The relative motion of the Earth to the C$\nu$B generates a relic neutrino wind with net directionality, such that each neutrino scattering event will transfer an average momentum $\Delta p_{\nu_i}$ to the target, which has already been estimated in~\eqref{eq:av2}. This induces a small macroscopic acceleration in a target with total mass $M$,
\begin{equation}
    a_N(\nu_{i,s}) = \frac{1}{M}\Gamma_N(\nu_{i,s}) \Delta p_{\nu_i},
\end{equation}
where $\Gamma_N = N_T \beta_{\nu_i} \sigma_N n_\nu$ is the neutrino scattering rate and $N_T$ is the total number of nuclei in the target. %To estimate the average momentum transfer to the test mass per scattering event, $\Delta p_{\nu_i}$, we consider the simple 1-dimensional setup sketched in Figure~\ref{fig:frames}, where the relative motion of the Earth to C$\nu$B generates an asymmetry in the flux and momentum of left and right travelling neutrinos. The resulting asymmetry leads to an average momentum transfer~\cite{Shergold:2021evs}
%\begin{equation}
%    \Delta p_{\nu_i} = \frac{\beta^+_{\nu_i} p^+_{\nu_i} - \beta^-_{\nu_i} p^-_{\nu_i}}{\beta^+_{\nu_i} + \beta^-_{\nu_i}} \simeq 2\beta_\Earth m_{\nu_i} + \mathcal{O}(\widetilde\beta_{\nu_i}^4,\beta_\Earth^3),
%\end{equation}
%where $\beta^\pm_{\nu_i}$ are given in~\eqref{eq:bpm}, whilst $p_{\nu_i}^\pm = \beta_{\nu_i}^\pm E_{\nu_i}^\pm$, with $E_{\nu_i}^\pm = \sqrt{(p_{\nu_i}^\pm)^2 + m_{\nu_i}^2}$. The far right equality holds only when both the C$\nu$B frame neutrino and Earth velocities are small, which is valid in the standard relic neutrino scenario. However, as we also consider the C$\nu$B scenarios with $T_{\nu_i} \gg T_{\nu,0}$, we will use the more general result going forward. 
After summing over all neutrino degrees of freedom, the total acceleration of a target with mass $M$ due to neutrino-nucleus scattering is
\begin{equation}\label{eq:accNoCoherenceD}
    a_{N,\mathrm{tot}}^D \simeq \frac{G_F^2}{8\pi}\frac{N_A}{A m_A}(Q_V^2+3Q_A^2)\sum_{\nu,i,s} E_{\nu_i} |\vec{p}_{\nu_i}| \Delta p_{\nu_i} \mathcal{A}_s(\beta_{\nu_i})n_\nu(\nu_{i,s}^D),
\end{equation}
for Dirac neutrinos, whilst for Majorana neutrinos
\begin{equation}\label{eq:accNoCoherenceM}
    a_{N,\mathrm{tot}}^M \simeq \frac{G_F^2}{4\pi}\frac{N_A}{A m_A}\sum_{\nu,i,s} E_{\nu_i} |\vec{p}_{\nu_i}| \Delta p_{\nu_i} \left(\beta_{\nu_i}^2 Q_V^2 + 3(2-\beta_{\nu_i}^2)Q_A^2\right)n_\nu(\nu_{i,s}^M),
\end{equation}
where $N_A$ and $m_A = 1\,\mathrm{g}\,\mathrm{mol}^{-1}$ are Avogadro's number and the `Avogadro mass', respectively. Akin to the PTOLEMY proposal, coherent scattering is sensitive to the helicity composition of the C$\nu$B. However, unlike PTOLEMY, the difference in the Dirac and Majorana neutrino scattering cross sections allows insight into the nature of neutrinos irrespective of whether the standard scenario is assumed. In practice, however, the number of uncertain quantities entering into~\eqref{eq:accNoCoherenceD} and~\eqref{eq:accNoCoherenceM} make the distinction incredibly difficult.

The results~\eqref{eq:accNoCoherenceD} and~\eqref{eq:accNoCoherenceM} apply when coherence can only be maintained over a single nucleus, \textit{i.e.} for neutrino wavelengths $\lambda_{\nu_i} = 2\pi/|\vec{p}_{\nu_i}|$ of order the nuclear radius. Coherent scattering on a large nucleus of radius $10\,\mathrm{fm}$ can therefore be achieved with neutrino momenta of order $|\vec{p}_{\nu_i}|\sim \mathcal{O}(0.1\,\mathrm{GeV})$, which far exceeds that of relic neutrinos. Clearly, relic neutrinos with macroscopic wavelengths $\lambda_\nu\sim \mathcal{O}(\mathrm{mm})$ should be capable of maintaining coherence over many nuclei, leading to vastly enhanced cross sections. 

To account for this, the scattering amplitudes should be augmented by a structure factor, $F(\vec{q})$, to give macroscopic coherent scattering cross sections proportional to $|F(\vec{q})|^2$, where $\vec{q} \sim \vec{p}_{\nu_i}$ is the recoil momentum of the scattered nucleus. For a large target consisting of many scattering centres, each located at position $\vec{x}_i$, the structure factor is given by
\begin{equation}\label{eq:structureFactor}
    F(\vec{q}) = \sum_i e^{-i\vec{q}\cdot \vec{x}_i} \implies |F(\vec{q})|^2 = \sum_{i,j}e^{-i\vec{q}\cdot(\vec{x}_i - \vec{x}_j)},
\end{equation}
which encodes the relative phase between each of the nuclei in the target. For small recoils $|\vec{q}|^{-1}\ll \langle |\vec{x}_i - \vec{x}_j|\rangle \simeq R$, where $R$ is the radius of the target, all nuclei are in phase the structure factor reduces to $N_T^2$. As such, if the target is chosen with $R\simeq \lambda_{\nu_i}$, the coherent scattering rate $\Gamma_N$ picks up an enhancement factor equal to the number of nuclei within a volume $\lambda_{\nu_i}^3$,
\begin{equation}\label{eq:coherenceFactor}
    N_{C,i} = \left(\frac{2\pi}{|\vec{p}_{\nu_i}|}\right)^3 \frac{N_A}{A\,m_A}\rho,
\end{equation}
where $\rho$ is the mass density of the target. The total acceleration of a test mass due to macroscopic coherent scattering is therefore given by
\begin{equation}\label{eq:accCoherenceD}
    a_{N,\mathrm{tot}}^{C,D} \simeq \pi^2 G_F^2\left(\frac{N_A}{A m_A}\right)^2 (Q_V^2+3Q_A^2)\rho \sum_{\nu,i,s} \frac{E_{\nu_i}}{|\vec{p}_{\nu_i}|^2} \Delta p_{\nu_i} \mathcal{A}_s(\beta_{\nu_i})n_\nu(\nu_{i,s}^D),
\end{equation}
for Dirac neutrinos, whilst the expression for Majorana neutrinos takes the form
\begin{equation}\label{eq:accCoherenceM}
    a_{N,\mathrm{tot}}^{C,M} \simeq 2\pi^2 G_F^2 \left(\frac{N_A}{A m_A}\right)^2\rho\sum_{\nu,i,s} \frac{E_{\nu_i}}{|\vec{p}_{\nu_i}|^2} \Delta p_{\nu_i} \left(\beta_{\nu_i}^2 Q_V^2 + 3(2-\beta_{\nu_i}^2)Q_A^2\right)n_\nu(\nu_{i,s}^M).
\end{equation}
These are significantly larger than their microscopically coherent counterparts~\eqref{eq:accNoCoherenceD} and~\eqref{eq:accNoCoherenceM} due to the scaling with $N_A^2$. Importantly, macroscopic coherent scattering naturally favours scenarios with small neutrino momenta, making it an ideal candidate for the detection of non-relativistic relic neutrinos. 

To avoid confusion, we comment on the divergent limit as $|\vec{p}_{\nu_i}| \to 0$. This is a result of the assumption that $R \simeq \lambda_{\nu_i}$, which becomes impossible to uphold as $\lambda_{\nu_i} \to \infty$. To account for this, one should make the replacement $2\pi/|\vec{p}_{\nu_i}| \to R$ in~\eqref{eq:coherenceFactor} for neutrino wavelengths much larger than the experiment. We discuss the case where only partial coherence can be obtained, $\lambda_{\nu_i} \ll R$, and give a derivation of the structure factor in Appendix~\ref{sec:structureFactors}. 

Neutrinos can also scatter from electrons in the target. Working in the mass basis, these can proceed in two ways; either `mass diagonal', in which both the incoming and final state neutrinos are the same mass eigenstate, or `mass changing', where the neutrinos differ. As the neutral current is both flavour and mass diagonal, this can only contribute to the mass diagonal processes, whilst charged current interactions can contribute to both. 

After working through the calculations given in Appendix~\ref{sec:polarisedXsecs}, we find the cross sections for neutrinos to scatter on electrons
\begin{align}
    \sigma_e(\nu_{i,s}^D \to \nu_j^D) &= \frac{G_F^2}{2\pi}E_{\nu_i}E_{\nu_j}\mathcal{A}_s(\beta_{\nu_i})K^D_{ij},\label{eq:nueXsecD}\\
    \sigma_e(\nu_{i,s}^M \to \nu_j^M) &= \frac{G_F^2}{\pi}E_{\nu_i}E_{\nu_j}K^M_{ij}\label{eq:nueXsecM},
\end{align}
for neutrino momenta much less than the electron mass, where the functions $K_{ij}$ depend on the electron vector and axial couplings, as well as elements of the PMNS matrix, and are given in Appendix~\ref{sec:polarisedXsecs}. The cross sections~\eqref{eq:nueXsecD} and~\eqref{eq:nueXsecM} should be augmented by structure factors when considering macroscopic coherent scattering. We also highlight that in order for the $\nu_i \neq \nu_j$ processes to contribute, the incident neutrino must be sufficiently energetic to produce mass eigenstate $j$. Explicitly, we require
\begin{equation}
    E_{\nu_i} \geq m_{\nu_j} + \frac{\Delta m_{ji}^2}{2m_e} \simeq m_{\nu_j},
\end{equation}
where $\Delta m_{ij}^2 = m_{\nu_i}^2 - m_{\nu_j}^2$ is the squared mass splitting between mass eigenstates $i$ and $j$.

Neutrino-electron scattering naively seems like a subleading effect compared to neutrino-nucleus scattering due to the absence of the nuclear vector and axial charges that appear in~\eqref{eq:nuNuclexSecD} and~\eqref{eq:nuNuclexSecM}. However, as noted in~\cite{Shergold:2021evs} there are $Z$ electrons for every nucleus in the target, such that the contribution from neutrino-electron scattering picks up a $Z^2$ enhancement when the scattering is fully coherent. In this limit, we also set $E_{\nu_j} = \sqrt{m_{\nu_j}^2 + |\vec{p}_{\nu_i}|^2}$. 

Once again assuming that an average momentum $\Delta p_{\nu_i}$ is transferred to the test mass by each scattering event, the total acceleration due to macroscopic coherent neutrino-electron scattering is given by
\begin{equation}
    a_{e,\mathrm{tot}}^{C,D} \simeq 4\pi^2 G_F^2 \left(\frac{N_A}{m_A}\frac{Z}{A}\right)^2 \rho \sum_{\nu,i,j,s}\frac{E_{\nu_j}\Delta p_{\nu_i}}{|\vec{p}_{\nu_i}|^2} K_{ij}^D \,\theta\!\left(E_{\nu_i} - m_{\nu_j}\right)\mathcal{A}_s(\beta_{\nu_i})n_{\nu}(\nu_{i,s}), 
\end{equation}
for Dirac neutrinos, and
\begin{equation}
    a_{e,\mathrm{tot}}^{C,M} \simeq 8\pi^2 G_F^2 \left(\frac{N_A}{m_A}\frac{Z}{A}\right)^2 \rho \sum_{i,j,s}\frac{E_{\nu_j}\Delta p_{\nu_i}}{|\vec{p}_{\nu_i}|^2} K_{ij}^M\,\theta\!\left(E_{\nu_i} - m_{\nu_j}\right)n_{\nu}(\nu_{i,s}), 
\end{equation}
for Majorana neutrinos, where $\theta(E_{\nu_i} - m_{\nu_j})$ is the Heaviside step function, ensuring that the incident neutrino has sufficient energy for the $\nu_i \to \nu_j$ process. We remind the reader that $E_{\nu_j} = \sqrt{m_{\nu_j}^2 + |\vec{p}_{\nu_i}|^2}$ for coherent scattering.

The size of the contribution from neutrino-electron scattering depends strongly on the properties of the material, specifically how well the electrons can transfer momentum to the bulk solid. For example, in a metallic target with many delocalised electrons, a fraction of the energy transferred from the neutrinos may instead be lost to bremsstrahlung radiation. On the other hand, a non-metallic target where the electrons are tightly bound to their host nucleus will recoil efficiently due to neutrino-electron scattering. We therefore choose to parameterise the total acceleration of a test mass due to the macroscopic coherent scattering of a neutrino wind as
\begin{equation}\label{eq:totalAccNoSub}
    a_\mathrm{tot}^C = a_{N,\mathrm{tot}}^C+ \varepsilon a_{e,\mathrm{tot}}^C,
\end{equation}
where $\varepsilon \in [0,1]$ is the efficiency of momentum transfer by neutrino-electron scattering. It has been argued in~\cite{Bahcall:1990nv} that even in good conductors, restoring forces between the ions and scattered electrons in the target suppress bremsstrahlung whilst strongly coupling the electron momentum to that of the bulk solid. In line with this reasoning, we will choose $\varepsilon=1$ when plotting the sensitivity coherent scattering experiments.

Once again assuming a sensitivity $a_0$ to accelerations of the target, and inverting~\eqref{eq:totalAccNoSub}, we find that a coherent neutrino scattering experiment could set the constraint
\begin{equation}\label{eq:cohLimitsD}
    \begin{split}
    \sum_{\nu,i,s}\frac{\Delta p_{\nu_i}}{|\vec{p}_{\nu_i}|^2}\Bigg[\left(\frac{Q_V^2}{A^2} + 3\frac{Q_A^2}{A^2}\right)E_{\nu_i}+&4\varepsilon\frac{Z^2}{A^2}\sum_j K_{ij}^D E_{\nu_j}\theta\!\left(E_{\nu_i} - m_{\nu_j}\right)\Bigg]\mathcal{A}_s(\beta_{\nu_i})\eta_{\nu}(\nu_{i,s}^D)\\
    &\lesssim \frac{4}{3\zeta(3)}\left(\frac{m_A}{N_A}\right)^2 \frac{a_0}{T_{\nu,0}^3 G_F^2}\frac{1}{\rho}\\
    &\simeq (6.70\times10^{14})\left[\frac{11.34\,\mathrm{g}\,\mathrm{cm}^{-3}}{\rho}\right]\left[\frac{a_0}{10^{-15}\,\mathrm{cm}\,\mathrm{s^{-2}}}\right],
    \end{split}
\end{equation}
on the Dirac neutrino overdensity, and
\begin{equation}\label{eq:cohLimitsM}
    \begin{split}
    \sum_{i,s}\frac{\Delta p_{\nu_i}}{|\vec{p}_{\nu_i}|^2}\Bigg[\bigg(\beta_{\nu_i}^2 \frac{Q_V^2}{A^2} + 3(2-\beta_{\nu_i}^2)&\frac{Q_A^2}{A^2}\bigg)E_{\nu_i}+4\varepsilon\frac{Z^2}{A^2}\sum_j K_{ij}^M E_{\nu_j}\theta\!\left(E_{\nu_i} - m_{\nu_j}\right)\Bigg]\eta_{\nu}(\nu_{i,s}^M)\\
    &\lesssim \frac{2}{3\zeta(3)}\left(\frac{m_A}{N_A}\right)^2 \frac{a_0}{T_{\nu,0}^3 G_F^2}\frac{1}{\rho}\\
    &\simeq (3.35\times10^{14})\left[\frac{11.34\,\mathrm{g}\,\mathrm{cm}^{-3}}{\rho}\right]\left[\frac{a_0}{10^{-15}\,\mathrm{cm}\,\mathrm{s^{-2}}}\right],
    \end{split}
\end{equation}
on the Majorana neutrino overdensity. As before, we have chosen $a_0 = 10^{-15}\,\mathrm{cm}\,\mathrm{s}^{-2}$ as our reference acceleration, whilst $\rho = 11.34\,\mathrm{g}\,\mathrm{cm}^{-3}$ corresponds to a lead target. 

Clearly, the scale of accelerations due to coherent scattering is much smaller than those from the Stodolsky effect, provided that there are asymmetries in the C$\nu$B. However, as first discussed in~\cite{Domcke:2017aqj} and further developed in~\cite{Lee:2020dcd}, there is also the possibility of observing coherent scattering as tiny strains at laser interferometer gravitational wave detectors, rather than as accelerations of \textit{e.g.} a torsion balance. The strain profile for a series of successive scattering events at times $t_{n}$ within a given sampling window, each transferring a momentum $\Delta p_{\nu_i}$, is
\begin{equation}\label{eq:strain}
    h(\omega) = \sum_{\nu,i,s} \sqrt{\frac{2\omega}{\pi}}\frac{\Delta p_{\nu_i}}{ML}\left|\frac{1}{\omega_r^2 - (\omega-i\omega_r\xi_\omega)^2}\right|\bigg|\sum_{t_{n}}e^{- i\omega t_{n}}\bigg|,
\end{equation}
where $\omega$ is the signal frequency, $\omega_r$ is the resonance frequency of the system, $L$ is the interferometer arm length and $\xi_\omega\ll 1$ is related to the damping of the oscillator, discussed in~\cite{Lee:2020dcd}. We have assumed in~\eqref{eq:strain} that the target is a single oscillator with one resonance frequency. In practice, laser interferometer mirrors are a set of coupled harmonic oscillators with several resonance frequencies, which may lead to cancellations in parts of the spectrum. More complicated setups are reviewed comprehensively in~\cite{Lee:2020dcd} and~\cite{Saulson:1990jc}.

The observant reader will notice that the sum appearing in~\eqref{eq:strain} is analogous to the structure factor~\eqref{eq:structureFactor} discussed thus far. The strains from successive scattering events will therefore add coherently when the signal frequency is much less than the mean scattering frequency, $\omega \ll \langle |t_n - t_m| \rangle^{-1}$. Supposing that neutrinos strike the target at regular intervals, such that $t_n = n/\Gamma_{\nu}$, with
\begin{equation}
    \Gamma_{\nu}(\nu_{i,s}) = N_T^2 \beta_{\nu_i} \left(\sigma_N(\nu_{i,s}) + \sum_j \sigma_e(\nu_{i,s} \to \nu_j)\right) n_{\nu}(\nu_{i,s})
\end{equation}
the fully coherent scattering rate and $n \in \mathbb{Z}$, we find that scattering events within a range
\begin{equation}
    \langle|n-m|\rangle \equiv n_\mathrm{coh} \simeq \frac{\Gamma_\nu}{\omega},
\end{equation}
of each other will add coherently. In these regions, the squared `structure factor' will scale as $n_\mathrm{coh}^2$. If the experiment has a sampling rate $\Gamma_\mathrm{exp} \ll \Gamma_\nu$, there will be $n_\mathrm{tot} = \Gamma_{\nu}/\Gamma_\mathrm{exp}$ total events within a given sampling window, of which a fraction $n_\mathrm{tot}/n_\mathrm{coh}$ will sum coherently. This allows us to make the replacement
\begin{equation}
    \bigg|\sum_{t_{n}}e^{- i\omega t_{n}}\bigg| \simeq \sqrt{n_\mathrm{tot}n_\mathrm{coh}} = \frac{\Gamma_\nu}{\sqrt{\omega \Gamma_\mathrm{exp}}}.
\end{equation}
Substituting this into~\eqref{eq:strain} and inverting, we find that a gravitational wave detector with strain sensitivity profile $h_0(\omega)$ can set the overdensity constraints
\begin{equation}\label{eq:gwD}
    \begin{split}
    \sum_{\nu,i,s}\frac{\Delta p_{\nu_i}}{|\vec{p}_{\nu_i}|^2}\left[\left(\frac{Q_V^2}{A^2} + 3\frac{Q_A^2}{A^2}\right)E_{\nu_i}+4\varepsilon\frac{Z^2}{A^2}\sum_j K_{ij}^D E_{\nu_j}\theta\!\left(E_{\nu_i} - m_{\nu_j}\right)\right]\mathcal{A}_s(\beta_{\nu_i})\eta_{\nu}(\nu_{i,s}^D)\\
    \lesssim \frac{2\sqrt{2\pi}}{3\zeta(3)}\left(\frac{m_A}{N_A}\right)^2 \frac{h_0(\omega)}{T_{\nu,0}^3 G_F^2}\frac{L}{\rho}\sqrt{\Gamma_\mathrm{exp}}\left|\omega_r^2 - (\omega-i\omega_r\xi_\omega)^2\right|,
    \end{split}
\end{equation}
on Dirac neutrinos, and
\begin{equation}\label{eq:gwM}
    \begin{split}
    \sum_{i,s}\frac{\Delta p_{\nu_i}}{|\vec{p}_{\nu_i}|^2}\Bigg[\bigg(\beta_{\nu_i}^2 \frac{Q_V^2}{A^2} + 3(2-&\beta_{\nu_i}^2)\frac{Q_A^2}{A^2}\bigg)E_{\nu_i}+4\varepsilon\frac{Z^2}{A^2}\sum_j K_{ij}^M E_{\nu_j}\theta\!\left(E_{\nu_i} - m_{\nu_j}\right)\Bigg]\eta_{\nu}(\nu_{i,s}^M)\\
    &\lesssim \frac{\sqrt{2\pi}}{3\zeta(3)}\left(\frac{m_A}{N_A}\right)^2 \frac{h_0(\omega)}{T_{\nu,0}^3 G_F^2}\frac{L}{\rho}\sqrt{\Gamma_\mathrm{exp}}\left|\omega_r^2 - (\omega-i\omega_r\xi_\omega)^2\right|,
    \end{split}
\end{equation}
on Majorana neutrinos. We stress, however, that the results~\eqref{eq:gwD} and~\eqref{eq:gwM} only apply when $\Gamma_{\nu} \gg \Gamma_\mathrm{exp}$. Otherwise,~\eqref{eq:strain} should be used with the structure factor set equal to unity and the sum over neutrino degrees of freedom omitted, in which case the strain profile is insensitive to the overdensity. In the standard scenario, the relic neutrino scattering rate is $\Gamma_{\nu}\sim \mathcal{O}(\mathrm{kHz})$~\cite{Shergold:2021evs}, whilst the land based interferometers LIGO and Virgo sample at rates $\Gamma_{\mathrm{exp}} \sim 4-16\,\mathrm{kHz}$~\cite{LIGOScientific:2019hgc}. As such, these are only capable of placing constraints on overdensities $\eta_{\nu} \gg 1$, for which $\Gamma_{\nu} \gg \Gamma_{\mathrm{exp}}$. Finally, we note that the signal from thermal noise can add coherently in the same manner as that from relic neutrinos, whilst also peaking at the same resonance frequencies, $\omega_r$. As such, increasing the exposure time may weaken the constraints on the relic neutrino overdensity through a reduced strain sensitivity, $h_0(\omega)$.
\subsection{Accelerator}\label{sec:accelerator}
Due to the low temperature of the C$\nu$B, there are very few methods with an energy threshold that are capable of detecting relic neutrinos. However, as pointed out in~\cite{Bauer:2021uyj}, the centre-of-mass frame (CoM) energy requirements for thresholded neutrino capture processes can be met by running an accelerated beam of ions through the C$\nu$B. This offers the additional advantage of being able to tune the neutrino energy to hit a resonance, in doing so significantly enhancing capture cross sections. Here we will largely follow the derivation given in~\cite{Bauer:2021uyj}, but extend it to include non-degenerate neutrino masses, in which case the contribution from each neutrino mass eigenstate must be considered separately. 

We consider the resonant bound beta decay (RB$\beta$) and resonant electron capture (REC) processes
\begin{align}
    {^A_Z P} + \nu_e \to {^A_{Z+1} D} + e^-\,(\mathrm{bound}), \label{eq:RBB}\\
    {^A_Z P} + e^-\,(\mathrm{bound}) + \bar\nu_e \to {^A_{Z-1} D}, \label{eq:REC}
\end{align}
where $P$ and $D$ are the parent and daughter states respectively, with mass number $A$ and atomic number $Z$. To maximise the capture rate, $P$ should be fully ionised for the RB$\beta$ process, and ionised down to a single electron for a REC process~\cite{Bauer:2021uyj}. This method is only sensitive to the electron neutrino component of the C$\nu$B through the processes~\eqref{eq:RBB} and~\eqref{eq:REC}. However, these are just two examples of resonant processes; one might also consider resonant capture on a muon, in which case this experiment would be sensitive to the muonic component of relic neutrinos. 

The energy of neutrino mass eigenstate $i$ in the rest frame of the high energy ion beam is
\begin{equation}
    E_{\nu_i}^b \simeq \frac{E_{\nu_i}}{M}E,
\end{equation}
where $M$ and $E$ are the beam ion mass and energy, respectively. For $E_{\nu_i}^b \ll M$, incoming neutrinos of mass eigenstate $i$ are captured on a beam ion with cross section
\begin{equation}\label{eq:relicAcceleratorXsec}
    \sigma_i = \frac{2\pi}{\left(E_{\nu_i}^b\right)^2}\left(\frac{2J_D+1}{2J_P+1}\right)\left[\frac{\Gamma_D^2/4}{\left(E_{\nu_i}^b-Q\right)^2+\Gamma_D^2/4}\right]|U_{ei}|^2 \mathcal{B}_{DP},
\end{equation}
for daughter and parent state spins $J_D$ and $J_P$, respectively, where $\Gamma_D$ is the daughter decay width and $\mathcal{B}_{DP} = \mathrm{Br}\left(D + e^-\left(\mathrm{bound}\right)\to P + \nu_e\right)$ or $\mathrm{Br}\left(D\to P+\nu_e +e^-\,\left(\mathrm{bound}\right)\right)$ is the branching ratio for the daughter state to decay back to the parent state. The threshold to resonantly capture a neutrino, $Q$, depends on several properties of the daughter and parent states and is discussed alongside the computation of $\mathcal{B}_{DP}$ at length in~\cite{Bauer:2021uyj}.

By inspection of~\eqref{eq:relicAcceleratorXsec}, we see that the capture rate of neutrino mass eigenstate $i$ is maximised when $E_{\nu_i}^b = Q$. However, due to the finite width of the neutrino and beam momentum distributions, $\Delta_{\nu_i}$ and $\Delta_b$ respectively, only a fraction of relic neutrinos will be captured resonantly. To estimate this fraction, we make the ansatz that the relic neutrino flux in the beam rest frame follows a Gaussian distribution, normalised appropriately
\begin{equation}
    \frac{d\phi_{\nu_i}^b}{dE_{\nu_i}^b} = \gamma_b \beta_b \sum_s n_\nu(\nu_{i,s}) \frac{1}{\Delta_{\nu_i}^b\sqrt{2\pi}}\exp\left[-\frac{1}{2}\left(\frac{E_{\nu_i}^b-\mu_i}{\Delta_{\nu_i}^b}\right)^2\right],
\end{equation}
where $\gamma_b = E/M$ and $\beta_b \simeq 1$ are the Lorentz factor and velocity of the ion beam, respectively, whilst $\mu_i$ is the mean neutrino energy in the beam rest frame. Ideally, the beam energy should be chosen such that these distributions will be centred on $\mu_i = Q$ for all three of the neutrino mass eigenstates, however, due to their different masses and temperatures, it is unlikely that more than one will be exactly on resonance. Explicitly, if $\mu_i = Q$, then $\mu_j = (E_{\nu_j}/E_{\nu_i})Q$ for $j \neq i$. The parameter $\Delta_{\nu_i}^b$ denotes the width of the neutrino momentum distribution in the beam rest frame, which by treating $\Delta_{\nu_i}$  and $\Delta_b$ as uncertainties in the lab frame momenta is given approximately by
\begin{equation}\label{eq:beamWidth}
    \Delta_{\nu_i}^b = \sqrt{\left(\Delta_{\nu_i}\frac{\partial E_{\nu_i}^b}{\partial p_{\nu_i}}\right)^2 + \left(\Delta_{b}\frac{\partial E}{\partial p}\right)^2} \simeq \mu_i \sqrt{\delta_{\nu_i}^2 + \delta_b^2},
\end{equation}
where we have introduced the fractional uncertainties $\delta_{\nu_i} = \Delta_{\nu_i}/E_{\nu_i}$ and $\delta_b = \Delta_b/E$, and $p \simeq E$ is the beam momentum.
Assuming a Fermi-Dirac distribution~\eqref{eq:fermiDirac} at temperature $T_{\nu_i}$ and taking the appropriate moments, $\Delta_{\nu_i}$ can be estimated as
\begin{equation}
    \Delta_{\nu_i} \simeq 0.291\,\mathrm{meV} \left(\frac{T_{\nu_i}}{T_{\nu,0}}\right),
\end{equation}
such that for non-relativistic neutrinos with $T_{\nu_i} = T_{\nu,0}$, $\delta_{\nu_i} \simeq 2.91 \times 10^{-3} \,(0.1\,\mathrm{eV}/m_{\nu_i})$. This is slightly smaller than the estimate of $\delta_{\nu_i}$ given in~\cite{Bauer:2021uyj}. By comparison, the ion beam at RHIC has $\delta_b \simeq 10^{-4}$~\cite{Luo:2016hgj}, and as a result we expect that the dominant contribution to~\eqref{eq:beamWidth} will come from $\delta_{\nu_i}$ for all but the largest allowed neutrino masses. By making the replacement
\begin{equation}
    \frac{\Gamma_D^2/4}{\left(E_{\nu_i}^b-Q\right)^2+\Gamma_D^2/4} \longrightarrow \frac{\pi}{2}\Gamma_D \,\delta\left(E_{\nu_i}^b-Q\right)
\end{equation}
in~\eqref{eq:relicAcceleratorXsec}, which is valid for narrow resonances satisfying $\Gamma_D \ll \Delta_{\nu_i}^b$, we find that the total lab frame neutrino capture rate per target ion on the beam is given by
\begin{equation}\label{eq:acceleratorCapRate}
    \begin{split}
    \frac{R}{N_T} &= \frac{1}{\gamma_b}\sum_i \int \sigma_i \frac{d\phi_{\nu_i}^b}{dE_{\nu_i}^b} dE_{\nu_i}^b\\
    &= \sqrt{\frac{\pi^3}{2}}\left(\frac{2J_D+1}{2J_P+1}\right) \frac{\Gamma_D}{Q^2} \mathcal{B}_{DP}\sum_{i,s} \frac{|U_{ei}|^2 \,n_{\nu}\left(\nu_{i,s}\right)}{\mu_i\sqrt{\delta_{\nu_i}^2 + \delta_b^2}}\exp\left[-\frac{1}{2}\left(\frac{Q-\mu_i}{\mu_i\sqrt{\delta_{\nu_i}^2 + \delta_b^2}}\right)^2\right].
    \end{split}
\end{equation}
Written in this form,~\eqref{eq:acceleratorCapRate} also encompasses the case where the neutrino energy is not known exactly, resulting in a beam energy is not centred exactly on resonance. If the experiment is set up assuming a neutrino energy $E_{\nu_i,p}$ but the true neutrino energy is $E_{\nu_i,t}$, then the mean beam rest frame neutrino energy transforms as $\mu_i \to (1-\delta_{E_i})^{-1}\mu_i$, where $\delta_{E_i} = (E_{\nu_i,t} - E_{\nu_i,p})/E_{\nu_i,t}$. The fractional uncertainty $\delta_{\nu_i} = \Delta_{\nu_i}/E_{\nu_i}$ should also be evaluated in terms of the true neutrino energy $E_{\nu_i,t} = (1-\delta_{E_i})^{-1} E_{\nu_i,p}$. It is advantageous to work in terms of $\delta_{E_i}$ rather than $E_{\nu_i,t}$, particularly for non-relativistic neutrinos with $E_{\nu_i} \simeq m_{\nu_i}$, as the former can be approximated by the fractional uncertainty in the measured value of the neutrino mass. 

The daughter states produced in the resonant processes~\eqref{eq:RBB} and~\eqref{eq:REC} are unstable, leading to a signal that decays over time. As a result, the neutrino capture rate~\eqref{eq:acceleratorCapRate} is not the best measure of performance for this experiment. Instead, we define the quality factor
\begin{equation}
    \begin{split}
    R_\tau &= \frac{\gamma_b}{\Gamma_D}\frac{R}{N_T}\\
    &= \sqrt{\frac{\pi^3}{2}}\left(\frac{2J_D+1}{2J_P+1}\right) \frac{\gamma_b}{Q^2} \mathcal{B}_{DP}\sum_{i,s} \frac{|U_{ei}|^2 \,n_{\nu}\left(\nu_{i,s}\right)}{\mu_i\sqrt{\delta_{\nu_i}^2 + \delta_b^2}}\exp\left[-\frac{1}{2}\left(\frac{Q-\mu_i}{\mu_i\sqrt{\delta_{\nu_i}^2 + \delta_b^2}}\right)^2\right],
    \end{split}
\end{equation}
which is the ratio of the neutrino capture rate to the effective daughter decay rate, $\Gamma_D/\gamma_b$. In terms of the quality factor, the number of daughter states on the beam at any one time is given by (see appendix C of~\cite{Bauer:2021uyj})
\begin{equation}
    N_D(x) = N_T R_\tau \left(1-e^{-x}\right) + \mathcal{O}\left(R_\tau^2\right),
\end{equation}
where $x = t/(\gamma_b \tau_D)$ parameterises the number of daughter lifetimes $\tau_D = 1/\Gamma_D$ that have elapsed in a lab frame time $t$ and $N_T$ is the initial number of parent states on the beam. We see that for $x > 1$, the number of daughter states quickly tends to its maximum value $N_T R_\tau$, at which time the rate of neutrino captures is equal to the number of daughter decays back to the parent state. This places an upper limit on what can be achieved with the systems~\eqref{eq:RBB} and~\eqref{eq:REC}; if $N > N_T R_\tau$ events are required in order to make a statistically significant discovery of the C$\nu$B, then no detection is possible with this method. 

To resolve this issue, we can instead consider 3-state RB$\beta$ and REC systems~\cite{Bauer:2021uyj}
\begin{align}
    {^A_Z P} + \nu_e \to {^A_{Z+1} D} + e^-\,(\mathrm{bound}) \to {^A_{Z+2} F} + 2e^-\,(\mathrm{bound}) + \bar\nu_e, \label{eq:RBB3}\\
    {^A_Z P} + 2e^-\,(\mathrm{bound}) + \bar\nu_e \to {^A_{Z-1} D} + e^-\,(\mathrm{bound})\to {^A_{Z-2} F} + \nu_e, \label{eq:REC3}
\end{align}
where the new final state $F$ is a stable decay product of the daughter state $D$ that differs from $P$. Similar to the 2-state systems, $P$ should be ionised down to two electrons for an RB$\beta$ process, or completely ionised for a REC process. With this modification, there is now a probability for each daughter state to decay to the stable $F$ state, where it will remain indefinitely. As a result, the number of $F$ states on the beam at large $x$ far exceeds $N_T R_\tau$, the maximum number of $D$ states. Explicitly, the number of states on the beam evolves according to
\begin{equation}\label{eq:NFstates}
    N_F(x) = N_T R_\tau \mathcal{B}_{DF} \left(x + e^{-x} - 1\right) + \mathcal{O}(R_\tau^2),
\end{equation}
where $\mathcal{B}_{DF} = \mathrm{Br}(D + e^-(\mathrm{bound}) \to F+  2e^-(\mathrm{bound})+ \bar\nu_e)$ or $\mathrm{Br}(D + e^-(\mathrm{bound}) \to F+ \nu_e)$ is the branching ratio for the daughter state to decay to the new final state. Including the $\mathcal{O}(R_\tau^2)$ terms in~\eqref{eq:NFstates}, the maximum number of states that can be converted to signal is now
\begin{equation}
    \lim_{x\to\infty}N_F(x) = \frac{N_T \mathcal{B}_{DF} \chi}{\chi - \mathcal{B}_{DP}} \gg N_T R\tau.
\end{equation}
Here, $\chi \in [0,1]$ accounts for the fraction of daughter states that decay to the wrong parent isomer.

As many parent ions as possible should be put on the beam in order to maximise the amount of signal. However, the synchrotron radiation emitted by a high energy ions can damage equipment, an effect which becomes significantly worse at the high energies required to perform this experiment. Following~\cite{Bauer:2021uyj}, we make the crude estimate that the maximum number of ions in ionisation state $I$ than be put on the beam before causing damage is
\begin{equation}
    N_{T,\mathrm{max}} = \frac{6\sqrt{2}\pi R_{c}^\frac{7}{2}\sqrt{r}}{\alpha I^2 a_p}\frac{1}{\gamma_b^5} \,q_{\mathrm{out}}(T_\infty,T_c),
\end{equation}
for an accelerator ring of radius $R_c$ and beampipe radius $r$, where $\alpha$ is the fine structure constant and $a_p\in[0,1]$ is the absorptance of the beam pipe that accounts for the incomplete absorption of synchrotron radiation. The function $q_\mathrm{out}(T_\infty,T_c)$ encodes the rate of heat loss by the beampipe in contact with a coolant at temperature $T_c$, assuming a safe equilibrium temperature $T_\infty$ can be attained. This is in turn given by
\begin{equation}
    q_\mathrm{out}(T_\infty, T_c) = \frac{\kappa_\mathrm{con}}{\Delta}(T_\infty - T_c) + \varepsilon_r \sigma(T_\infty^4 + T_c^4),
\end{equation}
where $\kappa_\mathrm{con}$, $\varepsilon_r$ and $\Delta$ are thermal conductivity, emissivity and thickness of the beampipe wall respectively, and $\sigma$ is the Stefan-Boltzmann constant.

We now have everything required to estimate the constraints that can be placed on the local C$\nu$B overdensity using this method. Assuming that $N$ events are required for statistical significance, and setting $\mu_i = (E_{\nu_i}/E_{\nu_j})Q$ and $\gamma_b = Q/E_{\nu_j}$, \textit{i.e.} choosing the beam energy such that mass eigenstate $j$ is precisely on resonance, we find the limit on the overdensity after an experimental runtime $x$
\begin{equation}\label{eq:acceleratorLimits}
    \begin{split}
        \sum_{i,s} \frac{E_{\nu_j}}{E_{\nu_i}}&\frac{|U_{ei}|^2  \,\eta_{\nu}(\nu_{i,s})}{\sqrt{\delta_{\nu_i}^2 + \delta_b^2}}\exp\left[-\frac{1}{2}\left(\frac{E_{\nu_j}-E_{\nu_i}}{E_{\nu_i}\sqrt{\delta_{\nu_i}^2 + \delta_b^2}}\right)^2\right] \\
        &\leq \frac{2\alpha I^2 a_p}{9\sqrt{\pi}\zeta(3)}\frac{N}{T_{\nu,0}^3}\left(\frac{2J_P+1}{2J_D+1}\right)\frac{Q^7}{E_{\nu_j}^4}\frac{1}{R_c^{\frac{7}{2}}\sqrt{r}}\frac{1}{q_{\mathrm{out}}(T_\infty,T_c)}\frac{1}{\mathcal{B}_{DP}X(x)}\\
        &\simeq (9.58\times 10^6) \,\frac{N I^2}{\mathcal{B}_{DP}} \left(\frac{2J_P+1}{2J_D+1}\right)\left[\frac{Q}{10\,\mathrm{keV}}\right]^7\left[\frac{10\,\mathrm{meV}}{E_{\nu_j}}\right]^4\left[\frac{1-e^{-1}}{X(x)}\right],
    \end{split}
\end{equation}
where
\begin{equation}
    X(x) = \begin{cases}
       1-e^{-x}, &\quad \text{2-state systems},\\
       \mathcal{B}_{DF}(x+e^{-x}-1), &\quad \text{3-state systems}.
    \end{cases}
\end{equation}
If the neutrino energy is not well known, recall that we must make the replacement $E_{\nu_i} \to (1-\delta_{E_i})^{-1}E_{\nu_i}$ in~\eqref{eq:acceleratorLimits}. For the reference scenario in~\eqref{eq:acceleratorLimits} we have chosen an LHC-sized ring with the choice of experimental parameters given in~\cite{Bauer:2021uyj}, using a two state system at time $x = 1$.

Perhaps most striking about~\eqref{eq:acceleratorLimits} is the $Q^7$ dependence, strongly emphasising the need for targets that have a small neutrino capture threshold to place any meaningful constraints. Reducing the threshold also decreases the beam energy requirements to hit a resonance, making the experiment easier to perform. Provided that the threshold can be kept small, however, it is clear that an accelerator experiment can set very competitive constraints on the C$\nu$B overdensity. Typical thresholds for REC and RB$\beta$ processes range from ten to a few hundred keV, requiring beam energies of a hundred to several thousand TeV. Fortunately, this can be alleviated somewhat by instead using excited states on the beam, which effectively reduces the threshold from $Q$ to $Q-E^*$, where $E^*$ is the excitation energy. With this method, keV or smaller thresholds are attainable~\cite{Keblbeck:2022twm,Gamage:2019xvx}, strengthening the bounds~\eqref{eq:acceleratorLimits} by many orders of magnitude. Unfortunately, using excited states comes at the cost of beam stability and increased experimental challenge, both of which are discussed at length in~\cite{Bauer:2021uyj}. It is also important to note this experiment could be performed with targets other than ions; any resonant process where the parent state can be accelerated on a beam, \textit{e.g.} a muon to pion system, can be used with the formalism developed here. 

It should be noted that we have not used polarised cross sections in this section as they do not change the bound on the overdensity. If we use polarised cross sections, then~\eqref{eq:relicAcceleratorXsec} should be appended with a factor of $\mathcal{A}_s(\beta_{\nu_i})$, which due to the relativstic nature of neutrinos in the beam rest frame equates to a global factor of two for beam frame left helicity neutrinos, and zero for right helicity neutrinos. However, replacing $\beta_{\Earth}$ in~\eqref{eq:helicityFlip} with the beam velocity, $\beta_b \simeq 1$, we see that any helicity asymmetry should be completely washed out by the relative motion of the beam to the C$\nu$B. As a result, the beam rest frame left helicity neutrino flux should be the average of the lab frame left and right helicity fluxes, cancelling the factor of two and recovering~\eqref{eq:acceleratorLimits}. Finally, we note that in the standard scenario, we expect the capture rate for Majorana neutrinos twice as large as for Dirac neutrinos due to the additional right helicity flux.

\subsection{Neutrino decay}\label{sec:lim}

There is now considerable evidence that at least two of the three neutrino states are massive. Consequently, massive neutrino states pick up an electromagnetic moment through loop induced effects, allowing for decays from heavier to lighter neutrinos through the emission of a photon. Considering only the degrees of freedom in the SM, along with a right chiral neutrino field that is required to generate a neutrino mass, the neutrino lifetime is predicted to be $\tau_{\nu_i} \simeq 2.4\,(10\,\mathrm{meV}/m_{\nu_i})^5\times 10^{46}\,\mathrm{y}$~\cite{Pal:1981rm}, which far exceeds the age of the universe. However, this could be significantly shorter in the presence of additional degrees of freedom, with the current strongest bounds allowing for neutrino lifetimes that satisfy $\sum_i |U_{e i}|^2 \tau_{\nu_i}/m_{\nu_i} \gtrsim 220\,\mathrm{y}\,\mathrm{eV}^{-1}$~\cite{Raffelt:1985rj}. 

The electromagnetic decay of neutrinos from the cosmic neutrino background would result in an background of photons, which in the rest frame of the decaying neutrino are emitted with energy
\begin{equation}
    E_\gamma^{ij}(m_{\nu_i}) = \frac{\Delta m_{ij}^2}{2m_{\nu_i}},
\end{equation}
for the decay $\nu_i \to \nu_j + \gamma$. It has therefore been suggested in~\cite{Bernal:2021ylz} that the spectral lines from relic neutrino decays could be observed using line intensity mapping (LIM), which could place competitive bounds on the neutrino lifetime and provide direct evidence for the cosmic neutrino background. The observables at a LIM experiment depend on the emitted photon luminosity density at each point $\vec{r}$, which for the decay $\nu_i \to \nu_j$ is given by
\begin{equation}
    \rho_\mathrm{L}^{ij}(\vec{r}) = \sum_{\nu,s} n_{\nu}(\nu_{i,s})(\vec{r}) \Gamma^{ij}_\nu E_\gamma^{ij}(m_{\nu_i}),
\end{equation}
where $\Gamma_\nu^{ij}$ is the partial decay width for the process. The signal therefore depends not only on the neutrino decay lifetime, but also the magnitude of the relic neutrino number density. In~\cite{Bernal:2021ylz} the authors assume the standard scenario, where each of the six populated neutrino states has a constant number density $n_{\nu,0}$, and forecast the sensitivity of several LIM experiments to the neutrino lifetime. If we instead fix the neutrino lifetime to a well-motivated value from theory, we can translate their forecasted sensitivities to the neutrino lifetime to an overdensity bound via
\begin{equation}
    \sum_{\nu,s}\eta_{\nu}(\nu_{i,s}) \leq 2 \frac{\Gamma_{\nu,0}^{ij}}{\Gamma_{\nu}^{ij}},
\end{equation}
\begin{figure}[tbp]
\centering 
\includegraphics[width=.49\textwidth]{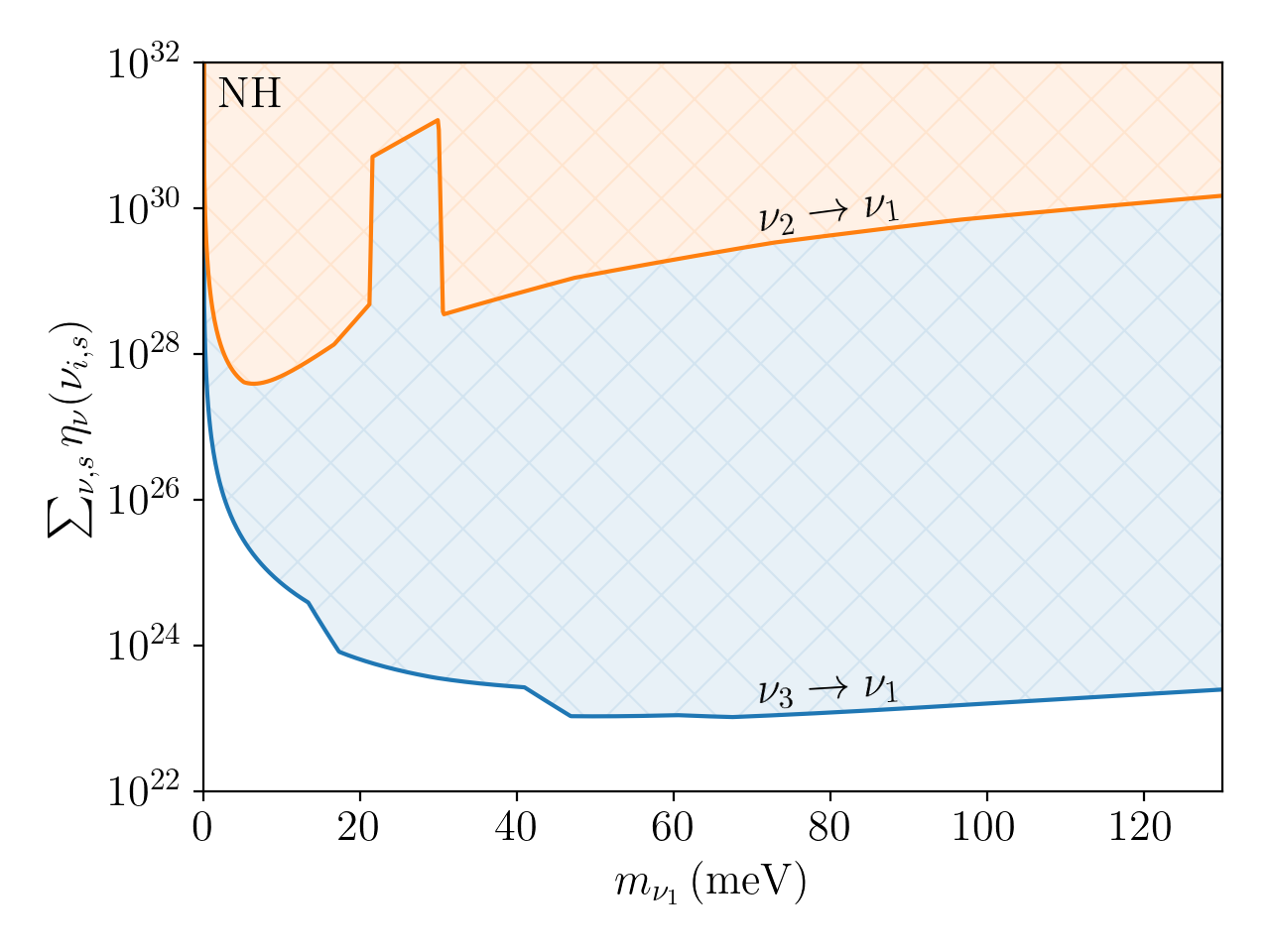}
\hfill
\includegraphics[width=.49\textwidth]{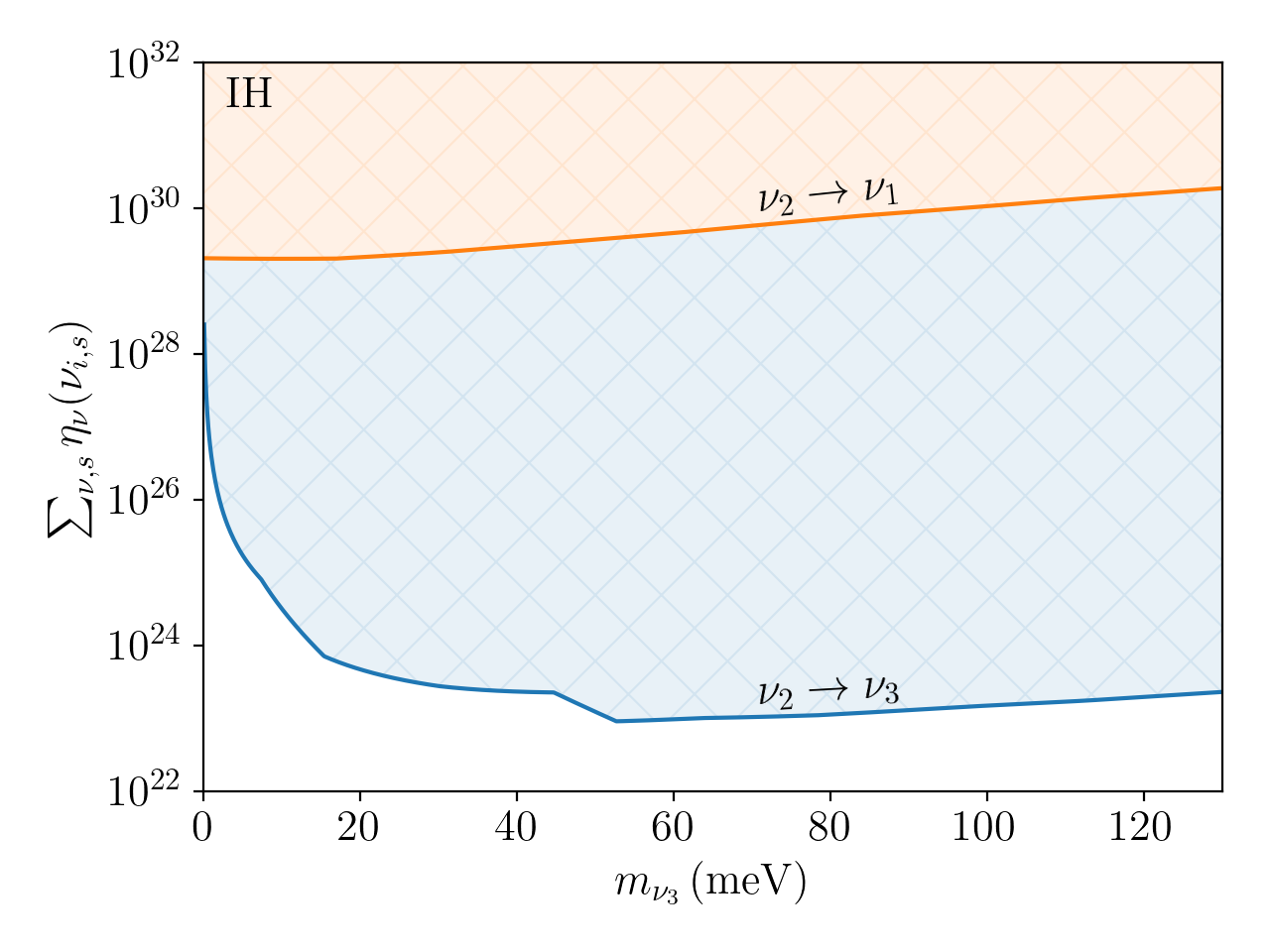}
\caption{\label{fig:lim}Bounds that could be set on the decaying relic neutrino overdensity by LIM experiments as a function of the lightest neutrino mass. For all values of the mass, we use the best experimental sensitivity of COMAP~\cite{COMAP:2018kem}, CCAT-prime~\cite{Choi:2019rrt} and AtLAST~\cite{AtLAST:2020} as given in~\cite{Bernal:2021ylz}, and assume the theoretical decay width~\eqref{eq:nutauTheory}, with effective electromagnetic moment~\eqref{eq:nuEMmoment}.  Left: In the normal mass hierarchy. Right: In the inverted mass hierarchy.}
\end{figure}
where $\Gamma_{\nu,0}^{ij}$ is the sensitivity projection given in~\cite{Bernal:2021ylz}. We note, however, that this process is unable to place any bounds on the radiatively-stable neutrino overdensities. The simplest choice for the neutrino decay width is to consider an uncharged neutrino\footnote{Millicharged neutrinos are possible with the introduction of an $\mathrm{SU}(2)_L$ singlet neutrino~\cite{Giunti:2014ixa}, also generating a Dirac mass. However, the charge of the neutrino is heavily constrained by measurements of the angular velocities of pulsars to satisfy $Q_{\nu} \lesssim 1.3\times 10^{-19}\,e$~\cite{Studenikin:2012vi}, where $e$ is the elementary charge.} with a non-zero effective electromagnetic moment $\mu_{ij}^\mathrm{eff}$, in which case the partial width is given by~\cite{Raffelt:1999gv}
\begin{equation}\label{eq:nutauTheory}
    \Gamma_{\nu}^{ij} = \frac{(\mu_{ij}^{\mathrm{eff}})^2}{8\pi}\left(\frac{\Delta m_{ij}^2}{m_{\nu_i}}\right)^3,
\end{equation}
leading to the overdensity bound
\begin{equation}
    \sum_{\nu,s}\eta_{\nu}(\nu_{i,s}) \leq 16\pi \frac{\Gamma_{\nu,0}^{ij}}{(\mu_{ij}^{\mathrm{eff}})^2} \left(\frac{m_{\nu_i}}{\Delta m_{ij}^2}\right)^3.
\end{equation}
In the SM, the effective electromagnetic moment is given by~\cite{Shrock:1982sc, Giunti:2014ixa}
\begin{equation}\label{eq:nuEMmoment}
    \begin{split}
        \mu_{ij}^\mathrm{eff} &\equiv \sqrt{|\mu_{ij}|^2 + |\epsilon_{ij}|^2}\\
        &\simeq (7.8\times 10^{-25}) \mu_\mathrm{B} \left(\frac{\sqrt{m_{\nu_i}^2 - \Delta m_{ij}^2}}{10\,\mathrm{meV}}\right),
    \end{split}
\end{equation}
where $\mu_{ij}$ and $\epsilon_{ij}$ are the neutrino transition magnetic and electric dipole moments, respectively, and $\mu_\mathrm{B}$ is the Bohr magneton. We will use~\eqref{eq:nuEMmoment} for the remainder of this section, although we note that the bounds on $\mu_{ij}^\mathrm{eff}$ are far weaker than the theoretical value, still allowing for neutrino transition electromagnetic moments satisfying~\cite{Raffelt:1999gv}
\begin{equation}\label{eq:transMomBound}
    \mu_{ij}^\mathrm{eff}\lesssim 3.2\times 10^{-16}\mu_\mathrm{B}\, \left(\frac{m_{\nu_i}}{10\,\mathrm{meV}}\right)^{9/4}.
\end{equation}
We plot the forecasted sensitivity to the C$\nu$B overdensity in Figure~\ref{fig:lim}, where the bounds for the $\nu_3 \to \nu_2$ (NH) and $\nu_1 \to \nu_3$ (IH) transitions are expected to be similar to for $\nu_3 \to \nu_1$ (NH) and $\nu_2 \to \nu_3$ due to the relative smallness of $\Delta m_{12}^2$. Clearly, if neutrinos have the lifetime and effective electromagnetic moment predicted by theory, LIM experiments will have very little sensitivity to the overdensity. However, for a neutrino electromagnetic moment saturating the experimental bound~\eqref{eq:transMomBound}, LIM experiments could set overdensity limits that are competitive with the other direct detection proposals. For completeness, we note that a similar proposal to detect relic neutrinos using radiative neutrino-neutrino scattering can be found in~\cite{Asteriadis:2022zmo}.

\section{Indirect detection proposals}\label{sec:indirect}

It is clear that the direct detection of relic neutrinos is incredibly challenging for any terrestrial experiment, requiring either extreme precision or energy in order to make an observation. Fortunately, the presence of the C$\nu$B may instead be deduced from the effect it has on visible matter, which naturally becomes stronger with larger overdensities. Here we discuss indirect detection proposals, where the effects of relic neutrinos are inferred from their effects on other observable parameters. 

\subsection{Cosmic ray neutrino attenuation}\label{sec:zburst}
The presence of the C$\nu$B may be inferred from measurements of cosmic rays reaching Earth, whose flux may be attenuated by scattering on relic neutrinos. This effect is expected to be most pronounced when the incident cosmic ray scatters from a relic neutrino resonantly, resulting in a narrow absorption line in the cosmic ray spectrum analogous to the Greisen-Zatsepin-Kuzmin (GZK) cutoff~\cite{Greisen:1966jv,Zatsepin:1966jv} for protons scattering on the CMB. 

In particular, we consider scattering of high energy neutrinos on the C$\nu$B at the $Z$-boson resonance~\cite{Eberle:2004ua,Crooks:2000jw,Fargion:1997ft,Weiler:1999ny,Fodor:2002hy} as well as the $\rho^0,\omega$ and $\phi$ vector meson resonances~\cite{Dev:2021tlo}. The $Z$-boson is chosen due to its large resonant cross section and the vector meson resonances due their significantly smaller mass, $m_V \ll M_Z$, which as a result require much lower neutrino energies to produced resonantly. We note that several other resonances exist; one may also consider the scattering of electrons on relic neutrinos at the $W$-resonance or into charged vector mesons, or neutrino scattering into neutral (pseudo)-scalar resonances. However, at ultrahigh energies, the cosmic ray electron flux is considerably smaller than the proton flux~\cite{Berat:2022iea}, which is comparable to the neutrino flux~\cite{Ahlers:2012rz}, whilst the absorption cross section for (pseudo-)scalar resonances is smaller than the (axial-)vector cross section by a factor $\sim m_\nu^2/m_S^2$, for scalar meson mass $m_S$. We therefore expect that the $Z$-boson and vector meson resonances are the most promising channels through which we can observe this effect.

The cross section for the resonant process $\nu_i + \nu_{i,\mathrm{C}\nu\mathrm{B}} \to R \to X$, where $R$ is the vector resonance under consideration and $X$ is some non-specific final state, takes the standard Breit-Wigner form
\begin{equation}\label{eq:resXsec}
    \sigma_R = \frac{3\pi}{k^2}\left[\frac{\bar E^2 \Gamma_R^2}{(\bar E^2-m_R^2)^2 +\Gamma_R^2 m_R^2}\right] \mathrm{Br}(R\to \nu_i \bar\nu_i),
\end{equation}
where $\bar E^2\simeq 2E_{\nu_i} E_\mathrm{CR}$ is the CoM energy given in terms of the cosmic ray energy, $E_\mathrm{CR}$, $m_R$ and $\Gamma_R$ are the mass and total decay width of the resonance, respectively, $k$ is the initial state momentum in the CoM frame and $\mathrm{Br}(R\to\nu_i\bar\nu_i)$ is the branching ratio for the resonance to decay back to a pair of neutrinos. These are in turn given by
\begin{align}
    \mathrm{Br}(Z\to\nu_i\bar\nu_i) &\simeq 0.067,\\
    \mathrm{Br}(\rho^0\to\nu_i\bar\nu_i) &\simeq 7.45\times 10^{-14},\\
    \mathrm{Br}(\omega\to\nu_i\bar\nu_i) &\simeq 7.99\times 10^{-14},\\
    \mathrm{Br}(\phi\to\nu_i\bar\nu_i) &\simeq 6.04\times 10^{-12},
\end{align}
for each of the resonances, where we have used the $Z$-boson branching ratio to neutrinos and total vector meson decay widths found in~\cite{ParticleDataGroup:2020ssz}, along width the theoretical vector meson decay widths to neutrinos derived in appendix~\ref{sec:mesonDecay}. 

Directly on resonance when $\bar E^2 = m_R^2$ and $k \simeq m_R/2$, the square bracketed term appearing in~\eqref{eq:resXsec} is equal to one, and the cross section takes its maximum value. By comparing the branching ratios and masses, we see that the cross section for the production of vector meson resonances is much smaller than that for $Z$-production. However, due to their lower masses the vector meson resonances can be produced with significantly lower energies than their $Z$-boson counterparts, requiring less extreme cosmic neutrinos sources to observe this effect. In general, the cosmic ray neutrino energy required to produce resonance $R$ by scattering from relic neutrino mass eigenstate $i$ is
\begin{equation}
    E_\mathrm{CR} \simeq \frac{m_R^2}{2E_{\nu_i}}, 
\end{equation}
such that for non-relativistic neutrinos with mass $m_{\nu_i}= 0.1\,\mathrm{eV}$, the $Z$-resonance requires $E_\mathrm{CR}\simeq 4\times 10^{10}\,\mathrm{TeV}$, whilst the vector meson resonances can be hit with much lower energies $E_\mathrm{CR}\simeq10^{6}\,\mathrm{TeV}$. Importantly, there is predicted to be a sizeable diffuse flux of neutrinos near the vector meson threshold at $10^6\,\mathrm{TeV}$, which rapidly drops off at higher energies~\cite{Ahlers:2012rz}. IceCube-Gen2 is expected to be able to probe this region~\cite{IceCube-Gen2:2020qha}, and a significant deviation from the predicted flux may be indicative of strong interactions between C$\nu$B and cosmic ray neutrinos. However, as the cosmic ray neutrinos at EeV and above energies are predicted to originate primarily from extragalactic sources~\cite{PierreAuger:2017pzq}, this would not necessarily indicate a local overdensity. This is particularly true for distant sources at large redshifts $z$, as the relic neutrino density should scale as $(1+z)^3$. For this reason, we will only consider local sources for the remainder of this section for which $n_{\nu} \neq n_{\nu}(z)$.

Now suppose that a source at a distance $L$ from Earth emits neutrinos of mass eigenstate $i$ with flux $\phi_\mathrm{CR}(E_\mathrm{CR},\ell = 0)$. The change in this flux along the line of sight due to attenuation by annihilation on relic neutrinos will satisfy
\begin{equation}
    \frac{d\phi_\mathrm{CR}}{d\ell} = - \phi_\mathrm{CR}(E_\mathrm{CR},\ell)\sum_{R,s}\sigma_R(E_\mathrm{CR})n_\nu(\nu_{i,s}),
\end{equation}
This is easily solved to find the flux of high energy cosmic ray neutrinos reaching the Earth
\begin{equation}\label{eq:CRattentuation}
    \phi_\mathrm{CR}(E_\mathrm{CR},L) = \phi_\mathrm{CR}(E_\mathrm{CR},0)\exp\left(-L\sum_{R,s} \sigma_R(E_\mathrm{CR}) n_\nu(\nu_{i,s}) \right).
\end{equation}
Considering the $Z$-resonance, the effective survival distance of cosmic ray neutrinos on resonance is therefore
\begin{equation}
    L_S = \frac{1}{\sigma_Z(E_\mathrm{CR})n_{\nu}(\nu_{i,s})} \simeq \left(\frac{1.29\times 10^4}{\sum_{s}\eta_\nu(\nu_{i,s})}\right)\,\mathrm{kpc}.
\end{equation}
Cosmic ray neutrinos originating from distances $L \gg L_S$ should have clear absorption lines in their spectra, which for $\eta_{\nu} \gtrsim 100$ extends to all extragalactic sources. It is also important to estimate the widths, $\Delta E_\mathrm{CR}$, of these absorption lines, as any cosmic ray detector with insufficient energy resolution $\Delta \gg \Delta E_\mathrm{CR}$ will be unable to clearly resolve the attenuation. The cross section~\eqref{eq:resXsec} receives a resonant enhancement when $(\bar E^2 - m_R^2)^2 \ll \Gamma_R^2 m_R^2$. This is satisfied on the interval
\begin{equation}
    \frac{m_R^2 - m_R\Gamma_R}{2E_{\nu_i}} \ll E_\mathrm{CR} \ll  \frac{m_R^2 + m_R\Gamma_R}{2E_{\nu_i}},
\end{equation}
which has width
\begin{equation}
    \Delta E_\mathrm{CR} \simeq \frac{m_R}{E_{\nu_i}}\Gamma_{R}.
\end{equation}
For the $Z$-resonance scenario with $E_{\nu_i} \simeq m_{\nu_i} = 0.1\,\mathrm{eV}$, this corresponds to a width $\Delta E_{\mathrm{CR}} \simeq 2\times 10^9\,\mathrm{TeV}$ at an energy $E_{\mathrm{CR}} = 4\times 10^{10}\,\mathrm{TeV}$, or equivalently a fractional energy resolution of around $5\%$. Achieving this energy resolution at such high energies is an incredible challenge; IceCube achieves an energy resolution of around $25\%$ at $\mathcal{O}(10\,\mathrm{GeV})$ energies~\cite{Ahlers:2018mkf}. Nevertheless, the attenuation may be still be visible as a small decrease in the number of events in the energy bin centred around the resonance. 

We can also translate the result~\eqref{eq:CRattentuation} to a limit on the local overdensity. Given that the initial flux of the source is well modelled and no attenuation is seen at Earth, we can place the following constraint on the local relic neutrino overdensity
\begin{equation}
    \sum_s \eta_\nu\left(\nu_{i,s}\right) \leq \frac{4\pi^2}{3T_{\nu,0}^3 \zeta(3)} \left(L\sum_{R}\sigma_R(E_\mathrm{CR})\right)^{-1}\ln\left(\frac{\phi_\mathrm{CR}(E_\mathrm{CR},0)}{\phi_\mathrm{CR}(E_\mathrm{CR},L)}\right).
\end{equation}
We also note that this process can be used to constrain the presence of resonances beyond the Standard Model (BSM) without modification. 

The role of resonant cosmic neutrino scattering in constraining C$\nu$B overdensities has been explored in detail in~\cite{Brdar:2022kpu}, assuming the standard scenario, where it has been estimated that IceCube-Gen2 will be able to place the experimental constraints $\eta_{\nu_i}(\nu_{i,s}) \lesssim 10^{10}$ after ten years of exposure. If realised, this constraint would improve upon the best existing experimental constraint from KATRIN, which we have previously discussed in Section~\ref{sec:existing}.

It has previously been suggested that the decays of resonances resulting from resonant $Z$-production could be responsible for the highest energy cosmic rays observed today, in particular those above the GZK cutoff~\cite{Kalashev:2001sh,Fodor:2001qy}. However, the required cosmic neutrino fluxes have since been ruled out by the ANITA experiment~\cite{ANITA:2005gdw}. Nevertheless, secondaries from such `$Z$-burst' scenarios could provide an additional window into relic neutrino detection. 

\subsection{Atomic de-excitation}\label{sec:renp}
An alternative method of detecting relic neutrinos using the Pauli exclusion principle has been suggested in~\cite{Yoshimura:2014hfa}. Due to the presence of the C$\nu$B, processes emitting neutrinos will have their phase space restricted by a factor $\sim (1-f_{\nu_i}(|\vec{p}_{\nu_i}|))$ for each final state neutrino, which becomes important in the regions where $|p_{\nu_i}|$ is comparable to the C$\nu$B momentum and leads to a modified emission spectrum. In the standard scenario, the maximum suppression of any region of phase space is $1/2^n$ for a process emitting $n$ neutrinos. This effect could be significantly larger if both relic neutrino helicity states are at least partially filled by some mechanism, such that $g_{\nu_i} > 1$. 

The authors of~\cite{Yoshimura:2014hfa} consider the radiative emission of neutrino pairs (RENP) by de-exciting  atomic states~\cite{Yoshimura:2006nd,Fukumi:2012rn,Dinh:2012qb,Yoshimura:2013wva,Sasao:2013xaa}
\begin{equation}\label{eq:RENP}
    |e\rangle \to |g\rangle + \gamma + \nu_i + \bar\nu_j,
\end{equation}
where $|e\rangle$ is an excited state of $|g\rangle$. The neutrino pair can either be emitted by the de-exciting valence electron or nucleus, with the rate of the latter expected to be significantly larger~\cite{Yoshimura:2013wva}. This process involves three states, $|e\rangle$, $|p\rangle$ and $|g\rangle$, where $|p\rangle$ and $|g\rangle$ are connected by an $\mathrm{E}1$ transition, whilst the transition from $|e\rangle$ and $|g\rangle$ involves a much weaker $\mathrm{E}1\times \mathrm{M}1$ operator. The result is a strongly suppressed rate of de-excitation by photon pair emission, $|e\rangle \to |g\rangle + \gamma + \gamma$, aiding in the measurement of RENP which proceeds through the electron spin operator from $|e\rangle$ to $|p\rangle$, followed by an $\mathrm{E}1$ transition from $|p\rangle$ to $|g\rangle$. Additionally, RENP gains a macroscopic coherent enhancement when the final state photon is emitted back to back with the neutrino pair, analogous to the process of paired superradiance discussed in~\cite{Dicke:1954zz,Yoshimura:2008ya,Yoshimura:2012tm}.

In the presence of the C$\nu$B, the shape of the outgoing photon energy spectrum will be modified due to the suppression of neutrinos emitted with momenta $|\vec{p}_{\nu_i}| \simeq T_{\nu_i}$. This process can also be used to determine the neutrino mass spectrum. At each photon energy threshold $\omega_{ij} = \epsilon_\mathrm{eg}/2 +  (m_{\nu_i} + m_{\nu_j})^2/2\epsilon_\mathrm{eg}$, where $\epsilon_\mathrm{eg}$ is the excitation energy of $|e\rangle$ relative to $|g\rangle$, there will a discontinuity in the event rate as the production of the neutrino pair $i$ and $j$ becomes available. For three neutrino mass eigenstates, we expect that there will be six thresholds for neutrino pair emission by valence electrons, or three for emission by the nucleus where only the neutral current contributes. Additionally, RENP has sensitivity to the Dirac or Majorana nature of neutrinos due to the possible interference between identical final state neutrinos~\cite{Yoshimura:2006nd}. In what follows, we will only explore the dominant mass diagonal contribution from pair emission by the nucleus. 

The rate at which photons with energy $\omega$ are produced by RENP is given by~\cite{Yoshimura:2013wva}
\begin{equation}
    \Gamma_{\gamma 2\nu}(\omega) = \Gamma_0 F^2(\omega) I(\omega) \eta_{\omega}(t),
\end{equation}
where $\eta_\omega(t)$ is a dynamical factor discussed at length in~\cite{Dinh:2012qb}, for which we will use the conservative estimate of $\eta_{\omega}(t) \simeq 10^{-6}$ following~\cite{Yoshimura:2013wva}. The remaining quantities are
\begin{align}
    \Gamma_0 &= \frac{3}{4}G_F^2 n_T^3 V \epsilon_{\mathrm{eg}},\\
    F(\omega) &= \frac{Q_V J_N}{\epsilon_\mathrm{pe}(\epsilon_\mathrm{pe}+\omega)(\epsilon_\mathrm{eg}-\omega)}\sqrt{\frac{\gamma_\mathrm{pg}}{\epsilon_\mathrm{pg}}},\label{eq:Fomega}\\
    I(\omega) &= \sum_i \Delta_{ii}(\omega)I_{ii}(\omega)\theta(\omega_{ii} - \omega),\\
    \Delta_{ii}(\omega) &= \sqrt{1-\frac{4m_{\nu_i}^2}{\epsilon_{\mathrm{eg}}(\epsilon_\mathrm{eg}-2\omega)}},
\end{align}
where $\epsilon_{ab}$ denotes the energy gap between states $|a\rangle$ and $|b\rangle$, $n_T$ and $V$ are the number density and volume of the target, respectively, $\gamma_\mathrm{pg}$ is the transition rate from state $|p\rangle$ to $|g\rangle$ and $Q_V = A - 2Z(1-2\sin^2\theta_W)$ is the vector charge of the nucleus. The factor $J_N$ in~\eqref{eq:Fomega} accounts for electromagnetic interactions between the nucleus and de-exciting valence electron, and is given approximately by
\begin{equation}
    J_N \simeq \frac{9}{5} \alpha^2 Z^{\frac{4}{3}} m_e,
\end{equation}
with $\alpha$ the fine structure constant. The remaining factor, $I_{ii}(\omega)$, is the one of most interest to us as it contains the phase space integral, which includes the suppression factor due to the presence of the C$\nu$B. This is defined by
\begin{equation}
    \begin{split}
    I_{ii}(\omega) \Delta_{ii}(\omega) \equiv \frac{1}{\omega}\int\displaylimits_{E_-}^{E_+} dE_{\nu_i} \left(-E_{\nu_i}^2 + E_{\nu_i}(\epsilon_{\mathrm{eg}}-\omega)-\frac{1}{4}\epsilon_{\mathrm{eg}}(\epsilon_\mathrm{eg}-2\omega) + \frac{m_{\nu_i}^2}{2}(1+\delta_M)\right)\\
    \times \left\{1-f_{\nu_i}\left(\sqrt{E_{\nu_i}^2 -m_{\nu_i}^2}\right)\right\}\left\{1-f_{\bar\nu_i}\left(\sqrt{(\epsilon_\mathrm{eg}-\omega-E_{\nu_i})^2 - m_{\nu_i}^2}\right)\right\},
    \end{split}
\end{equation}
where the limits of the integral are $E_{\pm} = (\epsilon_{\mathrm{eg}}-\omega \pm \omega \Delta_{ii}(\omega))/2$ and $\delta_M$ is zero for Dirac neutrinos, and one for Majorana neutrinos. In the absence of Pauli blocking, we find that
\begin{equation}
    I_{ii}(\omega)\Big|_{f_{\nu_i}=0} = \frac{\omega^2}{6}+\frac{\omega^2}{3}\frac{m_{\nu_i}^2}{\epsilon_\mathrm{eg}(\epsilon_\mathrm{eg}-2\omega)}+\frac{m_{\nu_i}^2}{2}(1+\delta_M),
\end{equation}
which is exactly a factor of two smaller than the result computed in~\cite{Yoshimura:2013wva}. The full integral including the distribution functions can in fact be evaluated analytically, however we do not give the expression here. 

The maximum suppression due to Pauli blocking is seen in the regions of phase space that are already heavily restricted by kinematics, namely near the thresholds $\omega_{ii}$. Near $\omega_{ii}$, the emitted neutrino pair is collinear, with each neutrino carrying momentum $\sim \omega_{ii}/2$. In order to maximise the suppression effect, we require that this momentum be of the same order as the neutrino temperature\footnote{This is true for unclustered neutrinos that follow their equilibrium distribution~\eqref{eq:fermiDirac}, whose momenta will be distributed around $T_{\nu_i}$. The phase space of clustered relic neutrinos may be populated differently.}, such that optimal energy gap $\epsilon_\mathrm{eg}$ to observe the Pauli blocking will satisfy
\begin{equation}\label{eq:optimalGap}
    \omega_{ii} \simeq 2T_{\nu_i} \implies \epsilon_\mathrm{eg} \simeq 2\left(T_{\nu_i}+\sqrt{m_{\nu_i}^2 + T_{\nu_i}^2}\right),
\end{equation}
which reduces to $\epsilon_{\mathrm{eg}} \simeq 2 m_{\nu_i}$ for non-relativistic relic neutrinos, $T_{\nu_i} \ll m_{\nu_i}$.
Unlike the proposals presented so far, we cannot use RENP to directly measure the local overdensity. However, a strong suppression in the spectrum could indicate at least a partial filling of the right helicity neutrino or left helicity antineutrino states, corresponding to $g_{\nu_i} > 1$ and a deviation from the standard scenario. Additionally, we may gain insight into the energy dependence of the local overdensity by use of the relation
\begin{figure}[tbp]
\centering 
\includegraphics[width=.49\textwidth]{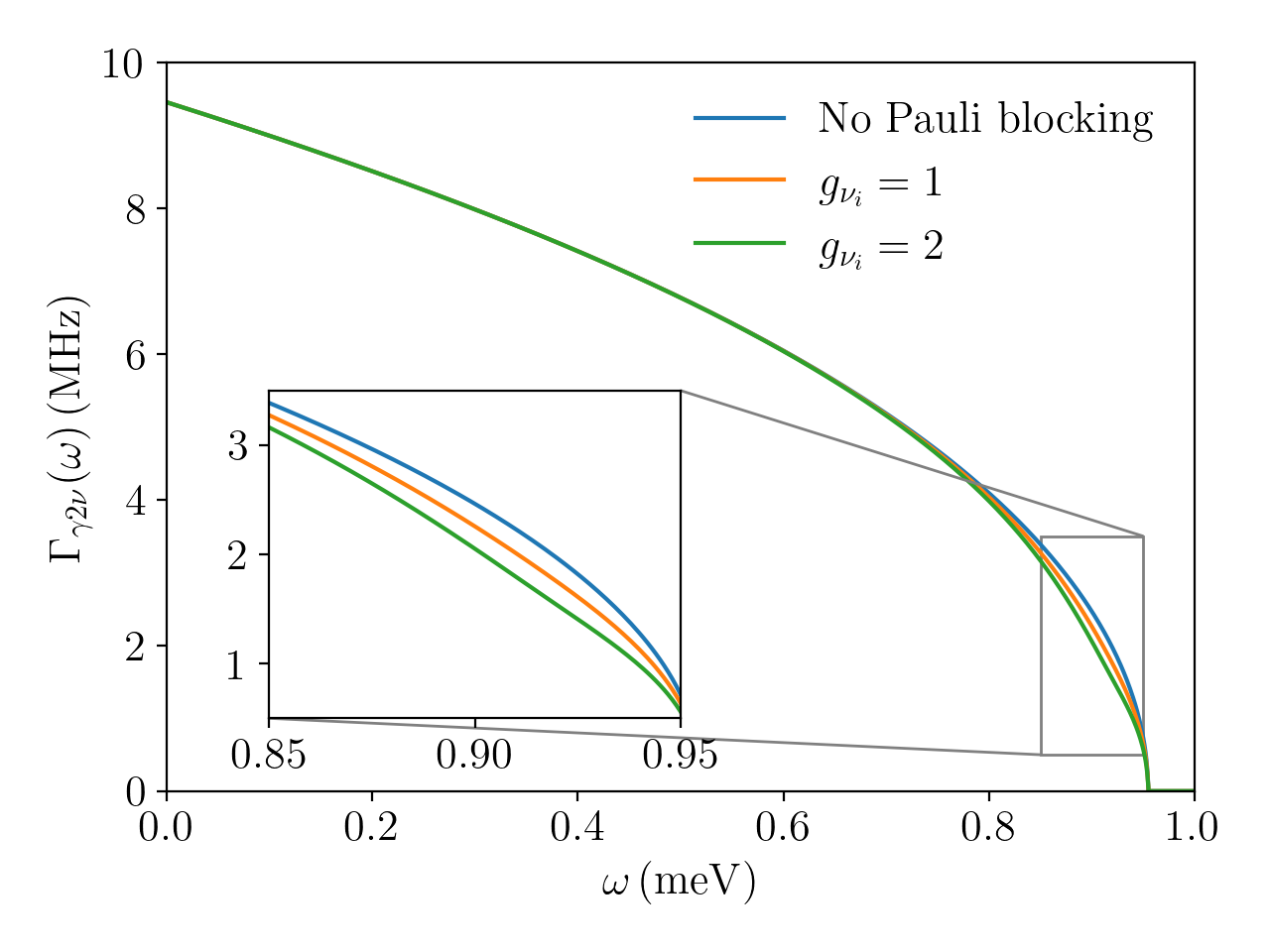}
\hfill
\includegraphics[width=.49\textwidth]{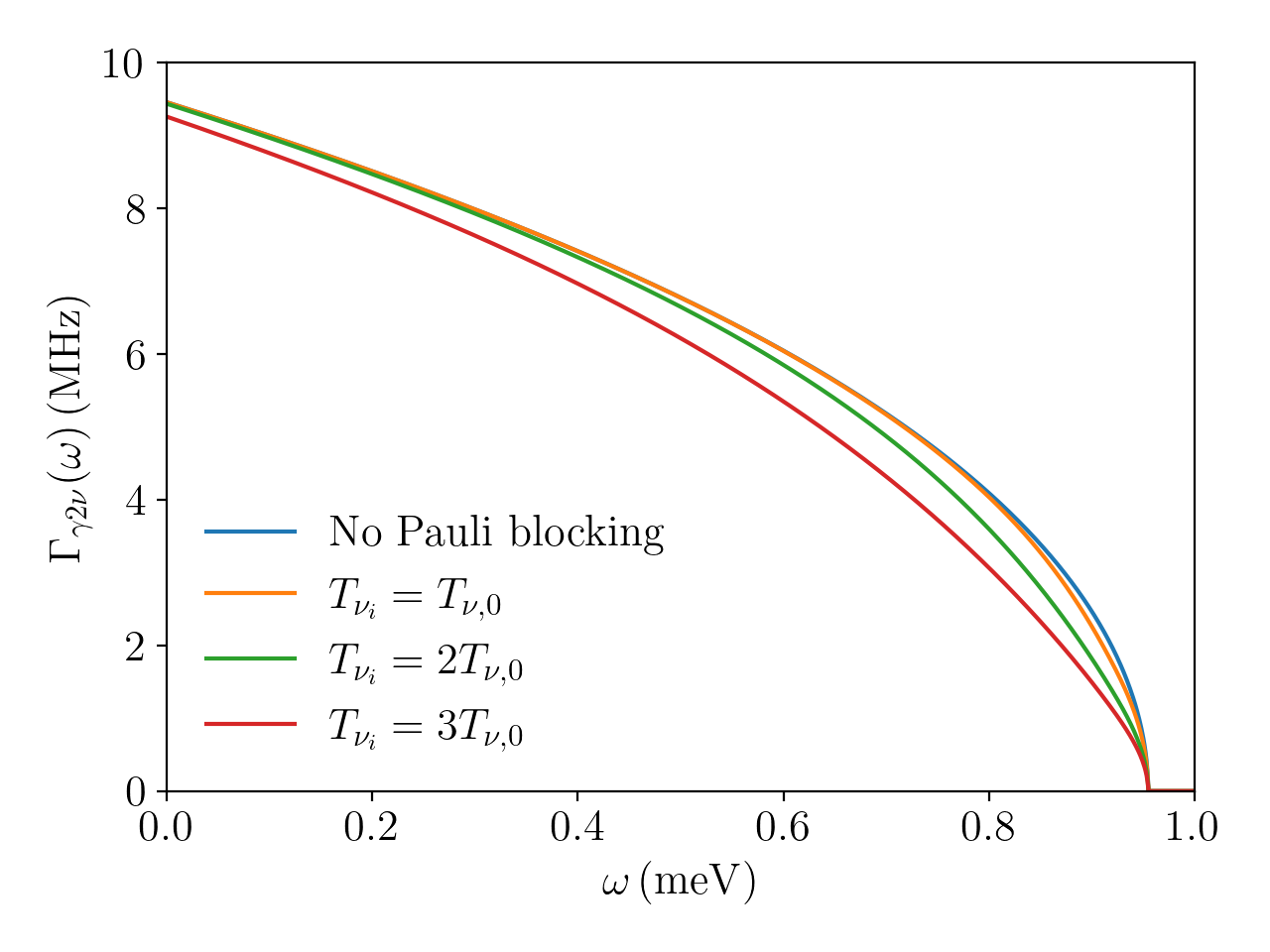}
\caption{\label{fig:renp}RENP photon energy spectrum in the presence of the C$\nu$B for the toy `Cs-like' system described in the text, where $n_T = 10^{21}\,\mathrm{cm}^{-3}$, $V = 100\,\mathrm{cm}^3$ and the lightest neutrino mass $m_{\nu_l} = 5\,\mathrm{meV}$. Left: Varying the neutrino degeneracy parameter, $g_{\nu_i}$, at constant temperature $T_{\nu_i} = T_{\nu,0}$. Right: Varying the neutrino temperature, $T_{\nu_i}$ in multiples of the standard temperature, $T_{\nu,0} = 0.168\,\mathrm{meV}$, with $g_{\nu_i} = 1$. In both plots, we assume Dirac neutrinos in the normal mass hierarchy.}
\end{figure}
\begin{equation}
    \sum_s n_{\nu}(\nu_{i,s}) = n_{\nu,0}\sum_s \eta_{\nu}(\nu_{i,s}) = \int \frac{d^3\vec{p}_{\nu_i}}{(2\pi)^3}\,f_{\nu_i}(|\vec{p}_{\nu_i}|),
\end{equation}
from which it follows that 
\begin{equation}
    f_{\nu_i}(|\vec{p}_{\nu_i}|) = \frac{3\zeta(3) T_{\nu,0}^3}{|\vec{p}_{\nu_i}|^2}\sum_s \frac{d\eta_\nu(\nu_{i,s})}{d|\vec{p}_{\nu_i}|},
\end{equation}
and similar for $f_{\bar\nu_i}$. As an illustrative example, we show the RENP photon spectrum with and without Pauli blocking in Figure~\ref{fig:renp} for a toy `Cs-like' system with $\epsilon_\mathrm{eg} = 11\,\mathrm{meV}$, $\epsilon_{pg} = 20\,\mathrm{meV}$ and $\epsilon_{pe} = 9\,\mathrm{meV}$, whilst $\gamma_\mathrm{pg} = 5\times 10^{-9}\,\mathrm{meV}$ and the lightest neutrino mass $m_{\nu_l} = 5\,\mathrm{meV}$ to approximately align with the condition~\eqref{eq:optimalGap}. This ratio of parameters is similar to that of the ${^{133}\mathrm{Cs}}$ system chosen in~\cite{Yoshimura:2013wva}, where the involved states are $|e\rangle = 6\,{^2\mathrm{P}_{1/2}}$, $|g\rangle = 6\,{^2\mathrm{S}_{1/2}}$ and $|p\rangle = 7\,{^2\mathrm{P}_{1/2}}$ in standard spectroscopic notation. For this system, the relevant parameters are $\epsilon_{eg} = 1.386\,\mathrm{eV}$, $\epsilon_{pg} = 2.699\,\mathrm{eV}$ and $\epsilon_{pe} = 1.313\,\mathrm{eV}$, whilst $\gamma_\mathrm{pg} = 5.226\times 10^{-10}\,\mathrm{eV}$~\cite{NIST:2021}.

Clearly, with the right choice of $\epsilon_\mathrm{eg}$, the effects of Pauli blocking can have a measurable impact on the neutrino emission spectrum which becomes more pronounced as we vary $g_{\nu_i}$. Interestingly, however, the Pauli blocking effect is also very sensitive to the neutrino temperature; as a large fraction of momentum states with $|\vec{p}_{\nu_i}| \ll T_{\nu_i}$ are filled, a higher temperature can lead to a drastically different photon spectrum. However, we expect that finding systems with gaps satisfying~\eqref{eq:optimalGap} and appropriate set of states $|g\rangle$, $|e\rangle$ and $|p\rangle$ will be a significant challenge, which also requires knowledge of the absolute neutrino mass. Nevertheless, this remains an interesting prospect for probing the C$\nu$B. We also note that indirectly observing relic neutrinos using Pauli blocking is not limited to RENP systems, a similar experiment could equally be performed using any process that emits low energy neutrinos.  
\section{Discussion}\label{sec:discussion}
\begin{figure}[tbp]
    \centering
    \includegraphics[width =.831\textwidth]{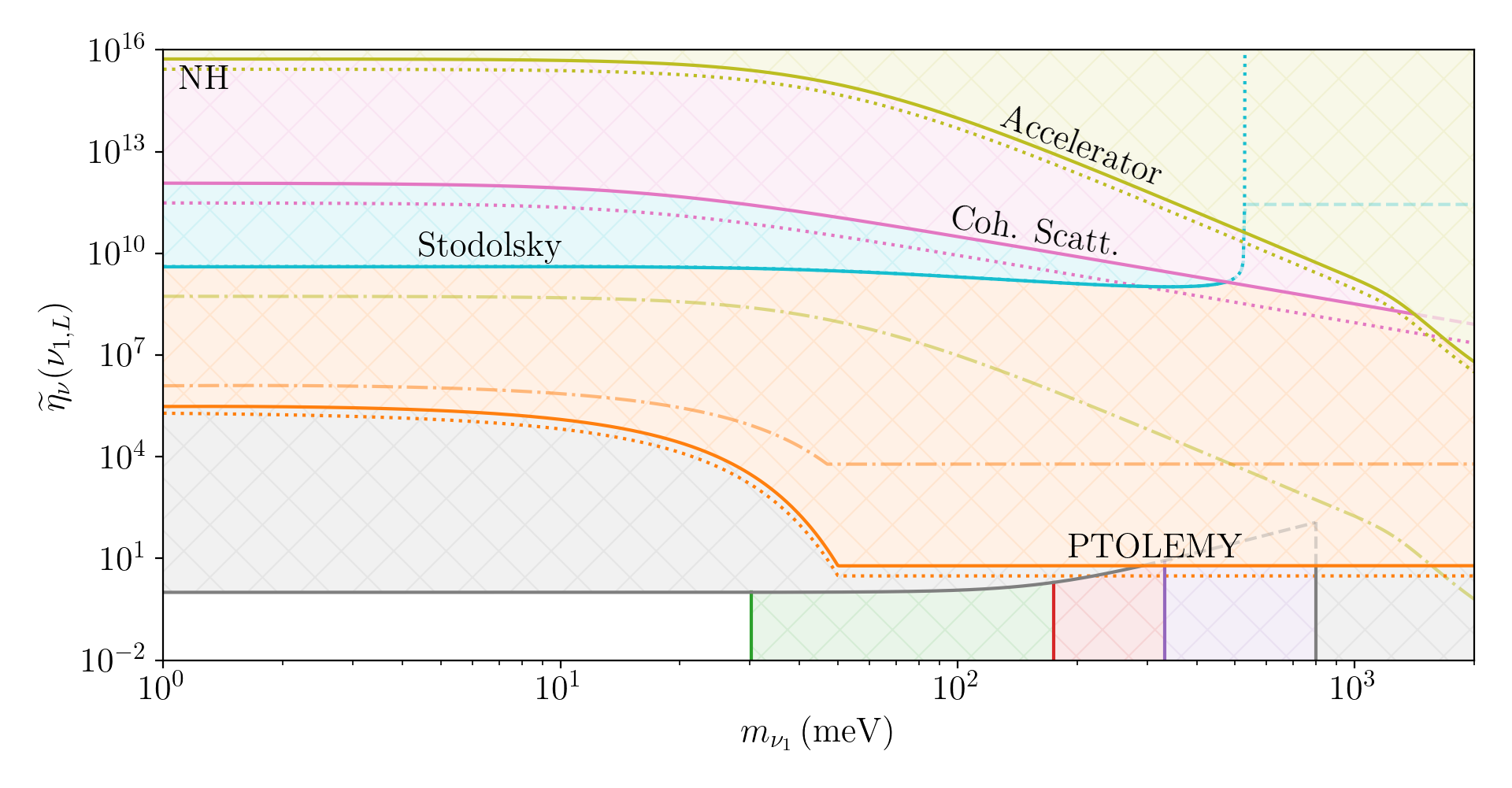}
    \hfill
    \includegraphics[width=.831\textwidth]{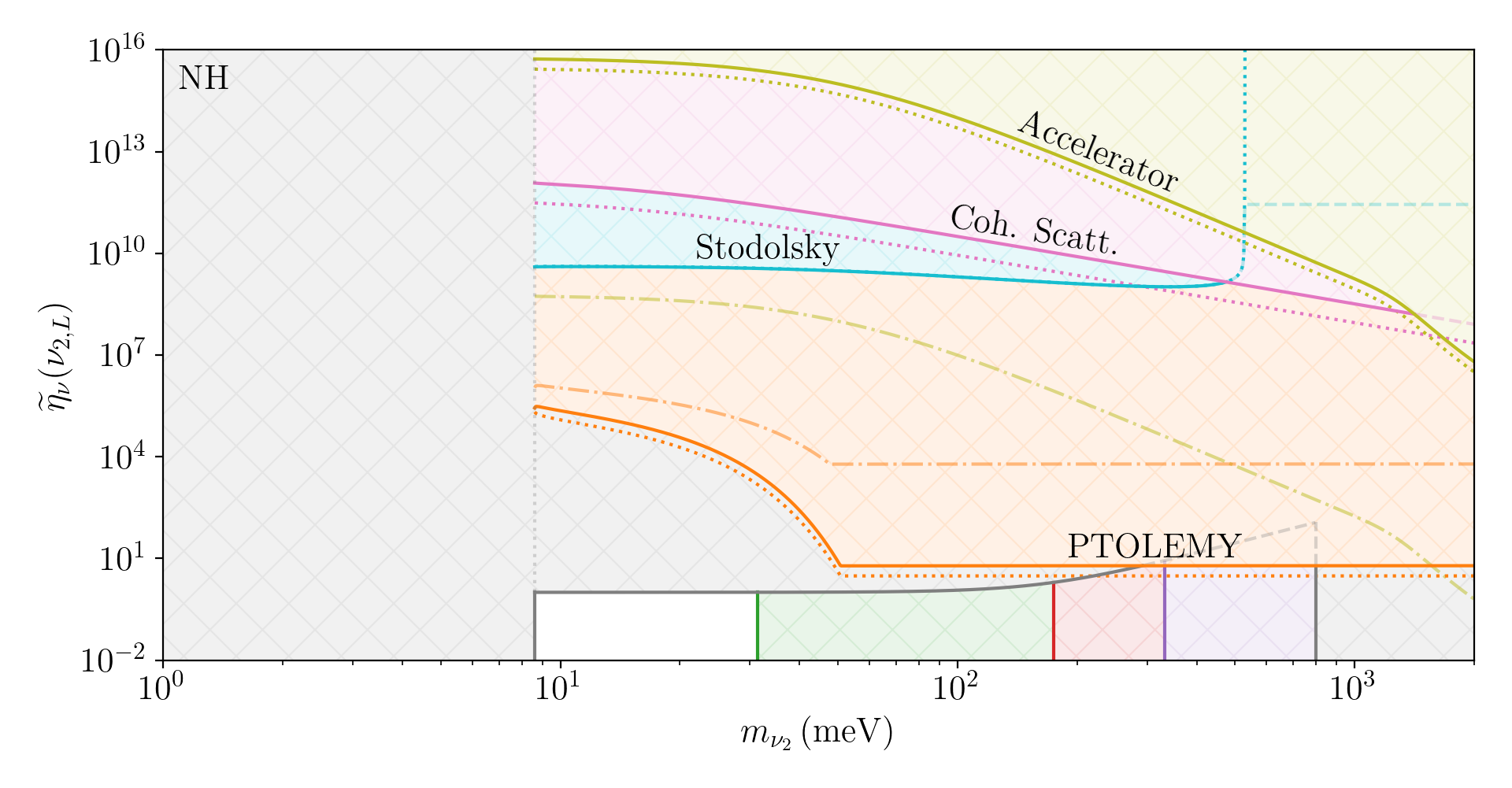}
    \hfill
    \includegraphics[width=.831\textwidth]{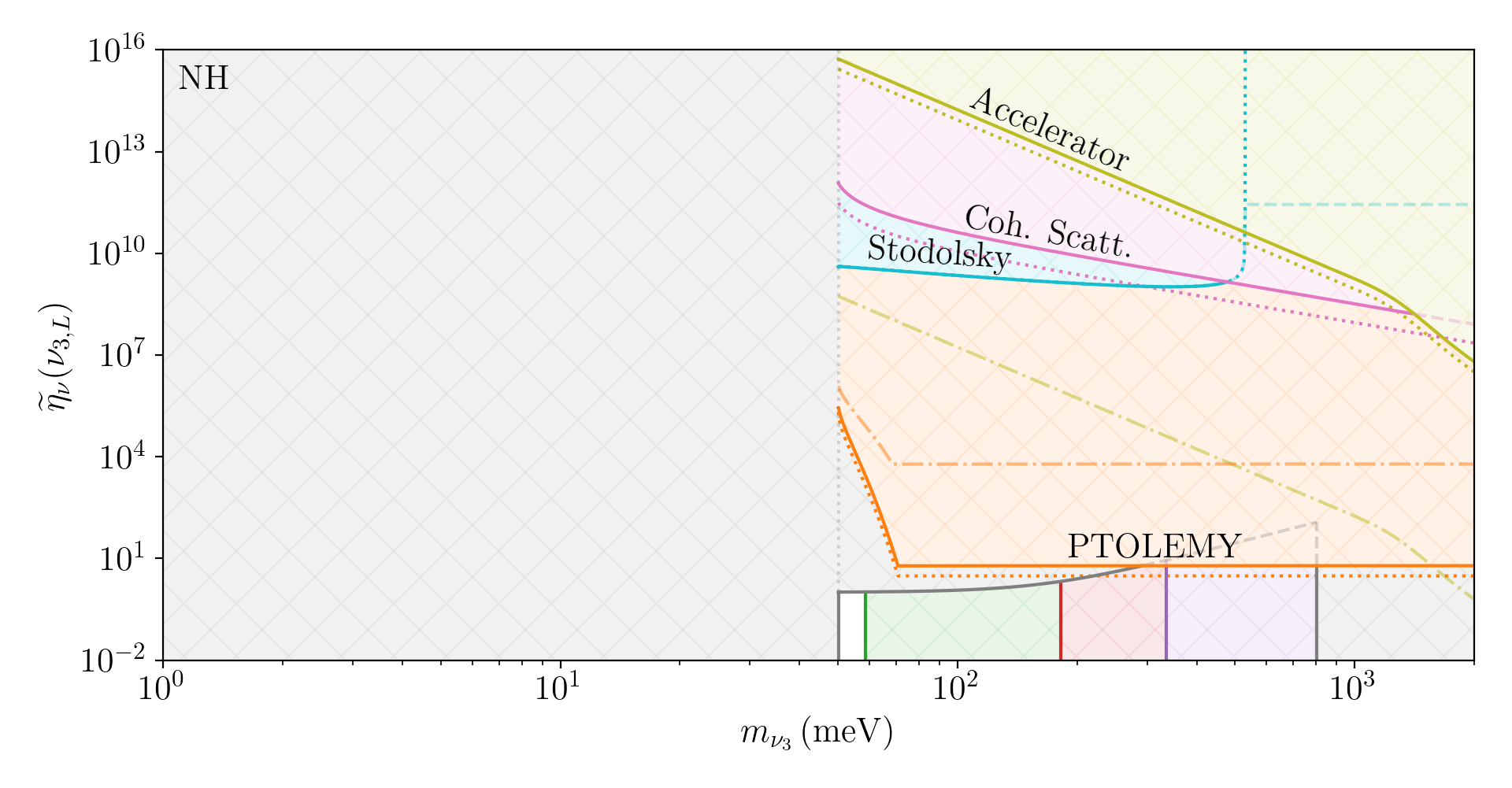}
    \caption{Sensitivity of each direct detection proposal to the C$\nu$B overdensity as a function of the neutrino mass, assuming $T_{\nu_i} = T_{\nu,0}$ and only left (right) helicity (anti)neutrinos in the normal mass hierarchy. See the text and Table~\ref{tab:linkTable} for a full description of the figure.} 
    \label{fig:NH_future}
\end{figure}

\begin{figure}[tbp]
    \centering
    \includegraphics[width =.831\textwidth]{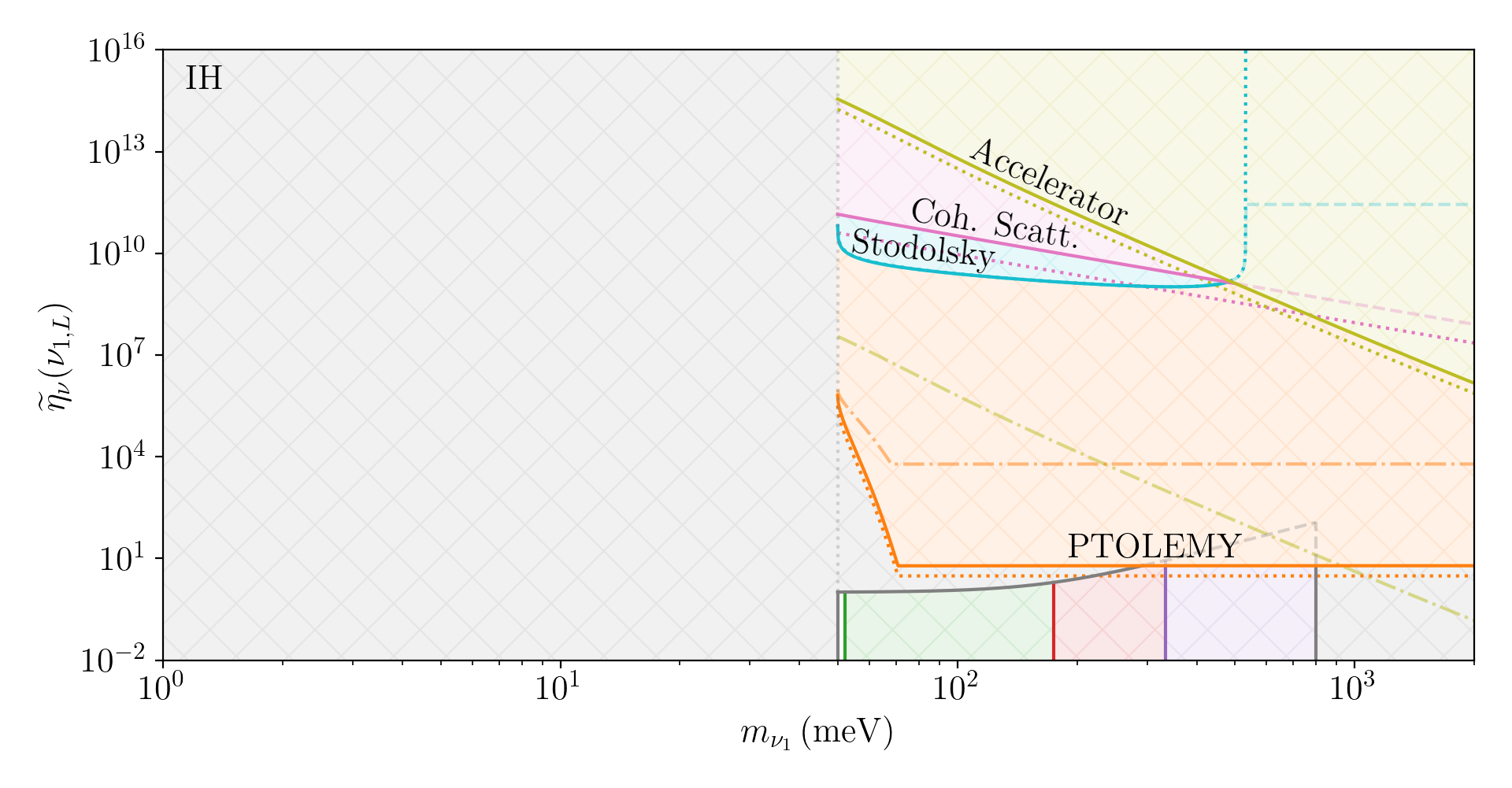}
    \hfill
    \includegraphics[width=.831\textwidth]{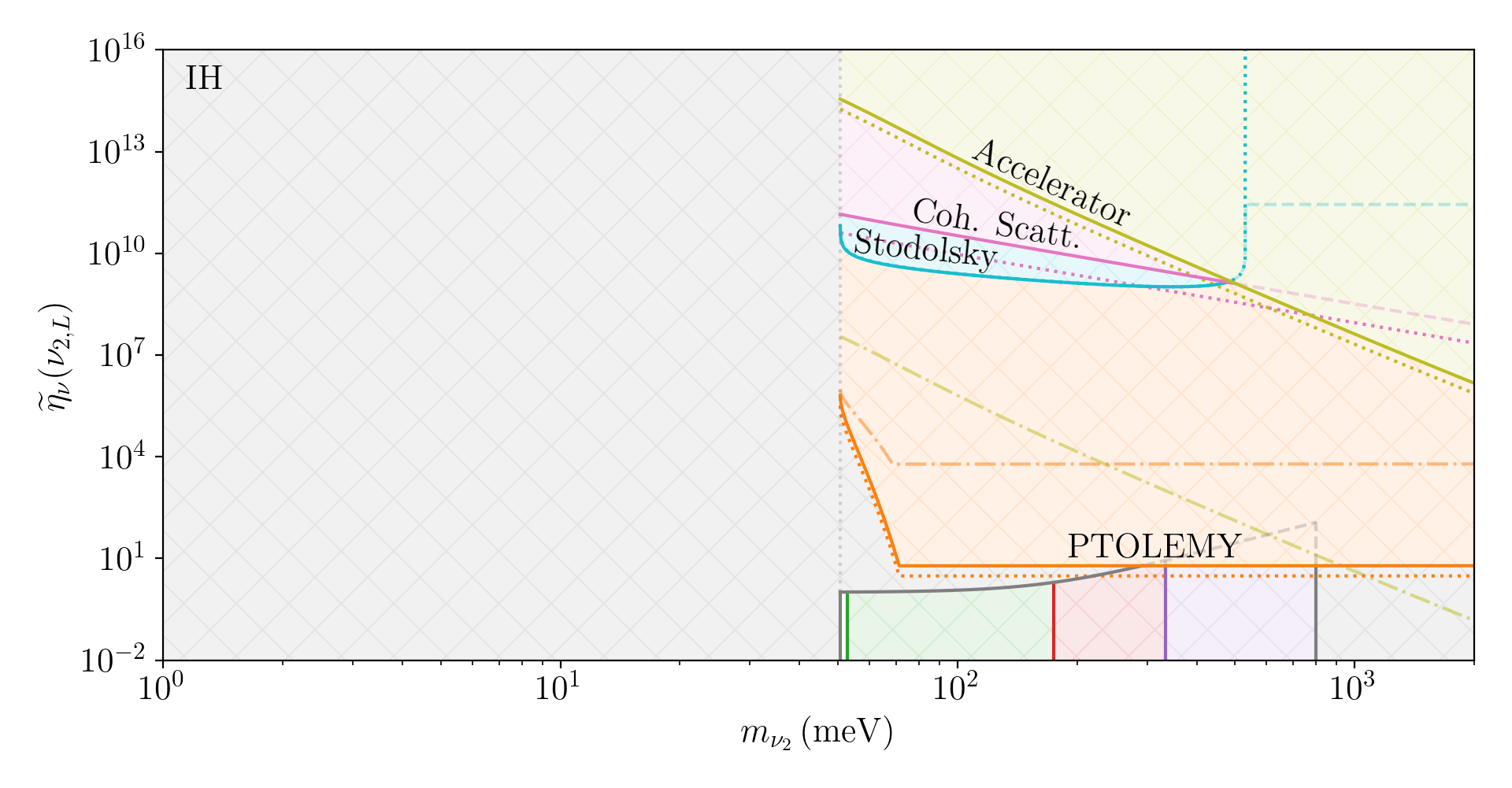}
    \hfill
    \includegraphics[width=.831\textwidth]{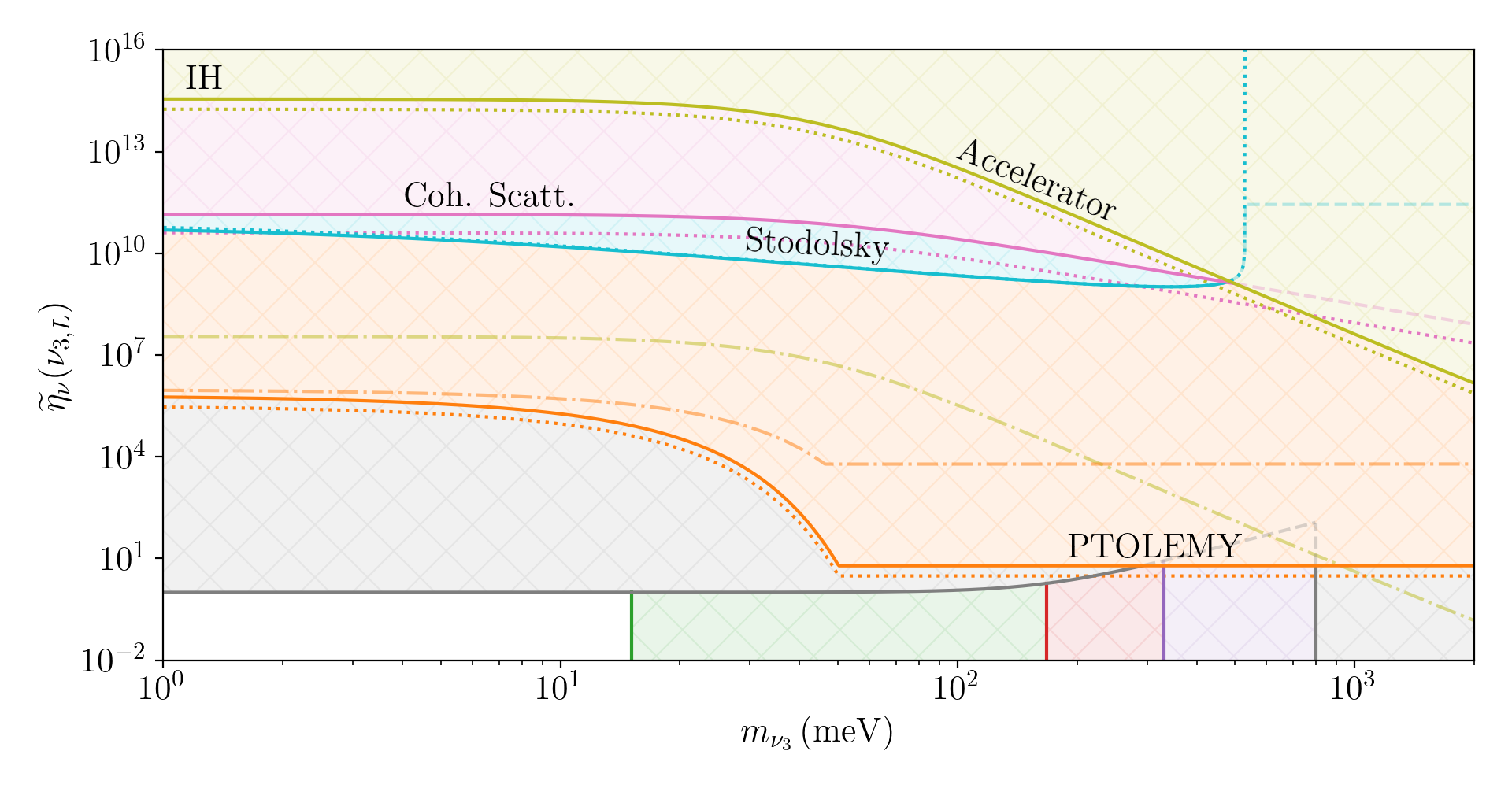}
    \caption{Sensitivity of each direct detection proposal to the C$\nu$B overdensity as a function of the neutrino mass, assuming $T_{\nu_i} = T_{\nu,0}$ and only left (right) helicity (anti)neutrinos in the inverted mass hierarchy. See the text and Table~\ref{tab:linkTable} for a full description of the figure.}  
    \label{fig:IH_future}
\end{figure}

We plot the constraints on the relic neutrino overdensity in the C$\nu$B frame and the reach of the direct detection proposals presented in Section~\ref{sec:direct} as a function of the neutrino mass, for both the NH and IH scenarios in Figures~\ref{fig:NH_future} and~\ref{fig:IH_future}, in all cases assuming the same overdensity for all three mass eigenstates, a constant temperature $T_{\nu_i} = T_{\nu_0}$ and unclustered neutrino momentum $|\vec{p}_{\nu_i}| \simeq 3.15\,T_{\nu_i}$. In addition, we assume only left (right) helicity Dirac (anti)neutrinos or both left and right helicity Majorana neutrinos, in line with the standard scenario presented in Section~\ref{sec:thermalHist}. Where a distinction can be made, all solid lines show the constraints on Dirac neutrinos, whilst the dotted lines show the constraints that could be set for Majorana neutrinos. The grey region shows the existing constraints on the C$\nu$B from the exclusion principle assuming $T_{\nu_i} = T_{\nu,0}$, as well as the mass bounds from KATRIN and oscillation experiments, whilst the purple, red and dark green regions are those that are potentially constrained by KamLAND-Zen and cosmology, which are discussed in Section~\ref{sec:existing}. Finally, we assume a runtime of $t = 1\,\mathrm{y}$, $N=25$ events, corresponding to $5\sigma$ significance, and a required signal noise ratio $r_{\mathrm{SN},0} = 1$ for all of the experimental constraints presented, where appropriate.

Clearly, PTOLEMY (orange) has the best sensitivity to the C$\nu$B overdensity for this parameter set, which rapidly becomes weaker as the neutrino mass approaches the proposed energy resolution $\Delta = 50\,\mathrm{meV}$. Importantly, PTOLEMY has the potential to probe part of the region currently unconstrained by the combination of existing constraints from KATRIN and the exclusion principle. We stress, however, that PTOLEMY is unable to set any constraints on the antineutrino overdensity, but note that the similar technique presented in~\cite{Cocco:2007za} and \cite{Cocco:2009rh} could fill this role. We also show using the sensitivity of PTOLEMY to Dirac neutrinos\footnote{The sensitivity to Majorana neutrinos using this method will differ by the same amount as the standard PTOLEMY technique.} using the time variation method with the orange dot-dashed curve. For neutrino masses below the proposed energy resolution of PTOLEMY, the sensitivity becomes comparable to that of the standard method for the combination of parameters considered here. However, if the signal-noise ratio can be improved further through advanced signal processing techniques, a longer runtime, or if the effect of the time dependence is significantly larger than $0.1\%$, this becomes the most sensitive technique to detect low mass relic neutrinos. 

For the Stodolsky effect sensitivity (cyan), we have additionally assumed that the C$\nu$B is composed entirely of left helicity neutrinos\footnote{We remind the reader that if the standard scenario is assumed, either $\widetilde\eta_\nu(\nu_{i,L}^D) = \widetilde\eta_\nu(\bar\nu_{i,R}^D)$ or $\widetilde\eta_\nu(\nu_{i,L}^M) = \widetilde\eta_\nu(\nu_{i,R}^M)$, such that the Stodolsky effect for Majorana neutrinos is expected to vanish identically, whilst the effect for Dirac neutrinos will only contain the helicity asymmetry term.} and used the reference scenario given in~\eqref{eq:stodLimitsD} for a ${^{208}\mathrm{Pb}}$ torsion balance. Under this assumption, the Stodolsky effect is significantly more sensitive to C$\nu$B overdensities than both coherent scattering (pink) and the accelerator proposal (light green) at low neutrino masses, but becomes considerably weaker at large masses when $\beta_{\nu_i} < \beta_\Earth$ and the helicity asymmetry is washed out by the relative motion of the Earth to the C$\nu$B frame. As shown by the dotted blue curve, there is no Stodolsky effect for Majorana neutrinos in this limit.

The sensitivity of coherent scattering, for which we assume the reference scenario in~\eqref{eq:cohLimitsD} and a ${^{208}\mathrm{Pb}}$ torsion balance, is almost uniform at small values of the lightest neutrino mass, where the contribution to the momentum transfer is dominated by the approximately constant mass of the heaviest neutrino state. In the quasi-degenerate mass regime ($m_{\nu_i} \gtrsim 0.1\,\mathrm{meV})$, the sensitivity improves quadratically with the neutrino mass and all three mass eigenstates contribute. Importantly, coherent scattering remains large in the region where the contribution to the Stodolsky effect from a net helicity asymmetry is zero. If the torsion balance proposal with sensitivity $a_0 = 10^{-23}\,\mathrm{cm}\,\mathrm{s^{-2}}$ can be realised~\cite{Hagmann:1999kf}, both the Stodolsky effect and coherent scattering become significantly more sensitive than PTOLEMY in the low mass regime. 

Finally, we show the sensitivity of an accelerator experiment in light green, where we have assumed the reference scenario presented in~\eqref{eq:acceleratorLimits} and the ${^{157}\mathrm{Gd}}$ target given in Table 1 of~\cite{Bauer:2021uyj}, for which $Q = 10.95\,\mathrm{keV}$ and $x \simeq 8.91\times 10^{-8}$ at $t = 1\,\mathrm{y}$. This proceeds via the RB$\beta$ process, and so is only sensitive to the neutrino overdensity. We note, however, that an accelerator experiment utilising a REC process would instead be sensitive to the antineutrino overdensity. For this choice of target, the required beam energy per nucleon is $E/A \simeq 100\,\mathrm{TeV}\,(0.1\,\mathrm{eV}/E_{\nu_i})$, assuming an LHC sized experiment. The accelerator experiment has the worst sensitivity at low masses, where the incredible beam energy requirement significantly reduces the number of ions that can be placed on the beam. At large masses, however, the sensitivity exceeds both coherent scattering and the Stodolsky effect. We also show the constraint that could be placed with a toy `${^{157}\mathrm{Gd}}$-like' target using a light green dot-dashed line, assuming a smaller $Q$-value of $1.095\,\mathrm{keV}$, from which it should be clear that the performance of the accelerator experiment can be enhanced considerably through a better choice of target.

\begin{figure}[tbp]
    \centering
    \includegraphics[width =.831\textwidth]{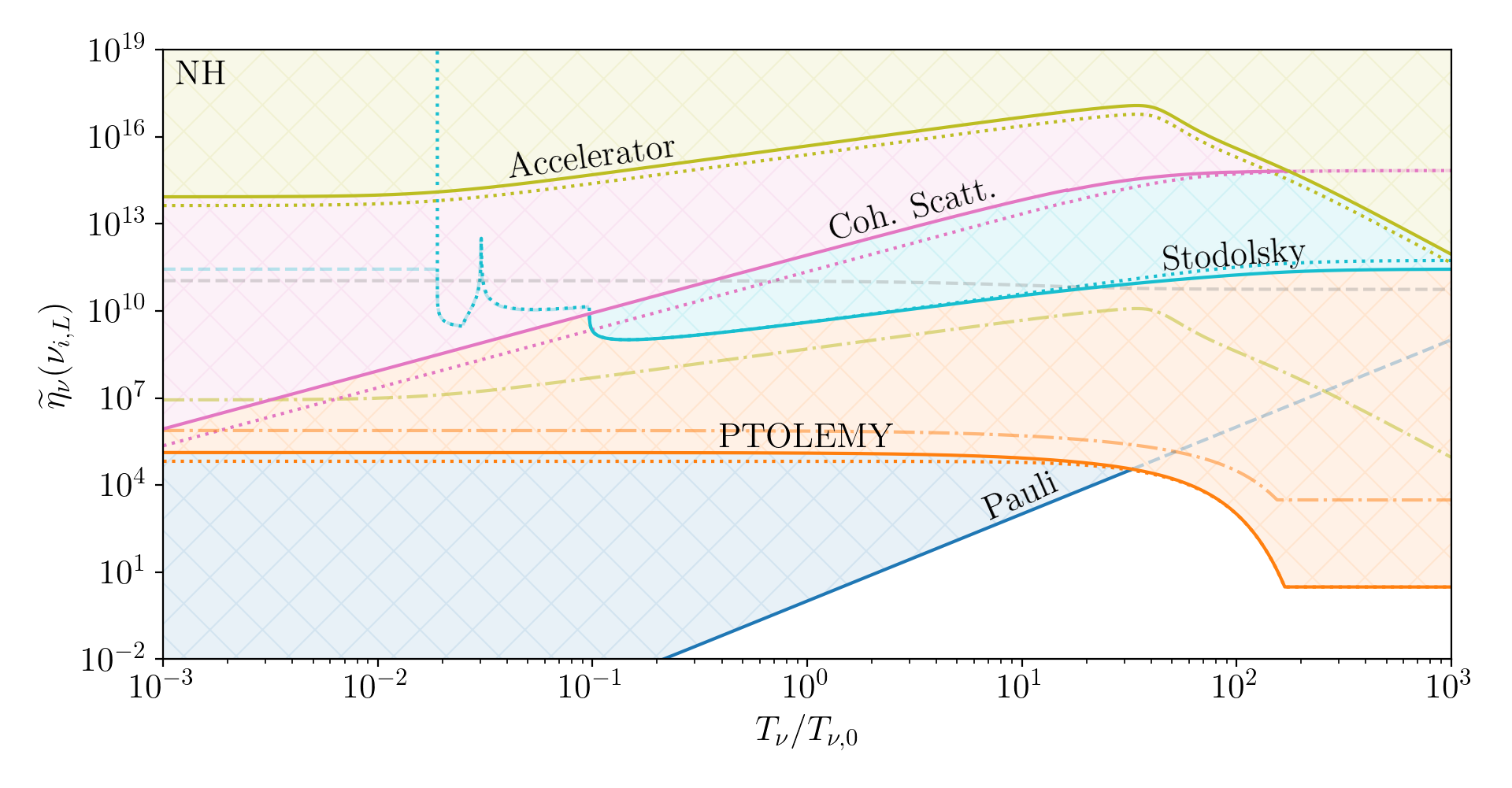}
    \caption{Sensitivity of each direct detection proposal to the C$\nu$B overdensity as a function of the neutrino temperature, assuming $m_{\nu_1} = 10\,\mathrm{meV}$ and only left (right) helicity (anti)neutrinos in the normal mass hierarchy. See the text and Table~\ref{tab:linkTable} for a full description of the figure.} 
    \label{fig:temp_NH}
\end{figure}

\begin{figure}[tbp]
    \centering
    \includegraphics[width=.831\textwidth]{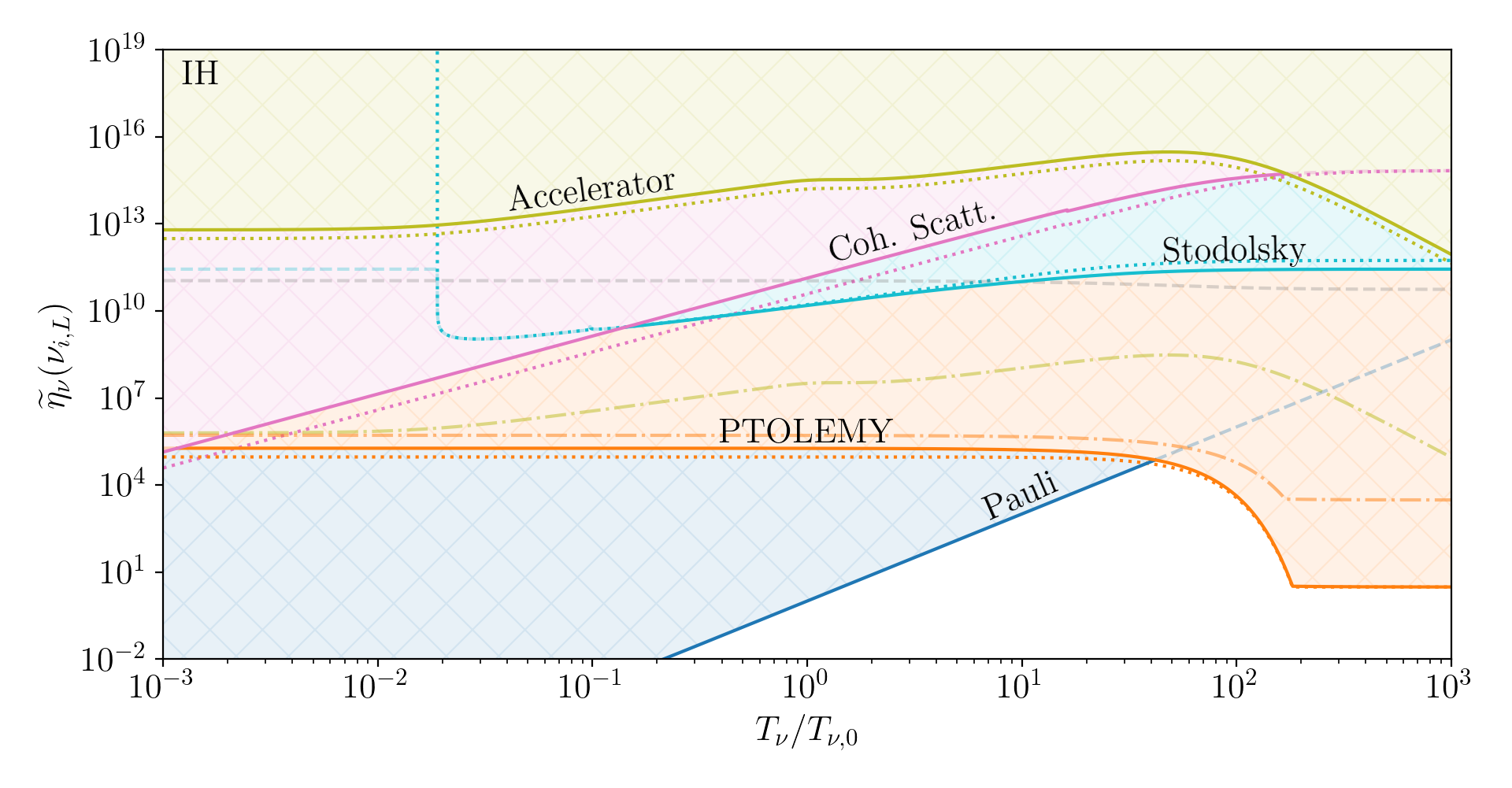}
    \caption{Sensitivity of each direct detection proposal to the C$\nu$B overdensity as a function of the neutrino temperature, assuming $m_{\nu_3} = 10\,\mathrm{meV}$ and only left (right) helicity (anti)neutrinos in the inverted mass hierarchy. See the text and Table~\ref{tab:linkTable} for a full description of the figure.}   
    \label{fig:temp_IH}
\end{figure}

In Figures~\ref{fig:temp_NH} and~\ref{fig:temp_IH} we instead show the constraints and sensitivities as a function of the neutrino temperature, $T_{\nu_i} = T_{\nu}$ for all three mass eigenstates, still with $\vec{p}_{\nu_i} \simeq 3.15\,T_{\nu_i}$, but now assuming a fixed lightest neutrino mass $m_{\nu_l} = 10\,\mathrm{meV}$, which is not presently excluded by any mass constraint. The most striking result from varying the temperature is that the Pauli exclusion principle constraint now allows for significantly larger neutrino overdensities, scaling as $T_{\nu}^3$. For this choice of parameters, the sensitivity of PTOLEMY is also considerably diminished, becoming unable to place strong constraints on the overdensity with either method until $T_{\nu} \gg \Delta = 50\,\mathrm{meV}$. Additionally, at low temperatures, the sensitivity of PTOLEMY to Majorana neutrinos is a factor of two better than the sensitivity to Dirac neutrinos due to additional flux of right helicity neutrinos. However, at high temperatures, the two sensitivities coincide as the right helicity flux becomes non-interacting. 

By plotting as a function of the temperature, we also see several interesting features of the Stodolsky effect. As the temperature decreases, the helicity asymmetry of each of the three mass eigenstates is washed out by the relative motion of the Earth, leading to three `threshold-like' steps in the sensitivity. The sharp peak in the Stodolsky sensitivity coincides with the point at which the acceleration due to the effect vanishes identically. This is due to the signs of the effective axial couplings, $A_{ii}$; as $A_{11} > 0$, whilst $A_{22}, A_{33} < 0$, it is possible for the contribution from $\nu_1$ to cancel exactly with those from $\nu_2$ and $\nu_3$.

As expected, the sensitivity of coherent scattering rapidly becomes stronger as $T_\nu$ decreases. This is a result of the coherent scattering volume increasing as $|\vec{p}_{\nu_i}|^{-3}$, but requires that the target size is increased by the same amount. However, for $T_\nu/T_{\nu,0} = 10^{-3}$, this still only requires a target radius of $R\sim\mathcal{O}(10\,\mathrm{cm})$, which is achievable by a tabletop experiment. At low temperatures, there is a clear distinction between the coherent scattering sensitivity to Dirac and Majorana neutrinos present, with a greater sensitivity to Majorana neutrinos by a factor $\sim 3.8$. This distinction vanishes at high temperatures, where the mass scale becomes irrelevant and both Dirac and Majorana neutrinos can be approximated as Weyl fermions.

Finally, the sensitivity of the accelerator experiment develops a `knee' at $T_{\nu} \simeq m_{\nu_i}$ when plotted in temperature space. Below this knee, the sensitivity becomes weaker with increasing temperature due to the broadening of the C$\nu$B momentum distribution, resulting in fewer relic neutrinos being captured on resonance. Above this temperature, the increased C$\nu$B momentum has a significant contribution to the neutrino energy and the $E_{\nu_i}^{4}$ scaling of the sensitivity dominates. This high neutrino energy scaling is largely due to the decreased beam energy requirements, and subsequently the larger number of ions that can be placed on the beam. For completeness, we also show the overdensity bound from KATRIN on Dirac neutrinos\footnote{We remind the reader that this constraint is stronger by a factor of two for Majorana neutrinos, assuming the standard scenario with $\widetilde{\eta}_{\nu}(\nu_{i,L}^M) = \widetilde{\eta}_{\nu}(\nu_{i,R}^M)$} using a dashed grey line, whilst the dot-dashed light green line corresponds the toy `${^{157}\mathrm{Gd}}$-like' target discussed previously. Additionally, the accelerator is a global factor of two more sensitive to Majorana neutrinos in the standard scenario, for reasons outlined in detail in Section~\ref{sec:accelerator}.
\begin{table}[tbp]
\centering
\begin{tabular}{M{0.1\textwidth}|M{0.65\textwidth}|M{0.15\textwidth}}

Curve & Description & Relevant text and equations \\
\hline\hline
Orange & PTOLEMY sensitivity to Dirac (solid) and Majorana (dotted) neutrinos, standard method. Time dependent method for Dirac neutrinos (dot-dashed). & Section~\ref{sec:ptolemy}, \eqref{eq:ptolemySignificance}, \eqref{eq:ptolemyResolution}, \eqref{eq:ptolemyResolutionTime}, \eqref{eq:ptolemySignificanceTime}.  \\
\hline
Cyan & Stodolsky effect sensitivity to Dirac (solid) and Majorana (dotted) neutrinos. & Section~\ref{sec:stodolsky}, \eqref{eq:stodLimitsD}, \eqref{eq:stodLimitsM}. \\
\hline
Pink & Coherent scattering sensitivity to Dirac (solid) and Majorana (dotted) neutrinos. & Section~\ref{sec:coherent}, \eqref{eq:cohLimitsD}, \eqref{eq:cohLimitsM}. \\
\hline
Light green & Accelerator sensitivity to Dirac (solid) and Majorana (dotted) neutrinos. Using an optimistic setup for Dirac neutrinos (dot-dashed). & Section~\ref{sec:accelerator}, \eqref{eq:acceleratorLimits}. \\
\hline
Grey & Excluded by theory and experiment for $T_{\nu_i} = T_{\nu,0}$ (solid, Figures~\ref{fig:NH_future} and~\ref{fig:IH_future}). Excluded by KATRIN (dashed, Figures~\ref{fig:temp_NH} and~\ref{fig:temp_IH}). & Sections~\ref{sec:pauli} and~\ref{sec:otherExisting}, \eqref{eq:pauliLimit}. \\
\hline
Blue & Excluded by Pauli exclusion principle for $T_{\nu_i} \neq T_{\nu,0}$. & Section~\ref{sec:pauli}, \eqref{eq:pauliLimit}. \\
\hline
Purple & Strongest mass bound on unstable Dirac neutrinos, from cosmology. & Section~\ref{sec:otherExisting}. \\
\hline 
Red & Strongest mass bound on unstable Majorana neutrinos, from KamLAND-Zen. & Section~\ref{sec:otherExisting}. \\
\hline 
Green & Strongest mass bound on stable neutrinos, from cosmology. & Section~\ref{sec:otherExisting}.
\end{tabular}
\caption{Descriptions of the curves in Figures~\ref{fig:NH_future},~\ref{fig:IH_future},~\ref{fig:temp_NH} and~\ref{fig:temp_IH}, along with links to the relevant text and equations.}
\label{tab:linkTable}
\end{table}
\clearpage
\section{Conclusions}\label{sec:conclusions}
Detecting relic neutrinos is an overwhelmingly difficult challenge due to their low energy and weakly interacting nature. A successful detection of the C$\nu$B could provide valuable insight into Big Bang nucleosynthesis and allow us to further improve the accuracy of cosmological models. Whilst many of the as yet unmeasured parameters such as the temperature and number density of the C$\nu$B can be predicted from theory, extended scenarios could result in significantly different values. As the success of many detection proposals depends heavily on these parameters, it is important to constrain the allowed, present day parameter space as much as possible to determine the most effective detection technique.

We have explored the constraints that can be set on the C$\nu$B overdensity from theory, experiment and cosmology, as well as the sensitivity of both direct and indirect relic neutrino detection proposals, for a range of neutrino temperatures and masses. Where they differ, we have calculated the limits for Dirac and Majorana neutrinos, in both the normal and inverted mass hierarchy scenarios. In all cases, we have worked in the mass basis and have allowed for non-degenerate neutrino masses. Additionally, we have accounted for a C$\nu$B reference frame that does not necessarily coincide with that of the Earth and transformed all quantities that depend on the relative frame kinematics appropriately. Finally, as the C$\nu$B is expected to consist entirely of left (right) helicity (anti)neutrinos, we have used polarised cross sections throughout.

The current experimental constraints on the relic neutrino overdensity are very weak, with the strongest constraint currently set by KATRIN at $\widetilde\eta_{\nu}(\nu_{i,L}) \lesssim 1.3\times 10^{11}$. By attributing deviations in the measured solar neutrino spectrum from theoretical predictions to relic neutrinos, we have demonstrated that Borexino strongly favours relic neutrinos with temperature $T_{\nu_i} \lesssim 5\,\mathrm{keV}$. 

Theory places much stronger constraints on relic neutrinos. At the neutrino temperature predicted by standard cosmology, $T_{\nu_i} = T_{\nu,0}$, we have shown that overdensities $\widetilde\eta_{\nu}(\nu_{i,s}) \gg 1$ are forbidden by the Pauli exclusion principle for neutrino masses $m_{\nu_i} \lesssim 0.15\,\mathrm{eV}$. At the upper mass bound set by KATRIN, $m_{\nu_i} \simeq 0.8\,\mathrm{eV}$, the exclusion principle bound becomes weaker, allowing for overdensities $\widetilde\eta_{\nu}(\nu_{i,s}) \lesssim 125$. However, if the neutrino temperature differs from the predicted value, the Pauli bound is modified by a factor $\sim (T_{\nu_i}/T_{\nu,0})^3$, allowing for significantly larger overdensities.

Cosmology also heavily restricts the allowed C$\nu$B parameter space. For a neutrino overdensity generated by the introduction of a chemical potential, we have demonstrated that the combination of constraints from BBN and $\Delta N_\mathrm{eff}$ limit the Dirac neutrino overdensities to $\sum_s \widetilde\eta_{\nu}(\nu_{1,s}^D) \lesssim 1.5$, $\sum_s \widetilde\eta_{\nu}(\nu_{2,s}^D) \lesssim 3.5$ and $\sum_s \widetilde\eta_{\nu}(\nu_{3,s}^D) \lesssim 3.5$, all assuming $T_{\nu_i} = T_{\nu,0}$ and $g_{\nu_i} = 1$. If we instead attribute the contribution to $\Delta N_{\mathrm{eff}}$ entirely to a modified neutrino temperature, the cosmological constraints become $T_{\nu_i} \leq 1.024\, T_{\nu,0}$ and $\sum_s \widetilde\eta_\nu(\nu_{i,s}) \leq 1.073$ for a mass eigenstate-independent temperature $T_{\nu_i} = T_{\nu}$, or $T_{\nu_i} \leq 1.351 \,T_{\nu,0}$ and $\sum_s \widetilde\eta_\nu(\nu_{i,s}) \leq 2.47$ in the most extreme scenario with two neutrinos at $T_{\nu,i} = 0$ and a third, hot neutrino state. Measured phase shifts in the baryon acoustic oscillation spectrum instead constrain the relic neutrino from below, permitting $\widetilde\eta_\nu(\nu_{i,s}) \geq 0.19$ at just below $\sim 3\sigma$.  However, these bounds do not apply today if the C$\nu$B has undergone significant changes since the radiation-dominated era, \textit{e.g.} as a result of neutrino decay or strong neutrino interactions with dark matter.

We have shown that the regions of overdensity space that can be probed by each direct detection proposal are strongly dependent on the C$\nu$B mass and temperature. For large relic neutrino masses or temperatures, $m_{\nu_i} \gtrsim 50\,\mathrm{meV}$ or $T_{\nu_i} \gtrsim 30\,\mathrm{meV}$, PTOLEMY could probe overdensities as small as $\widetilde{\eta}_\nu(\nu_{i,L}) \gtrsim 6$ with $5\sigma$ significance after one year, assuming only left helicity neutrinos. For very low energy neutrinos, however, this rapidly diminishes to $\widetilde{\eta}_\nu(\nu_{i,L}) \lesssim 3\times 10^5$, but may be improved by measuring the time dependence of the signal. PTOLEMY is also completely unable to observe relic antineutrinos, but may be complimented by an experiment using a $\beta^+$-decaying target~\cite{Cocco:2007za, Cocco:2009rh}. For a conservative setup, torsion balance experiments utilising the Stodolsky effect or coherent scattering could observe overdensities of $\widetilde\eta_{\nu}(\nu_{i,L}) \gtrsim 10^{10}$ and $\widetilde\eta_{\nu}(\nu_{i,L}) \gtrsim 10^{12}$ in the small neutrino mass regime, respectively. The sensitivities of these torsion balance experiments could be up to eight orders of magnitude stronger with an optimistic experimental setup, surpassing the potential of PTOLEMY. However, the Stodolsky effect relies on either a net neutrino-antineutrino or helicity asymmetry in the C$\nu$B, and as such is expected to vanish identically for Majorana neutrinos in the standard scenario, whilst only containing the helicity asymmetry contribution for Dirac neutrinos. Both torsion balance experiments are also capable of distinguishing between Dirac and Majorana neutrinos, with the Stodolsky effect for Majorana neutrinos expected to vanish completely at large neutrino masses or low neutrino temperatures. On the other hand, the sensitivity of a coherent scattering experiment improves at a rate $T_{\nu_i}^{-2}$ with decreasing neutrino temperature. We have also shown that an accelerator experiment using a conservative setup is sensitive to relic neutrino overdensities $\widetilde{\eta}_{\nu}(\nu_{i,L}) \gtrsim 10^9$ at the KATRIN mass bound, $m_{\nu_i} = 0.8\,\mathrm{eV}$, rapidly becoming less sensitive at lower masses. For $T_{\nu_i} > m_{\nu_i}$, however, the sensitivity improves as $T_{\nu_i}^4$, whilst for a more optimistic setup the sensitivity could be up to seven orders of magnitude stronger. Finally, we showed in Section~\ref{sec:lim} that line intensity mapping experiments searching for radiative C$\nu$B neutrino decay could be sensitive to overdensities $\eta_\nu(\nu_{i,s}) \gtrsim 10^{23}$ assuming only Standard Model processes, but would be many orders of magnitude more sensitive for a neutrino transition electromagnetic moment approaching the experimental bound. 

Indirect searches could also yield interesting constraints; we have demonstrated that large overdensities could lead to the severe attenuation of extragalactic neutrino fluxes at high energies. By searching for this attenuation, IceCube-Gen2 could be sensitive to overdensities $\widetilde\eta_{\nu}(\nu_{i,s})\geq 10^{10}$. Additionally, the exclusion principle could heavily suppress processes emitting exclusively low energy neutrinos. Similar to the direct detection proposals, the extent of these effects depends heavily on the properties of the C$\nu$B.

In summary, the magnitude of any relic neutrino overdensity is heavily constrained by the Pauli exclusion principle and cosmology, whilst PTOLEMY is expected to have the best sensitivity of any C$\nu$B detection proposal in most scenarios. However, we have demonstrated that the other direct detection proposals could provide insight into the Dirac or Majorana nature of neutrinos, the helicity profile C$\nu$B and the mass hierarchy, whilst scenarios differing significantly from the standard cosmological history could see them outperform PTOLEMY.

\acknowledgments

We thank Martin Spinrath for bringing the bound from baryon acoustic oscillations to our attention. We would also like to thank Peter Denton for highlighting a minor mistake in a previous draft of this work. Jack D. Shergold is supported by an STFC studentship under the STFC training grant ST/T506047/1. Martin Bauer acknowledges support by the Future Leader Fellowship `DARKMAP'.

\appendix
\section{Electron energy shifts}\label{sec:energyShift}
Here we derive the energy shift of each electron spin state due to the Stodolsky effect. We begin by calculating the expectation values~\eqref{eq:nuME},~\eqref{eq:antinuME} and~\eqref{eq:majnuME}. The electron energy shifts due to Dirac neutrinos can be rewritten in terms of traces as
\begin{equation}
    \begin{split}
    \langle e_{s_e},\nu_{i,s}| \mathcal{H}^D|e_{s_e},\nu_{i,s}\rangle = \frac{G_F}{\sqrt{2}}&\mathrm{Tr}\left[u(p_{\nu_i},s)\bar u(p_{\nu_i},s)\gamma_\mu (1-\gamma^5)\right]\\
    \times &\mathrm{Tr}\left[u(p_{e},s_{e})\bar u(p_{e},s_{e})\gamma^{\mu}(V_{ii}-A_{ii}\gamma^5)\right],
    \end{split}
\end{equation}
\begin{equation}
    \begin{split}
    \langle e_{s_e},\bar\nu_{i,s}| \mathcal{H}^D|e_{s_e},\bar\nu_{i,s}\rangle = -\frac{G_F}{\sqrt{2}}&\mathrm{Tr}\left[v(p_{\nu_i},s)\bar v(p_{\nu_i},s)\gamma_\mu (1-\gamma^5)\right]\\
    \times &\mathrm{Tr}\left[u(p_{e},s_{e})\bar u(p_{e},s_{e})\gamma^{\mu}(V_{ii}-A_{ii}\gamma^5)\right],
    \end{split}
\end{equation}
whilst for Majorana neutrinos we instead have
\begin{equation}
    \begin{split}
    \langle e_{s_e},\nu_{i,s}| \mathcal{H}^M|e_{s_e},\nu_{i,s}\rangle = -\sqrt{2}G_F&\mathrm{Tr}\left[u(p_{\nu_i},s)\bar u(p_{\nu_i},s)\gamma_\mu \gamma^5\right]\\
    \times &\mathrm{Tr}\left[u(p_{e},s_{e})\bar u(p_{e},s_{e})\gamma^{\mu}(V_{ii}-A_{ii}\gamma^5)\right].
    \end{split}
\end{equation}
To evaluate the outer products of spinors in a basis independent way, we use the identities
\begin{align}
    u(p,s)\bar{u}(p,s) &= \frac{1}{2}(\slashed{p}+m)(1+\gamma^5 \slashed{S}),\label{eq:ubu}\\
    v(p,s)\bar{v}(p,s) &= \frac{1}{2}(\slashed{p}-m)(1+\gamma^5 \slashed{S}),\label{eq:vbv}
\end{align}
where $\slashed{A} \equiv \gamma_{\mu}A^\mu$  and the spin vector for massive and massless particles are defined, respectively, by
\begin{equation}\label{eq:spinVec}
    S^\mu = s\left(\frac{\vec{p}\cdot \vec{n}}{m},\vec{n}+\frac{(\vec{p}\cdot \vec{n})\vec{p}}{m(E+m)}\right), \quad S^\mu = s\left(1,\frac{\vec{p}}{|\vec{p}|}\right),
\end{equation}
for a particle with spin $s$, where the unit vector $\vec{n}$ denotes the spin orientation of the particle in its rest frame. The choice $\vec{n} = \vec{p}/|\vec{p}|$ picks out the component of $\vec{S}$ along the momentum direction, allowing us to instead identify $s$ with the particle helicity. This also considerably simplifies $S^\mu$ to
\begin{equation}\label{eq:spinVecHelicity}
    S^\mu = s\left(\frac{|\vec{p}|}{m},\frac{E}{m}\frac{\vec{p}}{|\vec{p}|}\right),
\end{equation}
where $s = \pm 1$ is now the particle helicity, with $+1$ corresponding to right helicity and $-1$ corresponding to left helicity. We note that~\eqref{eq:spinVecHelicity} is not valid for particles at rest, and instead~\eqref{eq:spinVec} should be used. With these definitions, we can perform the traces to find the energy shift
\begin{equation}
    \begin{split}\label{eq:diracShift}
    \Delta E_e^D(\vec{p}_e,s_e) = \frac{G_F}{\sqrt{2}} \frac{m_e}{E_e}\sum_{i,s}\frac{A_{ii}}{E_{\nu_i}}\Big[ m_{\nu_i}&(S_e\cdot S_{\nu_i})(n_{\nu}(\nu_{i,s}^D) + n_{\nu}(\bar\nu_{i,s}^D))  \\
    - &(S_e\cdot p_{\nu_i})(n_{\nu}(\nu_{i,s}^D) - n_{\nu}(\bar\nu_{i,s}^D))\Big] + f(V_{kk}),
   %\langle e_{s_e},\nu_{i,s}| \mathcal{H}^D|e_{s_e},\nu_{i,s}\rangle &= 2\sqrt{2}G_F m_e\left[ m_{\nu_i}(S_e\cdot S_{\nu_i})-(S_e\cdot p_{\nu_i})\right] + f(V_{kk}),\\
    %\langle e_{s_e},\bar\nu_{i,s}| \mathcal{H}^D|e_{s_e},\bar\nu_{i,s}\rangle &= 2\sqrt{2}G_F m_e\left[ m_{\nu_i}(S_e\cdot S_{\nu_i})+(S_e\cdot p_{\nu_i})\right] + f(V_{kk}),\\
    \end{split}
\end{equation}
for Dirac neutrinos, whilst for Majorana neutrinos the shift is given by
\begin{equation}\label{eq:majoranaShift}
    \Delta E_e^M(\vec{p}_e,s_e)=\sqrt{2}G_F \frac{m_e}{E_e}\sum_{i,s}A_{ii} \frac{m_{\nu_i}}{E_{\nu_i}} (S_e\cdot S_{\nu_i})n_{\nu}(\nu_{i,s}) + f(V_{kk}),
    %\langle e_{s_e},\nu_{i,s}| \mathcal{H}^M|e_{s_e},\nu_{i,s}\rangle &= 4\sqrt{2}G_F m_e m_{\nu_i}(S_e\cdot S_{\nu_i}) + f(V_{kk}).
\end{equation}
where we have made the replacement $\int d^3x \to V$ as no terms depend on position, and $f(V_{kk})$ contains terms that do not depend on electron spin that will cancel when we take the difference between the energy of the two spin states. We also see that in all cases the term containing $(S_e\cdot S_{\nu_i})$ is proportional to $m_{\nu_i}$, such that we only need to evaluate it using~\eqref{eq:spinVec} for massive neutrinos. 

As the Stodolsky effect depends on all neutrinos in the background, we should work in terms of $\vec{p}_{\nu_i,\mathrm{true}}$ and perform the flux-weighted averaging outlined in Section~\ref{sec:kinematics} at the end. However, before performing the lab frame calculation where the averaging is more complicated, we compute the averages of $(S_e\cdot S_{\nu_i})$ and $(S_e\cdot p_{\nu_i})$ in the C$\nu$B frame as a cross check. In the C$\nu$B frame the neutrino momentum is given by~\eqref{eq:pCnuB}, whilst the electron momentum is generated due to the relative motion of the Earth and is given by
\begin{equation}
    \widetilde{p}_e = \frac{m_e}{\sqrt{1-\beta_\Earth^2}}\left(\begin{array}{c}
        1   \\
        0   \\
        0   \\
        -\beta_\Earth
    \end{array}\right).
\end{equation}
Using this, we find the C$\nu$B frame averages
\begin{align}
    \frac{1}{4\pi}\int (\widetilde S_e\cdot \widetilde S_{\nu_i})\,d\widetilde\Omega &= \frac{s_e s\beta_\Earth}{\sqrt{1-\beta_\Earth^2}}\frac{|\vec{p}_{\nu_i}|}{m_{\nu_i}}~\label{eq:spinspinCNB},\\
    \frac{1}{4\pi}\int (\widetilde S_e\cdot \widetilde p_{\nu_i})\,d\widetilde\Omega &= \frac{s_e \beta_\Earth}{\sqrt{1-\beta_\Earth^2}} E_{\nu_i},
\end{align}
where $s$ is the neutrino helicity. Importantly, these are all proportional to the asymmetry parameter, $\beta_\Earth$. As such, we expect that the Stodolsky effect will also be proportional to $\beta_\Earth$ in the lab frame, where the electrons are at rest and the asymmetry is instead generated by the neutrino wind. If we did not make the assumption of isotropy in the C$\nu$B frame then Stodolsky effect would not necessarily be proportional to $\beta_\Earth$, but would still require some non-zero asymmetry parameter.

For electrons polarised along $z$ in the lab frame, $S_e^\mu = s_e(0,0,0,1)$, we find the flux-weighted averages
\begin{align}
    \left\langle\frac{(S_e\cdot S_{\nu_i,\mathrm{true}})}{E_{\nu_i,\mathrm{true}}}\right\rangle &= -\frac{\beta_\Earth s_e s}{3m_{\nu_i}}\frac{1}{\beta_{\nu_i}}\left(3 -\beta_{\nu_i}^2\right) + \mathcal{O}(\beta_\Earth^2),\label{eq:spinspinav}\\
    \left\langle\frac{(S_e\cdot p_{\nu_i,\mathrm{true}})}{E_{\nu_i,\mathrm{true}}}\right\rangle &= -\frac{2\beta_\Earth s_e}{3}(2-\beta_{\nu_i}^2)+ \mathcal{O}(\beta_\Earth^2),
\end{align}
which are all proportional to $\beta_\Earth$ as expected. We note that~\eqref{eq:spinspinav} diverges in the limit $\beta_{\nu_i}\to 0$. However, as~\eqref{eq:spinspinCNB} is finite, this divergence must be a consequence of the frame transformation and should cancel elsewhere to leave a finite result. As a consequence of~\eqref{eq:helicityAsymmetry}, this is indeed the case. 

After substituting the averages into the energy shifts~\eqref{eq:diracShift} and~\eqref{eq:majoranaShift}, we find an energy splitting between the two electron spin states
\begin{equation}
    \begin{split}
    \Delta E_e^D &= \frac{\sqrt{2}G_F}{3}\beta_\Earth\sum_i A_{ii}\Big[2\sum_{s}(2-\beta_{\nu_i}^2)(n_\nu(\nu_{i,s}^D)-n_\nu(\bar\nu_{i,s}^D)) \\
    &+ \frac{1}{\beta_{\nu_i}}\left(3 -\beta_{\nu_i}^2\right)(n_\nu(\nu_{i,L}^D)-n_\nu(\nu_{i,R}^D) + n_\nu(\bar\nu_{i,R}^D)-n_\nu(\bar\nu_{i,L}^D))\Big],
    \end{split}
\end{equation}
for Dirac neutrinos. Importantly, the contributions from right helicity neutrinos and left helicity antineutrinos vanish in the limit $\beta_{\nu_i}\to 1$. For Majorana neutrinos, where only the spin-spin term contributes, we instead find a splitting
\begin{equation}
    \Delta E_e^M = \frac{2\sqrt{2}G_F}{3}\beta_\Earth \sum_{i}\frac{A_{ii}}{\beta_{\nu_i}}\left(3 -\beta_{\nu_i}^2\right)(n_\nu(\nu_{i,L}^M)-n_\nu(\nu_{i,R}^M)),
\end{equation}
as given in the main text. 
\section{Helicity flipping probability}\label{sec:helicity}
In this appendix, we derive the helicity flipping probability, $P_F(\beta_\Earth)$, when transforming from the C$\nu$B to laboratory frame, assuming that relic neutrinos are isotropic in their own reference frame.

We begin by noting that the helicity of a relic neutrino will flip if its velocity changes sign when going between frames. In order for this to happen, two conditions must be met. First, the velocity of the relic neutrino must contain a component that is initially antiparallel to the velocity of the Earth. If the neutrino velocity only contains a parallel component, then the relative motion of the two frames will increase the magnitude of the relic neutrino velocity but can never change its sign. We therefore require
\begin{equation}
    P_F(\beta_\Earth) + P_{DNF}(\beta_\Earth) + P_{CNF} = 1,
\end{equation}
where $P_{DNF}(\beta_\Earth)$ is the probability that a neutrino can flip helicity, but does not, whilst $P_{CNF}$ accounts for the neutrinos that cannot flip helicity as their velocity is initially antiparallel to that of the Earth. For an isotropic background, $P_{CNF} = 1/2$. Second, the Earth must be moving faster than the relic neutrino along its direction of travel, 
\begin{equation}
    \beta_\Earth \cos\theta_h \geq \widetilde\beta_{\nu_i},
\end{equation}
where $\theta_h \in [0,\pi/2]$ is the angle between the velocity of the relic neutrino and the Earth. We restrict $\theta_h$ to this domain, as $\theta_h >\pi/2$ corresponds to neutrinos with only parallel velocity components, which are already accounted for in $P_{CNF}$.

Given this, we define the function
\begin{equation}
    S(\theta_h) = \mathrm{sgn}\left(\beta_\Earth \cos\theta_h - \widetilde\beta_{\nu_i}\right),
\end{equation}
which outputs $+1$ if neutrinos flip helicity during the frame transformation, or $-1$ if the helicity retains its sign. Its integral
\begin{equation}
    \int\displaylimits_{0}^{\pi/2}S(\theta_h)\, d\theta_h = A_{DNF}(\beta_\Earth) - A_{F}(\beta_\Earth),
\end{equation}
outputs the difference between the areas of two unit height rectangles, $A_F(\beta_\Earth)$ and $A_{DNF}(\beta_\Earth)$, with lengths proportional to $P_F(\beta_\Earth)$ and $P_{DNF}(\beta_\Earth)$ respectively. Their sum must be equal to the length of the domain, $A_F(\beta_\Earth) + A_{DNF}(\beta_\Earth) = \pi/2$, such that given $P_F(\beta_\Earth) + P_{DNF}(\beta_\Earth) = 1/2$, we must necessarily have $A_{F}(\beta_\Earth) = \pi P_F(\beta_\Earth)$ and $A_{DNF}(\beta_\Earth) = \pi P_{DNF}(\beta_\Earth)$. Finally, rewriting $P_{DNF}(\beta_\Earth) = 1/2 - P_F(\beta_\Earth)$, we find
\begin{equation}
    P_F(\beta_\Earth) = \frac{1}{4} - \frac{1}{2\pi}\int\displaylimits_{0}^{\pi/2}S(\theta_h) \,d\theta_h
    = \begin{cases}
       \frac{1}{\pi}\arcsin\left(\frac{\beta_\Earth}{\widetilde\beta_{\nu_i}}\right), \quad & \beta_\Earth < \widetilde\beta_{\nu_i},\\
        \frac{1}{2},\quad & \beta_\Earth \geq \widetilde\beta_{\nu_i},
    \end{cases}
\end{equation}
as given in~\eqref{eq:helicityFlip}.
\section{Polarised neutrino scattering cross sections}\label{sec:polarisedXsecs}
\begin{figure}[tbp]
\centering 
\includegraphics[width=.25\textwidth]{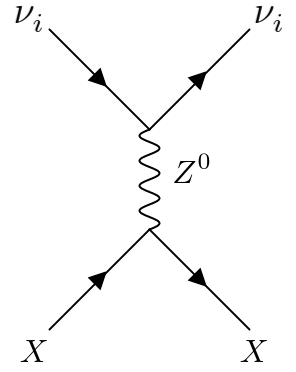}
\hspace{0.25\textwidth}
\includegraphics[width=.25\textwidth]{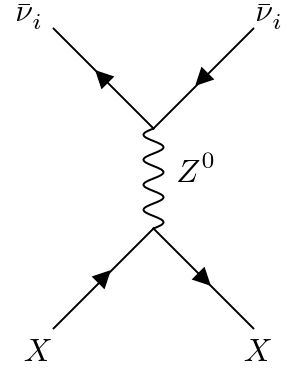}
\caption{\label{fig:ncNuclFeynman}Tree level diagrams contributing to the scattering of (anti)neutrino mass eigenstate $i$ on nucleus $X$. The Majorana neutrino amplitude is found by taking the difference between these two diagrams.}
\end{figure}

Here we calculate the polarised neutrino scattering cross sections on nuclei and electrons, working in the mass basis, for both Dirac and Majorana neutrinos. We begin with the neutrino-nucleus cross section, assuming full coherence over nuclear distances such that any details of substructure can be ignored. From left to right, the Feynman diagrams given in Figure~\ref{fig:ncNuclFeynman} correspond to the amplitudes
\begin{align}
    i\mathcal{M}_{N}(\nu_{i,s}^D) &= \frac{iG_F}{\sqrt{2}}\bar{u}(q_{\nu_i},s')\gamma_\mu (1-\gamma^5)u(p_{\nu_i},s)j^\mu_X,\label{eq:nuNuclAmpl}\\
    i\mathcal{M}_{N}(\bar\nu_{i,s}^D) &= \frac{iG_F}{\sqrt{2}}\bar{v}(p_{\nu_i},s')\gamma_\mu (1-\gamma^5)v(q_{\nu_i},s)j^\mu_X\label{eq:anuNuclAmpl},
\end{align}
for incoming neutrinos and antineutrinos with helicity $s$, where $p_{\nu_i}$ and $q_{\nu_i}$ are the four-momenta of the incoming and outgoing neutrino, respectively. The nuclear current is written in terms of the vector and axial charges, $Q_V = A-2Z(1-2\sin^2\theta_W)$ and $Q_A = A-2Z$, as
\begin{equation}
    j^{\mu}_X = -\frac{1}{2} \bar{u}(q_X,r')\gamma^\mu(Q_V - Q_A \gamma^5)u(p_X,r),
\end{equation}
for a nucleus with incoming and outgoing four-momenta $p_X$ and $q_X$. For Dirac neutrinos and antineutrinos, the scattering amplitudes on nuclei are given by~\eqref{eq:nuNuclAmpl} and~\eqref{eq:anuNuclAmpl}, respectively. Conversely, the Majorana scattering amplitude is found by taking the difference
\begin{equation}
    \begin{split}
     i\mathcal{M}_{N}(\nu_{i,s}^M) &= i\mathcal{M}_{N}(\nu_{i,s}^D) - i\mathcal{M}_{N}(\bar\nu_{i,s}^D)\\
     &= -i\sqrt{2}G_F\bar{u}(q_{\nu_i},s')\gamma_\mu \gamma^5 u(p_{\nu_i},s)j^\mu_X,
     \end{split}
\end{equation}
where in going from the first to second line we have applied the Majorana condition $v(p,s) = \mathcal{C}\bar{u}(p,s)^T$, with $\mathcal{C}$ the charge conjugation matrix.  Squaring the amplitude, averaging over target nucleus spins, and summing over the final state helicities yields
\begin{align}
    \left\langle\left|\mathcal{M}_{N}(\nu_{i,s}^D)\right|^2\right\rangle &= \frac{1}{2}\sum_{s',r,r'}\left|\mathcal{M}_{N}(\nu_{i,s}^D)\right|^2 = \frac{G_F^2}{16} \mathcal{T}_{\alpha\beta}\mathcal{T}^{\alpha\beta}_X,\\
    \left\langle\left|\mathcal{M}_{N}(\bar\nu_{i,s}^D)\right|^2\right\rangle &= \frac{G_F^2}{16} \widebar{\mathcal{T}}_{\alpha\beta}\mathcal{T}^{\alpha\beta}_X,\\
    \left\langle\left|\mathcal{M}_{N}(\nu_{i,s}^M)\right|^2\right\rangle &= \frac{G_F^2}{4} \mathcal{U}_{\alpha\beta}\mathcal{T}^{\alpha\beta}_X,
\end{align}
where the traces are defined by
\begin{align}
    \mathcal{T}^{\alpha\beta} &= \mathrm{Tr}\left[\gamma^\alpha(\slashed{q}_{\nu_{i}}+m_{\nu_i})\gamma^\beta(1-\gamma^5)u(p_{\nu_i},s)\bar u(p_{\nu_i},s)(1+\gamma^5)\right],\label{eq:trNuNC}\\
    \widebar{\mathcal{T}}^{\alpha\beta} &= \mathrm{Tr}\left[\gamma^{\alpha}v(p_{\nu_i},s)\bar v(p_{\nu_i},s)\gamma^\beta(1-\gamma^5)(\slashed{q}_{\nu_i}-m_{\nu_i})(1+\gamma^5)\right],\\
    \mathcal{U}^{\alpha\beta} &= \mathrm{Tr}\left[\gamma^\alpha \gamma^5(\slashed{q}_{\nu_{i}}+m_{\nu_i}) \gamma^\beta\gamma^5 u(p_{\nu_i},s)\bar u(p_{\nu_i},s)\right],\\
    \mathcal{T}^{\alpha\beta}_X &= \mathrm{Tr}\left[\gamma^\alpha(\slashed{q}_X + m_X)\gamma^\beta(Q_V-Q_A\gamma^5)(\slashed{p}_X + m_X)(Q_V+Q_A\gamma^5)\right],
\end{align}
whilst $m_X$ is the mass of the target nucleus. 
To evaluate the spinor product appearing in~\eqref{eq:trNuNC} we use the identities~\eqref{eq:ubu},~\eqref{eq:vbv} and~\eqref{eq:spinVec}, where we note that we once again only require the massive version of~\eqref{eq:spinVec} and $S^{\mu}$ is always multiplied by $m_{\nu_i}$. With these definitions, the traces are readily evaluated using a computer algebra package such as FeynCalc~\cite{Mertig:1990an,Shtabovenko:2016sxi,Shtabovenko:2020gxv}. 

As all momenta in the problem are small relative to the target mass, the lab frame approximately coincides with the CoM frame. This considerably simplifies the phase space integration, yielding a cross section
\begin{equation}\label{eq:nuclearPhaseSpace}
        \sigma^D_{N} \simeq \frac{1}{32\pi \bar{E}^2}\int\displaylimits_{-1}^1 d \cos\theta \left\langle\left|\mathcal{M}_{N}\right|^2\right\rangle,\\
\end{equation}
where $\bar E^2 = m_{\nu_i}^2 + m_X^2 + 2m_X E_{\nu_i}$ is the CoM energy and $\theta$ is the angle between the incoming and outgoing neutrino. Evaluating the traces, and performing the phase space integral~\eqref{eq:nuclearPhaseSpace}, we find the polarised cross sections
\begin{align}
    \sigma_N(\nu_{i,s}^D) &= \frac{G_F^2}{8\pi}(Q_V^2 + 3Q_A^2)(1-s\beta_{\nu_i})E_{\nu_i}^2,\\
    \sigma_N(\bar\nu_{i,s}^D) &= \frac{G_F^2}{8\pi}(Q_V^2 + 3Q_A^2)(1+s\beta_{\nu_i})E_{\nu_i}^2,\\
    \sigma_N(\nu_{i,s}^M) &= \frac{G_F^2}{4\pi}(\beta_{\nu_i}^2 Q_V^2 + 3(2-\beta_{\nu_i}^2)Q_A^2)E_{\nu_i}^2.\label{eq:majCohNM}
\end{align}
Naturally, the cross sections for right helicity neutrinos and left helicity antineutrinos vanish as  $\beta_{\nu_i}\to 1$.

\begin{figure}[tbp]
\centering 
\includegraphics[width=.22\textwidth]{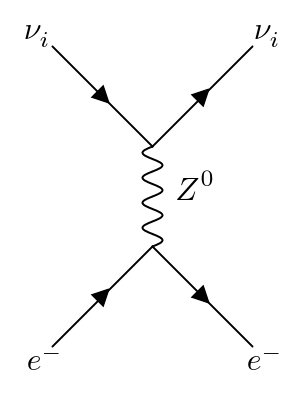}
\includegraphics[width=.22\textwidth]{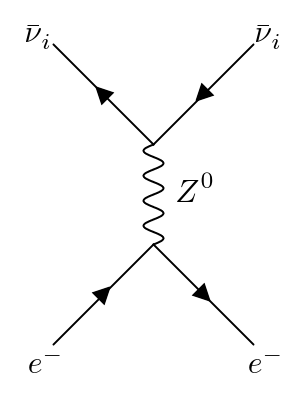}
\raisebox{-0.02\height}{\includegraphics[width=.23\textwidth]{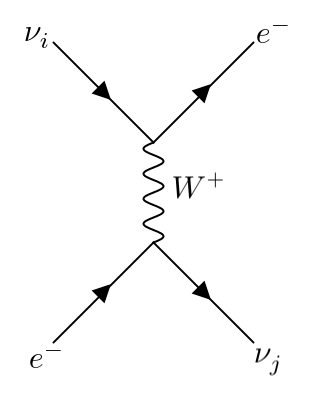}}
\raisebox{0.15\height}{\includegraphics[width=.3\textwidth]{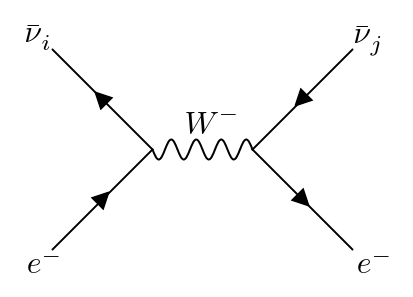}}
\caption{\label{fig:eFeynman}Tree level diagrams contributing to the scattering of neutrino mass eigenstate $i$ on electrons.}
\end{figure}

We now repeat the process for scattering on electrons, noting that both mass diagonal ($\nu_i \to \nu_i$) and mass changing ($\nu_i \to \nu_j$) processes can contribute to the total scattering rate through the diagrams given in Figure~\ref{fig:eFeynman}. Once again from left to right, the Feynman diagrams in Figure~\ref{fig:eFeynman} correspond to the amplitudes
\begin{align}
    i\mathcal{M}_e^{\mathrm{NC}} &= \frac{iG_F}{\sqrt{2}}\bar{u}(q_{\nu_i},s')\gamma_\mu(1-\gamma^5)u(p_{\nu_i},s)j_e^\mu,\\
    i\widebar{\mathcal{M}}_e^{\mathrm{NC}} &= \frac{iG_F}{\sqrt{2}}\bar{v}(p_{\nu_i},s')\gamma_\mu (1-\gamma^5)v(q_{\nu_i},s)j^\mu_e,\\
    i\mathcal{M}_{e,ij}^{\mathrm{CC}} &= \frac{iG_F}{\sqrt{2}}U_{ei} U_{ej}^* \bar{u}(q_e,r')\gamma_\mu(1-\gamma^5)u(p_{\nu_i},s)\bar{u}(q_{\nu_j},s')\gamma^\mu(1-\gamma^5)u(p_e,r),\\
    i\widebar{\mathcal{M}}_{e,ij}^{\mathrm{CC}} &= \frac{iG_F}{\sqrt{2}}U_{ei} U_{ej}^* \bar{v}(p_{\nu_i},s)\gamma_\mu(1-\gamma^5)u(p_e,r)\bar{u}(q_e,r')\gamma^\mu(1-\gamma^5)v(q_{\nu_j},s'),
\end{align}
where the electron current is given in terms of its vector and axial-vector couplings to the $Z$-boson, $g_V^e = -1/2 + 2\sin^2\theta_W$ and $g_A^e = -1/2$, by
\begin{equation}
    j^\mu_e = \bar{u}(q_e,r')\gamma^\mu(g_V^e-g_A^e\gamma^5)u(p_e,r).
\end{equation}
The neutral current amplitudes can only to contribute to the mass diagonal scattering rate, whilst the mass changing scattering rate receives a contribution from both neutral and charged current processes. The amplitudes for Dirac neutrinos to scatter on electrons are therefore given by
\begin{align}
    i\mathcal{M}_e(\nu_{i,s}^D\to \nu_j^D) &= i\mathcal{M}_e^{\mathrm{NC}}\delta_{ij} - i\mathcal{M}_{e,ij}^{\mathrm{CC}},\label{eq:eScatt1}\\
    i\mathcal{M}_e(\bar\nu_{i,s}^D\to \bar\nu_j^D) &= i\widebar{\mathcal{M}}_e^{\mathrm{NC}}\delta_{ij} - i\widebar{\mathcal{M}}_{e,ij}^{\mathrm{CC}},\label{eq:eqScatt2}
\end{align}
where $\delta_{ij}$ is the Kronecker delta. The relative sign between the neutral and charged current amplitudes in~\eqref{eq:eScatt1} and~\eqref{eq:eqScatt2} arises when permuting the external fermion fields. This can be calculated algorithmically using the permutation rules given in~\cite{Gluza:1991wj}. For Majorana neutrinos, all four diagrams in Figure~\ref{fig:eFeynman} contribute to the total scattering amplitude, which following the permutation rules to find the relative sign between diagrams is given by
\begin{equation}
    i\mathcal{M}_e(\nu_{i,s}^M\to \nu_j^M) = (i\mathcal{M}_e^{\mathrm{NC}} - i\widebar{\mathcal{M}}_e^{\mathrm{NC}})\delta_{ij} + i\widebar{\mathcal{M}}_{e,ij}^{\mathrm{CC}} - i\mathcal{M}_{e,ij}^{\mathrm{CC}}.
\end{equation}
From here, the procedure follows that of the nuclear scattering case, where we square the amplitudes, average over the initial state electrons spins and sum over final state helicities to find for Dirac neutrinos
\begin{equation}
        \left\langle\left|\mathcal{M}_e(\nu_{i,s}^D\to \nu_j^D)\right|^2\right\rangle = \frac{G_F^2}{4}\Big\{|U_{ei}|^2|U_{ej}|^2 \mathcal{T}_{1,\alpha\beta}\mathcal{T}_2^{\alpha\beta} + \delta_{ij}\left(\mathcal{T}_{\alpha\beta}\mathcal{T}_e^{\alpha\beta} - 2|U_{ei}|^2 \mathrm{Re}\left(\mathcal{T}\right)\right)\Big\},
\end{equation}
\begin{equation}
        \left\langle\left|\mathcal{M}_e(\bar\nu_{i,s}^D\to \bar\nu_j^D)\right|^2\right\rangle = \frac{G_F^2}{4}\Big\{|U_{ei}|^2|U_{ej}|^2 \widebar{\mathcal{T}}_{1,\alpha\beta}\widebar{\mathcal{T}}_2^{\alpha\beta} + \delta_{ij}\left(\widebar{\mathcal{T}}_{\alpha\beta}\mathcal{T}_e^{\alpha\beta} - 2|U_{ei}|^2 \mathrm{Re}\left(\widebar{\mathcal{T}}\right)\right)\Big\}
\end{equation}
where it is understood that we take $q_{\nu_i}\to q_{\nu_j}$ in the traces $\mathcal{T}_{\alpha\beta}$ and $\widebar{\mathcal{T}}_{\alpha\beta}$ where appropriate, and the two index traces are given by
\begin{align}
    \mathcal{T}_1^{\alpha\beta} &= \mathrm{Tr}\left[\gamma^{\alpha}(\slashed{q}_{e}+m_e)\gamma^{\beta}(1-\gamma^5)u(p_{\nu_i},s)\bar{u}(p_{\nu_i},s)(1+\gamma^5)\right],\\
    \mathcal{T}_2^{\alpha\beta} &= \mathrm{Tr}\left[\gamma^{\alpha}(\slashed{q}_{\nu_j}+m_{\nu_j})\gamma^\beta (1-\gamma^5)(\slashed{p}_e+m_e)(1+\gamma^5)\right],\\
    \widebar{\mathcal{T}}_1^{\alpha\beta} &= \mathrm{Tr}\left[\gamma^{\alpha}v(p_{\nu_i},s)\bar{v}(p_{\nu_i},s)\gamma^{\beta}(1-\gamma^5)(\slashed{p}_e+m_e)(1+\gamma^5)\right],\\
    \widebar{\mathcal{T}}_2^{\alpha\beta} &= \mathrm{Tr}\left[\gamma^{\alpha}(\slashed{q}_{e}+m_{e})\gamma^\beta (1-\gamma^5)(\slashed{q}_{\nu_j}-m_{\nu_j})(1+\gamma^5)\right],\\
    \mathcal{T}_e^{\alpha\beta} &= \mathrm{Tr}\left[\gamma^\alpha (\slashed{q}_e+m_e)\gamma^\beta (g_V^e-g_A^e\gamma^5)(\slashed{p}_e+m_e)(g_V^e+g_A^e\gamma^5)\right],
\end{align}
whilst the fully contracted traces arising from the interference terms are
\begin{align}
    \begin{split}
        \mathcal{T}&=\mathrm{Tr}\Big[\gamma^{\alpha}(\slashed{q}_{\nu_j}+m_{\nu_j})\gamma^\beta(1-\gamma^5)u(p_{\nu_i},s)\bar{u}(p_{\nu_i},s)(1+\gamma^5)\\
        &\times\gamma_\rho(\slashed{q}_e+m_e)\gamma_{\sigma}(g_V^e-g_A^e\gamma^5)(\slashed{p}_e+m_e)(1+\gamma^5)\Big],
    \end{split}\\
    \begin{split}
        \widebar{\mathcal{T}}&=\mathrm{Tr}\Big[\gamma^{\alpha}(\slashed{q}_{e}+m_{e})\gamma^\beta(g_V^e-g_A^e\gamma^5)(\slashed{p}_e+m_e)(1+\gamma^5)\\
        &\times\gamma_\alpha v(p_{\nu_i},s)\bar{v}(p_{\nu_i},s)\gamma_{\beta}(1-\gamma^5)(\slashed{q}_{\nu_j}-m_{\nu_j})(1+\gamma^5)\Big].
    \end{split}
\end{align}
Computing the averaged Majorana amplitude requires significantly more work, in particular for the interference term between the two charged current diagrams. After repeated applications of the Majorana condition, we find
\begin{equation}
    \begin{split}
    \left\langle\left|\mathcal{M}_e(\nu_{i,s}^M\to \nu_j^M)\right|^2\right\rangle &= \frac{G_F^2}{4}\Big\{ |U_{ei}|^2|U_{ej}|^2\left(\mathcal{T}_{1,\alpha\beta}\mathcal{T}_2^{\alpha\beta} + \widebar{\mathcal{T}}_{1,\alpha\beta}\widebar{\mathcal{T}}_2^{\alpha\beta} - 2\mathrm{Re}\left(\mathcal{U}_{1}\right) \right)\\
    &+ 4\delta_{ij}\left(\mathcal{U}_{\alpha\beta}\mathcal{T}_e^{\alpha\beta} + |U_{ei}|^2\mathrm{Re}\left(\mathcal{U}_{2} + \mathcal{U}_{3}\right) \right)\Big\},
    \end{split}
\end{equation}
where once more we have $q_{\nu_i}\to q_{\nu_j}$ in $\mathcal{U}_{\alpha\beta}$, and the fully contracted traces appearing in the Majorana amplitude are
\begin{align}
    \begin{split}
        \mathcal{U}_{1} &= \mathrm{Tr}\Big[\gamma^{\alpha}(1+\gamma^5)(\slashed{p}_e-m_e)(1-\gamma^5)\gamma^\beta u(p_{\nu_i},s)u(p_{\nu_i},s)\\
        &\times (1+\gamma^5)\gamma_{\alpha}(\slashed{q}_e + m_e) \gamma_\beta (1-\gamma^5)(\slashed{q}_{\nu_j} - m_{\nu_j})\Big],
    \end{split}\\
    \begin{split}
        \mathcal{U}_{2} &= \mathrm{Tr}\Big[\gamma^{\alpha}(\slashed{q}_e + m_e)\gamma^{\beta}(g_V^e-g_A^e\gamma^5)(\slashed{p}_e+m_e)(1+\gamma^5)\\
        &\times \gamma_{\alpha}v(p_{\nu_i},s)\bar{v}(p_{\nu_i},s)\gamma_\beta \gamma^5(\slashed{q}_{\nu_j}-m_{\nu_j})(1+\gamma^5)\Big],
    \end{split}\\
    \begin{split}
        \mathcal{U}_{3} &= \mathrm{Tr}\Big[\gamma^{\alpha}(\slashed{q}_{\nu_j}+m_{\nu_j})\gamma^{\beta}\gamma^5 u(p_{\nu_i},s)\bar{u}(p_{\nu_i},s)(1+\gamma^5)\\
        &\times \gamma_\alpha (\slashed{q}_e+m_e)\gamma_\beta (g_V^e-g_A^e\gamma^5)(\slashed{p}_e+m_e)(1+\gamma^5)\Big].
    \end{split}
\end{align}
As all momenta and neutrino mass splittings are small compared to the electron mass, we can use the analogous expression to~\eqref{eq:nuclearPhaseSpace} to compute the neutrino-electron scattering cross sections. This yields
\begin{align}
    \sigma_e(\nu_{i,s}^D \to \nu_j^D) &= \frac{G_F^2}{2\pi}E_{\nu_i}E_{\nu_j}(1-s\beta_{\nu_i})K^D_{ij},\\
    \sigma_e(\bar\nu_{i,s}^D \to \bar\nu_j^D) &= \frac{G_F^2}{2\pi}E_{\nu_i}E_{\nu_j}(1+s\beta_{\nu_i})K^D_{ij},\\
    \sigma_e(\nu_{i,s}^M \to \nu_j^M) &= \frac{G_F^2}{\pi}E_{\nu_i}E_{\nu_j}K^M_{ij}\label{eq:majCoheM},
\end{align}
to leading order in small quantities, where
\begin{equation}\hspace{-12pt}
    K^D_{ij}=4|U_{ei}|^2|U_{ej}|^2 + \delta_{ij}\left(3(g_A^e)^2 + (g_V^e)^2 + 2|U_{ei}|^2(3g_A^e + g_V^e)\right)\\
\end{equation}
\begin{equation}
    \begin{split}
        K^M_{ij}&=\left(4+2\sqrt{1-\beta_{\nu_i}^2}\sqrt{1-\beta_{\nu_j}^2}\right)|U_{ei}|^2|U_{ej}|^2 \\
        &+ \delta_{ij}\Big[\left(3(g_A^e)^2 + (g_V^e)^2 + 2|U_{ei}|^2(3g_A^e + g_V^e)\right) \\
        &+ \sqrt{1-\beta_{\nu_i}^2}\sqrt{1-\beta_{\nu_j}^2}\left(3(g_A^e)^2 - (g_V^e)^2 + 2|U_{ei}|^2(3g_A^e - g_V^e)\right)\Big].
    \end{split}
\end{equation}
Once more, the left helicity neutrino and right helicity antineutrino cross sections vanish in the relativistic limit. In the same limit, $\beta_{\nu_i}\to 1$, the sum of the scattering cross sections for Dirac neutrinos and antineutrinos is equal to the Majorana neutrino cross section, as both follow the same equation of motion in the massless regime. We also note that as all terms proportional the vector couplings, $Q_V$ and $g_V^e$, in the Majorana cross sections~\eqref{eq:majCohNM} and~\eqref{eq:majCoheM} tend to zero in the non-relativistic limit, $\beta_{\nu_i} \to 0$. This is a consequence of the pure axial neutral current vertex for Majorana neutrinos, which couples to the electron vector current at $\mathcal{O}(\beta_{\nu_i})$ and the axial current at $\mathcal{O}(1)$. 

\section{Coherent scattering structure factors}\label{sec:structureFactors}
In this section we will derive the $N_T^2$ enhancement in coherent scattering, and generalise it to the case where only partial coherence can be obtained. The cross section for a $2\to n$ scattering process is proportional to the scattering probability 
\begin{equation}\label{eq:1probability}
    \mathcal{P} = |{_{\mathrm{out}}}\langle \phi_1 \phi_2 \dots \phi_n|\phi_A \phi_B\rangle{_{\mathrm{in}}} |^2,
\end{equation}
where the labels in and out refer to states that are at time $t = \mp \infty$ respectively. We implicitly assume in~\eqref{eq:1probability} that the incoming state $B$ scatters off of a single scattering centre $A$, located at position $\vec{x}$. The states themselves are represented by wavepackets
\begin{equation}
    |\phi\rangle= \int d\Pi\,\phi(\vec{p})|\vec{p}\rangle, \quad d\Pi = \frac{d^3p}{(2\pi)^3} \frac{1}{\sqrt{2E_{p}}},
\end{equation}
 where $|\vec{p}\rangle$ is a one particle state of momentum $\vec{p}$, and $\phi(\vec{p})$ is the Fourier transformed spatial wavefunction, explicitly
\begin{equation}
    \phi(\vec{p}) = \int d^3x \,e^{-i\vec{p}\cdot \vec{x}} \phi(\vec{x}).
\end{equation}
The matrix element for a $2\to n$ scattering process is defined in terms of the states of definite momentum by
\begin{equation}\label{eq:1centreAmplitude}
    {_{\mathrm{out}}}\langle \vec{p}_1 \vec{p}_2 \dots \vec{p}_n|\vec{p}_A \vec{p}_B\rangle{_{\mathrm{in}}} =  (2\pi)^4 \delta^{(4)}(p_A + p_B - \sum_f p_f) \,i \mathcal{M}_{AB\to f},
\end{equation}
where the sum runs over the final states. 

\begin{figure}[tbp]
\centering 
\includegraphics[width=\textwidth]{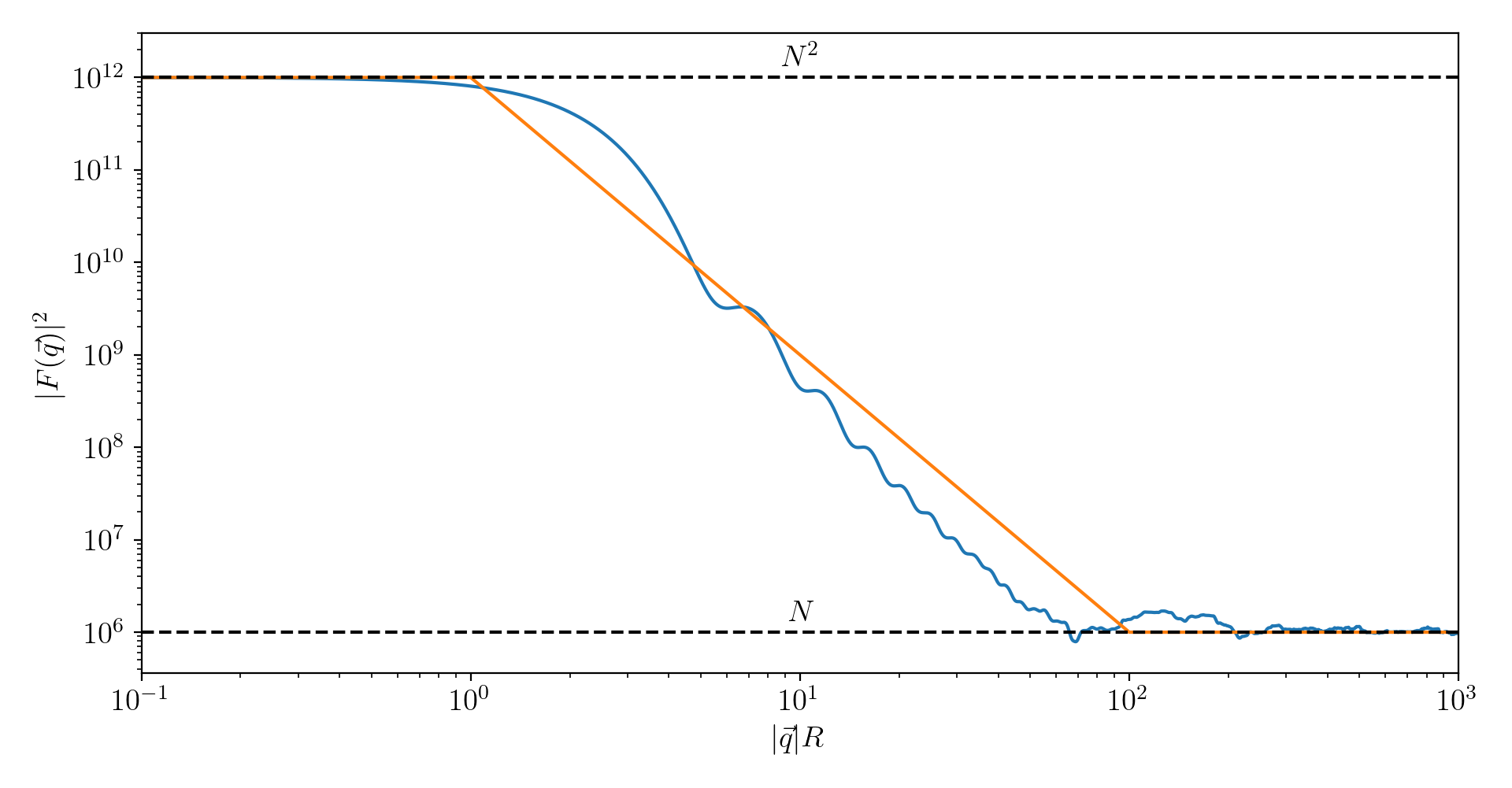}
\caption{\label{fig:structureFactor}Structure factor for a spherical target of radius $R$ containing $N = 10^6$ randomly placed scattering centres, smoothed using a moving average. For simplicity, we choose $\vec{q} = (0,0,|\vec{q}|)$. The true value~\eqref{eq:sFacTrue} is plotted in blue, whilst the approximation~\eqref{eq:sFacApprox} is shown in orange. The wiggles in the true value at $|\vec{q}|R > 1$ correspond to the Bragg condition, $\vec{q}\cdot (\vec{x}_i - \vec{x}_j) = n\pi$ for $n\in \mathbb{Z}$, being satisfied, where coherence is partially restored.}
\end{figure}

Now suppose that we instead have an incoming state $B$ with the potential to scatter from one of many centres $A_i$, each located at position $\vec{r}_i = \vec{x}-\vec{x}_i$ with the same momentum $\vec{p}_A$. The cross section will now be proportional to
\begin{equation}
    \begin{split}
    \mathcal{P}^C &= \left|\sum_i {_{\mathrm{out}}\langle} \phi_{1_i} \phi_2 \dots \phi_n|\phi_{A_i} \phi_B\rangle{_{\mathrm{in}}}\right |^2\\
    &= \Bigg|\sum_i \Bigg(\prod_{f=2}^n \int d\Pi_f \,\phi^*_f(\vec{p}_f) \int d\Pi_1 \,\phi^*_{1_i}(\vec{p}_1)\int d\Pi_A\,\phi_{A_i}(\vec{p}_A)\\
    &\times\int d\Pi_B\,\phi_B(\vec{p}_B) \,{_{\mathrm{out}}}\langle \vec{p}_1 \vec{p}_2 \dots \vec{p}_n|\vec{p}_A \vec{p}_B\rangle{_{\mathrm{in}}}\Bigg)\Bigg|^2,
    \end{split}
\end{equation}
where the superscript $C$ is used to denote the coherent quantities that we wish to compute, and we have assumed that each final state $\langle\phi_{1_i}|$ is produced at the site $\vec{r}_i$ with momentum $\vec{p_1}$. The Fourier transform of the field operator, $\phi_{i}(\vec{p})$ is
\begin{equation}
   \phi_i(\vec{p}) = \int d^3x\,e^{-i\vec{p}\cdot \vec{x}} \phi(\vec{x} - \vec{x}_i) = \phi(\vec{p}) e^{-i\vec{p}\cdot\vec{x}_i}, 
\end{equation}
such that the coherent scattering probability is given by
\begin{equation}\label{eq:coherentProbability}
    \mathcal{P}^C = \left|\prod_{f,i} \int d\Pi_f \,\phi^*_f(\vec{p}_f)\int d\Pi_i\,\phi_{i}(\vec{p}_i) \,{_{\mathrm{out}}}\langle \vec{p}_1 \vec{p}_2 \dots \vec{p}_n|\vec{p}_A \vec{p}_B\rangle{_{\mathrm{in}}}\, \sum_j e^{-i(\vec{p}_A - \vec{p}_1)\cdot \vec{x}_j}\right |^2,
\end{equation}
where the products over $i$ and $f$ now run over all initial and final states respectively, whilst the sum over $j$ runs over each scattering centre. We immediately see that going from $\mathcal{P} \to \mathcal{P}^C$ is equivalent to making the transformation
\begin{equation}
    {_{\mathrm{out}}}\langle \vec{p}_1 \vec{p}_2 \dots \vec{p}_n|\vec{p}_A \vec{p}_B\rangle{_{\mathrm{in}}} \to {_{\mathrm{out}}}\langle \vec{p}_1 \vec{p}_2 \dots \vec{p}_n|\vec{p}_A \vec{p}_B\rangle{_{\mathrm{in}}}\, \sum_i e^{-i(\vec{p}_A - \vec{p}_1)\cdot \vec{x}_i},
\end{equation}
allowing us to make the convenient definition of the coherent scattering amplitude
\begin{equation}
    i \mathcal{M}^C_{AB\to f} = i\mathcal{M}_{AB\to f} \,F(\vec{q}),
\end{equation}
where $F(\vec{q})$ is the structure factor, defined in terms of the momentum transfer $\vec{q} = \vec{p}_A - \vec{p}_1$ by 
\begin{equation}\label{eq:sFacTrue}
    F(\vec{q}) = \sum_i e^{-i\vec{q}\cdot \vec{x}_i}.
\end{equation}
The coherent scattering cross section will therefore be proportional to $\left|F(\vec{q})\right|^2$. As a result, in the low momentum regime $|\vec{q}|^{-1} \ll \langle |\vec{x}_i - \vec{x}_j|\rangle \simeq R$, where $R$ is the radius of the target, it follows that the scattering cross section is enhanced by a factor
\begin{equation}
    |F(\vec{0})|^2 = \sum_{i,j} e^{-i\vec{0}\cdot(\vec{x}_i - \vec{x}_j)} = N_T^2,
\end{equation}
for a system of $N_T$ scattering centres. We make the important note that the total scattering rate is not multiplied by an additional factor of $N_T$, as the incoming state scatters coherently on a single target containing $N_T$ centres, rather than $N_T$ targets incoherently. If coherence can only be maintained over a volume $|\vec{q}\,|^{-3} < R^3$, each containing $N_C < N_T$ centres, the coherent cross section will instead be proportional to $N_V N_C^2$, where $N_V = N_T/N_C$ is the number of coherent volumes in the target. The overall enhancement factor is therefore well approximated by
\begin{equation}\label{eq:sFacApprox}
    |F(\vec{q})|^2 \simeq N_T N_C = N_T \times 
    \begin{cases}
    N_T, \quad & (R|\vec{q}|)^3 \leq 1,\\[2pt]
    \displaystyle \left(\frac{1}{R|\vec{q}|}\right)^3 N_T, \quad & 1 < (R |\vec{q}|)^3 \leq N_T,\\[2pt]
    1,\quad & (R|\vec{q}|)^3 > N_T,
    \end{cases}
\end{equation}
which is computationally efficient at large $N_T$. We plot the structure factor in Figure~\ref{fig:structureFactor}, along with its approximation~\eqref{eq:sFacApprox}.

More generally, one can use the definition of the Dirac delta function to write
\begin{equation}
    (2\pi)^3 \delta^{(3)}(\vec{q}) = \int e^{-i\vec{q}\cdot \vec{x}} \,d^3x \simeq \frac{V}{N_T} \sum_i e^{-i\vec{q}\cdot \vec{x}_i},
\end{equation}
where the far right equality holds for large $N_T$, and $V$ is the volume of the target. Performing a double integral therefore yields
\begin{equation}\label{eq:sfacSqDelta}
    \begin{split}
    \int \int e^{-i\vec{q}\cdot (\vec{x}-\vec{x}')} \,d^3x\,d^3x' &\simeq \frac{V^2}{N_T^2} \sum_{i,j} e^{-i\vec{q}\cdot (\vec{x}_i-\vec{x}_j)}\\
    &= \left[(2\pi)^3 \delta^{(3)}(\vec{q})\right]\left[(2\pi)^3\delta^{(3)}(\vec{q})\right].
    \end{split}
\end{equation}
As the delta function is only defined under an integral, there is implicitly an integral over $\vec{q}$ being performed that effectively picks out the value $\delta(\vec{q} =0) = V/(2\pi)^3$ for the second delta function. Identifying the squared structure factor on right hand side of~\eqref{eq:sfacSqDelta} and rearranging, we therefore find that
\begin{equation}\label{eq:strucFacDelta}
   \left|F(\vec{q})\right|^2 \simeq \frac{N_T^2}{V}(2\pi)^3 \delta^{(3)}(\vec{q}).
\end{equation}
This is the form of the structure factor using to compute the RENP rate in Section~\ref{sec:renp}, which recovers the $N_T^2$ enhancement for $|\vec{q}| = 0$. For a comprehensive review of macroscopic coherent scattering, see~\cite{Akhmedov:2018wlf}.

\section{Meson decay}\label{sec:mesonDecay}
Here we derive the vector meson decay widths to neutrinos used in~\ref{sec:zburst}. The amplitude for the decay of vector meson $V\in\left\{\rho^0,\omega,\phi\right\}$ into two outgoing neutrinos with momenta $p_1$ and $p_2$ is\footnote{As a neutral current process, this must be mass diagonal.}
\begin{equation}\label{eq:mesonDecayAmplitude}
    i\mathcal{M}(V\to \nu_i \bar\nu_i)=-i\sqrt{2}G_F \bar{u}(p_1)\gamma^\mu(g_V^\nu - g_A^\nu \gamma^5)v(p_2)\left\langle0\right|j_\mu^Z\left|V(p_V)\right\rangle,
\end{equation}
where $g_V^x = T_3^x - 2Q_{\mathrm{EM}}^x \sin^2\theta_W$ and $g_A^x = T_3^x$ are the vector and axial-vector couplings of species $x$, given in terms of their weak isospin $T_3$ and electric charge $Q_{\mathrm{EM}}$, whilst $j_\mu^Z$ is the weak neutral current. 

We note that the hadronic matrix element appearing in~\eqref{eq:mesonDecayAmplitude} must be gauge invariant since it is just a number. Since the external state $\left|V\right\rangle$ transforms as a vector, the neutral current $j_\mu^Z$ must transform in the same way to leave the amplitude invariant. As such, $j_\mu^Z$ can only contain quark vector currents, and so we define in analogy with~\cite{Bharucha:2015bzk}
\begin{equation}\label{eq:vecCurrent}
    j_\mu^Z = g_V^s j_\mu^\phi + \frac{g_V^u + g_V^d}{\sqrt{2}} j_\mu^\omega + \frac{g_V^u - g_V^d}{\sqrt{2}}j_\mu^{\rho^0},
\end{equation}
where the individual vector currents are written in terms of the constituent quark fields as
\begin{equation}
    j^{\omega,\rho^0}_\mu = \frac{1}{\sqrt{2}}\left(\bar u \gamma_\mu u \pm \bar d \gamma_\mu d\right), \quad j^{\phi}_\mu = \bar s \gamma_\mu s,
\end{equation}
with the $+(-)$ sign chosen for the $\omega\left(\rho^0\right)$ meson. Due to mixing between the three mesons, amplitudes of the form $\left\langle V \right|j_\mu^{V'}\left|0\right\rangle$ with $V \neq V'$ are also non-zero. This is discussed at length in~\cite{Bharucha:2015bzk}; here we simply quote the results for each of the three mesons in terms of the decay constants $f_V^q$
\begin{equation}
    \left\langle\rho^0\right|j_\mu^Z\left|0\right\rangle = \frac{\varepsilon_\mu m_{\rho^0}}{\sqrt{2}}\left(g_V^u f_{\rho^0}^{u} - g_V^d f_{\rho^0}^{d}\right) \equiv \varepsilon_\mu m_{\rho^0} g_{\rho^0}^\mathrm{eff} f_{\rho^0}^{\mathrm{eff}},
\end{equation}
\begin{equation}
    \left\langle\omega\right|j_\mu^Z\left|0\right\rangle = \frac{\varepsilon_\mu m_{\omega}}{\sqrt{2}}\left[g_V^u f_\omega^u + g_V^d f_\omega^d - \eta^{\omega\phi}g_V^s\left(\frac{f_\omega^u + f_\omega^d}{\sqrt{2}}\right)\right]\equiv \varepsilon_\mu m_{\omega} g_{\omega}^\mathrm{eff} f_{\omega}^{\mathrm{eff}},
\end{equation}
\begin{equation}
    \left\langle\phi\right|j_\mu^Z\left|0\right\rangle = \varepsilon_\mu m_{\phi}f_\phi\left[g_V^s + \eta^{\omega\phi}\left(\frac{g_V^u + g_V^d}{\sqrt{2}}\right)\right]\equiv \varepsilon_\mu m_{\phi} g_{\phi}^\mathrm{eff} f_{\phi}^{\mathrm{eff}},
\end{equation}
where $\eta^{\omega\phi}\simeq 0.05$ accounts for $\omega-\phi$ mixing and $\varepsilon_\mu$ is the polarisation vector of the decaying meson. With these definitions and the values of $f_V^q$ given in appendix C of~\cite{Bharucha:2015bzk}, we find
\begin{equation}
    g_{\rho^0}^\mathrm{eff} f_{\rho^0}^{\mathrm{eff}} = 81.3\,\mathrm{MeV} , \quad g_{\omega}^\mathrm{eff} f_{\omega}^{\mathrm{eff}} = -19.8\,\mathrm{MeV}, \quad g_{\phi}^\mathrm{eff} f_{\phi}^{\mathrm{eff}} = -81.9\,\mathrm{MeV}.
\end{equation}
Setting $g_V^\nu = g_A^\nu = 1/2$ in~\eqref{eq:mesonDecayAmplitude}, we find the spin and polarisation averaged meson decay amplitude
\begin{equation}
    \left\langle\left|\mathcal{M}(V\to \nu_i \bar\nu_i)\right|^2\right\rangle = \frac{4 G_F^2}{3} m_V^4 \left(g_V^{\mathrm{eff}} f_V^{\mathrm{eff}}\right)^2+ \mathcal{O}\!\left(\frac{m_{\nu_i}^2}{m_V^2}\right), 
\end{equation}
such that the decay width for the vector meson to decay to a pair of neutrinos is
\begin{equation}
    \Gamma(V \to \nu_i \bar\nu_i)  = \frac{G_F^2}{12\pi} m_V^3 \left(g_V^{\mathrm{eff}} f_V^{\mathrm{eff}}\right)^2. 
\end{equation}
The result is identical to leading order in $m_{\nu_i}/m_V$ for both Dirac and Majorana neutrinos. 

% The bibliography will probably be heavily edited during typesetting.
% We'll parse it and, using the arxiv number or the journal data, will
% query inspire, trying to verify the data (this will probalby spot
% eventual typos) and retrive the document DOI and eventual errata.
% We however suggest to always provide author, title and journal data:
% in short all the informations that clearly identify a document.

\end{document}